\documentclass[twoside,12pt]{article}
\usepackage{epsfig}
\usepackage{url}
\usepackage{multirow}
\usepackage[bookmarks, pagebackref=false]{hyperref}
\usepackage{color}
	\definecolor{rossoCP3}{cmyk}{0,.88,.77,.40}
		\definecolor{graa}{rgb}{0.8,0.8,0.8}
		\definecolor{blaa}{rgb}{0.2,0.2,0.6}
		\definecolor{gron}{RGB}{0,150,0}
		\hypersetup{
			colorlinks,
			bookmarksopen,
			bookmarksnumbered,
			citecolor=blaa, 		
			linkcolor=black, 
			urlcolor=blaa,			
			}

\newcommand{\eqref}[1]{(\ref{#1})}

\def\Journal#1#2#3#4{{#1} {#2} (#4) #3 }

\def\NC{\em Nuovo Cimento}

\def\NPB{{\em Nucl. Phys.} B}

\def\PLB{{\em Phys. Lett.} B}

\def\PRL{\em Phys. Rev. Lett.}
\def\PREV{\em Phys. Rev.}
\def\PREP{\em Phys. Rep.}

\def\PRD{{\em Phys. Rev.} D}

\def\ZPC{{\em Z. Phys.} C}
\def\ZPA{{\em Z. Phys.} A}
\def\ANNP{\em Ann. Phys. (N.Y.)}

\def\HPA{\em Helv. Phys. Acta}
\def\DANS{\em Dok. Akad. Nauk SSSR}
\def\CMP{\em Commun. Math. Phys.}
\def\JPG{\em J. Phys. G}

\def\CPC{\em Comput. Phys. Commun.}
\def\JHEP{\em JHEP}

\def\PPNP{\em Prog. Part. Nucl. Phys.}
\def\EPJC{{\em Eur. Phys. J.} C}
\def\TMP{\em Theor. Math. Phys.}

\def\JETPL{\em JETP Lett.}

\def\SJNP{\em Sov. J. Nucl. Phys.}
\def\JETP{\em Sov. Phys. JETP}
\def\PR{\em Phys. Rev.}

\def\JCP{\em J. Comp. Phys.}
\def\ACM{\em Adv. Comput. Math.}
\def\ZP0{\em Zeit. Phys.}
\def\RPP{\em Rept. Prog. Phys.}
\def\CHNPC{\em Chin. Phys. C}
\def\FP{\em Front. Phys.}
\def\FPh{\em Fortsch. Phys.}
\def\JPA{\em J. Phys. A}

\newcommand{\ree}{R_{e^+e^-}}
\newcommand{\rtau}{R_{\tau}}
\newcommand{\hbb}{\Gamma_{H\to b\bar{b}}}

\begin{document}

\title{ \vspace{1cm} The QCD Renormalization Group Equation and the Elimination of Fixed-Order Scheme-and-Scale Ambiguities Using the Principle of Maximum Conformality}

\author{Xing-Gang Wu$^{1}$\footnote{email: wuxg@cqu.edu.cn}, Jian-Ming Shen$^{1,2}$\footnote{email: cqusjm@cqu.edu.cn}, Bo-Lun Du$^{1}$\footnote{email: dblcqu@cqu.edu.cn}, Xu-Dong Huang$^{1}$\footnote{email: hxud@cqu.edu.cn}, \\ Sheng-Quan Wang$^3$\footnote{email: sqwang@cqu.edu.cn} and Stanley J. Brodsky$^3$\footnote{email: sjbth@slac.stanford.edu} \\
$^1$ Department of Physics, Chongqing University, Chongqing 401331, P.R. China \\
$^2$ School of Physics and Electronics, Hunan University, \\
Changsha 410082, P.R. China \\
$^3$ SLAC National Accelerator Laboratory, Stanford University, \\
Stanford, California 94039, USA}

\maketitle

\begin{abstract}

The conventional scale setting approach to fixed-order perturbative QCD (pQCD) predictions is based on a guessed renormalization scale, usually taking as the one to eliminate the large log-terms of the pQCD series, together with an arbitrary range to estimate its uncertainty. This {\it ad hoc} assignment of the renormalization scale causes the coefficients of the QCD running coupling at each perturbative order to be strongly dependent on the choices of both the renormalization scale and the renormalization scheme, which leads to conventional renormalization scheme-and-scale ambiguities. However, such ambiguities are not necessary, since as a basic requirement of renormalization group invariance (RGI), any physical observable must be independent of the choices of both the renormalization scheme and the renormalization scale. In fact, if one uses the {\it Principle of Maximum Conformality} (PMC) to fix the renormalization scale, the coefficients of the pQCD series match the series of conformal theory, and they are thus scheme independent. The PMC predictions also eliminate the divergent renormalon contributions, leading to a better convergence property. It has been found that the elimination of the scale and scheme ambiguities at all orders relies heavily on how precisely we know the analytic form of the QCD running coupling $\alpha_s$. In this review, we summarize the known properties of the QCD running coupling and its recent progresses, especially for its behavior within the asymptotic region. Conventional schemes for defining the QCD running coupling suffer from a complex and scheme-dependent renormalization group equation (RGE), or the $\beta$-function, which is usually solved perturbatively at high orders due to the entanglement of the scheme-running and scale-running behaviors. These complications lead to residual scheme dependence even after applying the PMC, which however can be avoided by using a $C$-scheme coupling $\hat\alpha_s$, whose scheme-and-scale running behaviors are governed by the same scheme-independent RGE. As a result, an analytic solution for the running coupling can be achieved at any fixed order. Using the $C$-scheme coupling, a demonstration that the PMC prediction is scheme-independent to all-orders for any renormalization schemes can be achieved. Given a measurement which sets the magnitude of the QCD running coupling at a specific scale such as $M_Z$, the resulting pQCD predictions, after applying the single-scale PMC, become completely independent of the choice of the renormalization scheme and the initial renormalization scale at any fixed-order, thus satisfying all of the conditions of RGI. An improved pQCD convergence provides an opportunity of using the resummation procedures such as the Pad\'e approximation (PA) approach to predict higher-order terms and thus to increase the precision, reliability and predictive power of pQCD theory. In this review, we also summarize the current progress on the PMC and some of its typical applications, showing to what degree the conventional renormalization scheme-and-scale ambiguities can be eliminated after applying the PMC. We also compare the PA approach for the conventional scale-dependent pQCD series and the PMC scale-independent conformal series. We observe that by using the conformal series, the PA approach can provide a more reliable estimate of the magnitude of the uncalculated terms. And if the conformal series for an observable has been calculated up to $n_{\rm th}$-order level, then the $[N/M]=[0/n-1]$-type PA series provides an important estimate for the higher-order terms.

\end{abstract}

\begin{description}
\item[PACS numbers] 12.38.Aw, 12.38.Bx, 11.10.Gh, 11.10.Hi
\item[Key words] perturbative QCD calculations, renormalization, principle of maximum conformality
\end{description}


\tableofcontents

\section{Introduction}
\label{sec:intro}

Quantum Chromodynamics (QCD) is the non-Abelian gauge field theory that describes the strong interactions of colored quarks and gluons, and it is the SU(3)-color component of the Standard Model. Due to its asymptotic freedom property~\cite{Gross:1973id, Politzer:1973fx}, the QCD running coupling $\alpha_s\sim {\cal O}(1)$ becomes numerically small at short distances, allowing perturbative calculations of cross sections for high momentum transfer physical processes. Renormalization was first developed in quantum electrodynamics (QED) and then applied to QCD to make sense of infinite integrals emerged in perturbation theory. The relevance of perturbative QCD (pQCD) for the description of the experimental data often relies on our ability to go beyond the one-loop approximation. Due to the complexity of the multi-loop calculations, perturbative calculations are only known at fixed-order, especially when the high-energy processes involving hadrons in which the renormalization and factorization effects are entangled with each other.

The fixed-order predictions for observables in QCD are usually assumed to suffer from an uncertainty in fixing the renormalization scale~\cite{Grunberg:1980ja, Grunberg:1982fw, Stevenson:1980du, Stevenson:1981vj, Stevenson:1982wn, Stevenson:1982qw, Brodsky:1982gc}. This ambiguity in making fixed-order predictions occurs because one usually assumes an arbitrary renormalization scale, (representing a typical momentum flow of the process which is assumed to be the effective virtuality of the strong interaction in that process), together with an arbitrary range to ascertain its uncertainty. This {\it ad hoc} assignment of the renormalization scale, however, causes the coefficients of the QCD running coupling at each perturbative order to be strongly dependent on the choice of the renormalization scale as well as the renormalization scheme. Moreover, we do not know how wide a range the renormalization scale and scheme parameters should vary in order to achieve reasonable predictions of their errors. In fact the error analysis assuming conventional procedure can be quite arbitrary and unreliable. It is usually assumed that at sufficiently high order, one will eventually achieve reliable predictions and minimal dependence on the guessed renormalization scale for global quantities such as a total cross-section or a total decay width. However, such a small scale-dependence for a global quantity could be caused by accidental cancelations among different orders; the scale uncertainty for the contributions at each order could still be very large. One then cannot decide whether the poor pQCD convergence is the intrinsic property of pQCD series, or whether it is simply due to improper choice of scale.

The QCD running coupling sets the strength of the interactions involving quarks and gluons, which is finite when the ultraviolet (UV) divergences are removed by renormalization. The running coupling depends on the scale at which one observes it, and the scale dependence of the QCD running coupling is governed by the {\it renormalization group equation} (RGE), or equivalently the $\beta$-function. The first formulation of the RGE was given by Stueckelberg and Petermann~\cite{Stueckelberg:1953dz, Peterman}, Gell-Mann and Low~\cite{GellMann:1954fq}, and Bogoliubov and Shirkov~\cite{Bogoliubov}. The conventional RGE is scheme-dependent due to the scheme-dependent $\{\beta_{i\ge2}\}$-functions. Thus if the $\{\beta_{i\ge2}\}$-terms of the pQCD series have large dependence on the scheme choice, the perturbative predictions based on some schemes could be unreliable~\footnote{As an explicit example, the next-to-leading order terms give unreasonably large contributions to the Pomeron intercept under the $\overline{\rm MS}$-scheme, which is however can be greatly suppressed by using the momentum space subtraction scheme and a reasonable prediction on the Pomeron intercept can be obtained after applying the BLM or PMC~\cite{Brodsky:1998kn, Zheng:2013uja, Hentschinski:2012kr, Caporale:2015uva}.}; the large expansion coefficients could make the truncation of the perturbative series useless. The resulting uncertainties thus would not be minimized by including more higher-order terms. Even worse, it is known that in general the pQCD series will suffer from the divergent renormalon contributions which grow as $\alpha_s^n \beta^n_0 n !$~\cite{Beneke:1994qe, Neubert:1994vb, Beneke:1998ui, Gardi:2001wg}~\footnote{The high-order $\beta_i$-term satisfies the approximation, $\beta_i \approx \beta_0^{i+1}$, which could be used for estimating the $\beta_0$-powers at each perturbative order.}. Thus even if a pQCD prediction based on a guessed scale agrees with a measurement, one cannot be certain that it is a reliable, accurate representation of the theory.

As a guiding principle, a valid perturbative prediction for any physical observable must be independent of the initial choices of the renormalization scale and the renormalization scheme; this is the central property of {\it renormalization group invariance} (RGI)~\cite{Stueckelberg:1953dz, Callan:1970yg, Symanzik:1970rt, peter2, Peterman:1978tb}. After applying the standard regularization and renormalization procedures, one will obtain a finite pQCD prediction free of UV divergences as well as scheme and scale ambiguity. Thus a remaining primary goal for testing pQCD reliably is how to set the renormalization scale such that one obtains accurate fixed-order predictions with maximum precision while satisfying the principle of RGI.

\begin{figure}[htb]
\epsfysize=9.0cm
\begin{center}
\begin{minipage}[t]{10 cm}
\epsfig{file=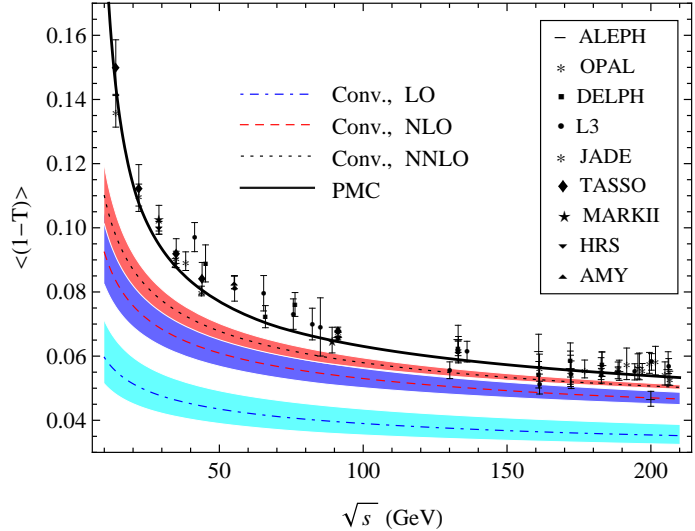,scale=1.0}
\end{minipage}
\begin{minipage}[t]{16.5 cm}
\caption{The thrust mean value $\langle (1-T) \rangle$ for three-jet events versus the center-of-mass energy $\sqrt{s}$ using the conventional (Conv.) and PMC scale settings~\cite{Wang:2019ljl}. The dot-dashed, dashed and dotted lines are the conventional results at LO, NLO and NNLO, respectively. The solid line is the PMC result. The PMC prediction eliminates the renormalization scale $\mu$ uncertainty. The bands for theoretical predictions are obtained by varying $\mu\in[\sqrt{s}/2,2\sqrt{s}]$. The experimental data points are taken from the ALEPH, DELPH, OPAL, L3, JADE, TASSO, MARKII, HRS and AMY experiments~\cite{Heister:2003aj, Abdallah:2003xz, Abbiendi:2004qz, Achard:2004sv, MovillaFernandez:1997fr, Pahl:2007zz, Braunschweig:1990yd, Petersen:1987bq, Bender:1984fp, Li:1989sn}. \label{convPMCmoment}}
\end{minipage}
\end{center}
\end{figure}

A review of various renormalization scale setting approaches which have been suggested in the literature can be found in Ref.\cite{Wu:2013ei}. It is noted that the solution of those ambiguities depends on how well we know the QCD running coupling and its renormalized value in the observables. In contrast to other scale setting approaches, such as the RG-improved effective coupling method (FAC)~\cite{Grunberg:1980ja, Grunberg:1982fw} and the {\it Principle of Minimum Sensitivity} (PMS)~\cite{Stevenson:1980du, Stevenson:1981vj, Stevenson:1982wn, Stevenson:1982qw}, the {\it Principle of Maximum Conformality} (PMC)~\cite{Brodsky:2011ta, Brodsky:2012sz, Brodsky:2011ig, Brodsky:2012rj, Mojaza:2012mf, Brodsky:2013vpa} determines the value of the renormalization scale of $\alpha_s$ consistent with all of the properties of RGE. The FAC and PMS are programmed to directly deal with the nature of the perturbative series, whose optimal scales are fixed by treating the total corrections as a whole. More explicitly, the FAC improves the perturbative series by requiring all higher-order terms vanish, or in another words, all higher-order terms are resummed into the leading-order $\alpha_s$ terms; and the PMS requires the fixed-order series satisfy the RGI at the optimal renormalization point. In distinction, the PMC is programmed to fix the behavior of the running coupling by absorbing only those contributions that are related to the renormalization of the running coupling. Because the scale setting methods, such as FAC, PMS and PMC, have quite different starting points, they can give strikingly different results in practical applications. As an example, Kramer and Lampe~\cite{Kramer:1987dd, Kramer:1990zt} have analyzed the application of various approaches for the prediction of jet production fractions in $e^+e^-$ annihilation in pQCD. They have showed that the predicted scales for FAC and PMS rise without bound at small values for the jet fraction, indicating the FAC and PMS do not have the right physical behavior (or correct momentum flow) in the limit of small jet energy, since they have summed physics into the running coupling not associated with renormalization. On the other hand, the BLM / PMC scale has the correct physical behavior~\cite{Brodsky:2011ig, Kramer:1987dd, Kramer:1990zt}. Lately, it has been found that such correct physical behavior for three-jet is important to achieve a reasonable thrust mean value $\langle(1-T)\rangle$~\cite{Wang:2019ljl, Gehrmann:2014uva}. Figure~\ref{convPMCmoment} shows that the PMC prediction of $\langle(1-T)\rangle$ versus the center-of-mass energy, which is greatly increased compared to the conventional predictions. The experimental data issued by the ALEPH, DELPH, OPAL, L3, JADE, TASSO, MARKII, HRS and AMY experiments~\cite{Heister:2003aj, Abdallah:2003xz, Abbiendi:2004qz, Achard:2004sv, MovillaFernandez:1997fr, Pahl:2007zz, Braunschweig:1990yd, Petersen:1987bq, Bender:1984fp, Li:1989sn} have also been presented as a comparison. Figure~\ref{convPMCmoment} suggests that the substantial deviation between the conventional predictions and the experimental data is caused by improper choice of the renormalization scale and the PMC provides a rigorous explanation for the experimental data.

The PMC provides the underlying principle for the well-known Brodsky-Lepage-Mackenzie (BLM) method~\cite{Brodsky:1982gc} and provides a rule for generalizing the BLM scales up to all orders~\footnote{Another suggestion of extending BLM to all orders has also been suggested in the literature, i.e. the seBLM approach~\cite{Mikhailov:2004iq, Kataev:2010du}, whose purpose is however to improve the pQCD convergence, but not to solve the conventional renormalization scheme-and-scale ambiguities. Moreover, its prediction is based on approximate solution of RGE~\cite{Shen:2016dnq} and obtained without distinguishing whether the $n_f$-terms are pertained to RGE or not, thus the seBLM results are inaccurate and sometimes will meet the very small scale problem~\cite{Wang:2013bla, Ma:2015dxa}.}. The BLM method is to deal with the $n_f$-power series, and all the features previously observed in the BLM literature are also adaptable to PMC with or without proper transformations; Most importantly, one needs to confirm that whether the $n_f$-terms have been correctly treated in previous BLM predictions, i.e. only those $n_f$-terms that are related to RGE should be adopted for setting the renormalization scale. The PMC shifts all the non-conformal $\{\beta_i\}$-terms into the running coupling at all orders, and it reduces to the standard scale setting procedure of Gell-Mann and Low (GM-L)~\cite{GellMann:1954fq} in the limit of small number of colors ($N_c \to 0$), i.e. the QED Abelian limit~\cite{Brodsky:1997jk}. Since the resultant pQCD series is identical to the series of a conformal theory with $\beta=0$~\cite{Mack:1969rr, Callan:1970ze, Gross:1970tb, Polyakov:1970}~\footnote{In recent years, there are some more discussions on the scheme transformation/invariant properties near or at the infrared fixed point $\beta=0$, cf. Refs.\cite{Ryttov:2012ur, Ryttov:2012nt, Shrock:2014qea, Gracey:2018oym}.}, the PMC prediction has the essential feature that it is scheme-independent at every finite order. After applying the PMC, one can obtain ``commensurate scale relations" among different approximants of the pQCD observables under different schemes~\cite{Brodsky:2013vpa, Brodsky:1994eh}, which also ensure the scheme independence of the PMC predictions.

One can demonstrate that the PMC prediction satisfies the self-consistency conditions of the renormalization group, such as reflectivity, symmetry and transitivity~\cite{Brodsky:2012ms}. The resulting PMC predictions thus satisfy all of the basic requirements of RGI. The transition scale between the perturbative and nonperturbative domains can also be determined by using the PMC~\cite{Deur:2014qfa, Deur:2016cxb, Deur:2016tte, Deur:2017cvd}, thus providing a procedure for setting the ``factorization" scale for pQCD evolution. It should be emphasized that the running coupling resums all of the $\{\beta_i\}$-terms by using the PMC, which naturally leads to a more convergent and renormalon-free pQCD series.

The PMC scales are achieved by applying the RGE of the QCD running coupling, i.e. by recursively applying the RGE, one can establish a perturbative $\beta$-pattern at each order in a pQCD expansion. For example, the usual scale-displacement relation for the running couplings at two different scales $Q_1$ and $Q_2$ can be deduced from the RGE, which reads
\begin{eqnarray}
a_{Q_2} &=& a_{Q_1}- \beta_{0} \ln\left(\frac{Q_2^{2}} {Q_1^2}\right) a_{Q_1}^2 +\left[\beta^2_{0} \ln^2 \left(\frac{Q_2^{2}}{Q_1^2}\right) -\beta_{1} \ln \left(\frac{Q_2^{2}} {Q_1^2}\right)\right] a_{Q_1}^3 \nonumber\\
&& + \left[- \beta_0^3\ln^3 \left(\frac{Q_2^{2}}{Q_1^2}\right) + \frac{5}{2}{\beta_0}{\beta_1} \ln^2 \left(\frac{Q_2^{2}}{Q_1^2}\right) - \beta_2 \ln \left(\frac{Q_2^{2}}{Q_1^2}\right) \right]a_{Q_1}^4 + \left[\beta_0^4 \ln^4 \left(\frac{Q_2^{2}}{Q_1^2}\right) \right. \nonumber\\
&& \left. - \frac{13}{3}{\beta_0^2}{\beta_1} \ln^3 \left(\frac{Q_2^{2}}{Q_1^2}\right) + \frac{3}{2}{\beta_1^2} \ln^2 \left(\frac{Q_2^{2}}{Q_1^2}\right) + 3{\beta_2}{\beta_0} \ln^2\left(\frac{Q_2^{2}}{Q_1^2}\right) - \beta_3 \ln \left(\frac{Q_2^{2}}{Q_1^2}\right) \right]a_{Q_1}^5 + \cdots, \label{scaledis}
\end{eqnarray}
where $a_{Q_i}= \alpha_s(Q_i)/\pi$, the functions $\beta_0, \beta_1, \cdots$ are generally scheme dependent, which correspond to the one-loop, two-loop, $\cdots$, contributions to the RGE, respectively. The PMC utilizes this perturbative $\beta$-pattern to systematically set the scale of the running coupling at each order in a pQCD expansion. The coefficients of the $\{\beta_i\}$-terms in the $\beta$-pattern can be identified by reconstructing the ``degeneracy relations"~\cite{Mojaza:2012mf, Brodsky:2013vpa} among different orders. The degeneracy relations, which underly the conformal features of the resultant pQCD series by applying the PMC, are general properties of a non-Abelian gauge theory~\cite{Bi:2015wea}. The PMC prediction achieved via this way resembles a skeleton-like expansion~\cite{Lu:1991yu, Lu:1991qr}. The resulting PMC scales reflect the virtuality of the amplitudes relevant to each order, which are physical in the sense that they reflect the virtuality of the gluon propagators at a given order, as well as setting the effective number ($n_f$) of active quark flavors. The correct momentum flow for the process involving three-gluon vertex can be achieved by properly dividing the total amplitude into gauge-invariant amplitudes~\cite{Binger:2006sj}. Specific values for the PMC scales are computed as a perturbative expansion, so they have small uncertainties which can vary order-by-order. It has been found that the PMC scales and the resulting fixed-order PMC predictions are to high accuracy independent of the initial choice of renormalization scale, e.g. the residual uncertainties due to unknown higher-order terms are negligibly small because of the combined suppression effect from both the exponential suppression and the $\alpha_s$-suppression~\cite{Mojaza:2012mf, Brodsky:2013vpa}.

Following the standard PMC procedures, different scales generally appear at each order, which is called as the PMC multi-scale approach and requires considerable theoretical analysis. To make the PMC scale setting procedures simpler and more easily to be automatized, a single-scale approach (PMC-s), which achieves many of the same PMC goals, has been suggested in Ref.\cite{Shen:2017pdu}. This method effectively replaces the individual PMC scale at each order by a single (effective) scale in the sense of a mean value theorem, e.g. it can be regarded as a weighted average of the PMC scales at each order derived under PMC multi-scale approach. The single ``PMC-s" scale shows stability and convergence with increasing order in pQCD, as observed by the $e^+e^-$ annihilation cross-section ratio $R_{e^+e^-}$ and the Higgs decay-width $\Gamma(H \to b \bar{b})$, up to four-loop level. Moreover, its predictions are explicitly independent of the choice of the initial renormalization scale. Thus the PMC-s approach, which requires a simpler analysis, can be adopted as a reliable substitute for the PMC multi-scale approach, especially when one does not need detailed information at each order.

The PMC prediction depends heavily on the properties of the RGE. A more precise solution for the RGE leads to a more precise determination of the running behavior of the QCD running coupling, and thus a more accurate determination of the optimal momentum flow (or simply, the optimal scale) of the process. When deriving a pQCD prediction, one has to follow the standard renormalization procedure of quantum field theory. A specific renormalization scheme need to be chosen, this defines the QCD coupling constant. The conventional RGE which determines the scale-running behavior of the QCD running coupling is thus scheme dependent. The QCD running coupling can be ``adiabatically" and continuously evolved not only in scales, but also in the choices of renormalization scheme by incorporating scheme-running equations, forming the so-called {\it extended RGEs}~\cite{Stevenson:1980du, Stevenson:1981vj, Stevenson:1982wn, Stevenson:1982qw}. Since along the evolution trajectory of the extended RGEs, no dissimilar scales/schemes are involved, reliable pQCD predictions can be achieved in this way~\cite{Lu:1992nt}. The extended RGEs provide a convenient way for estimating both the scale- and scheme- dependence of the pQCD prediction for a physical observable. The scheme-running equations can be solved perturbatively, which can also be used to estimate how the uncalculated higher-order terms contribute to the final result~\cite{Lu:1992nt}. The solution of the scale-running equation can also be solved perturbatively via an iterative process, which is equivalent to the standard analysis~\cite{Bardeen:1978yd, Furmanski:1981cw, Chetyrkin:1997sg} by using a proper integration constant~\cite{Brodsky:2011ta}.

It has been found that by utilizing the {\it $C$-scheme} coupling suggested by Boito, Jamin and Miravitllas~\cite{Boito:2016pwf}, its scheme-and-scale running behaviors are governed by a single RGE which is free of scheme-dependent $\{\beta_{i\ge2}\}$-functions. Using the $C$-scheme coupling, it is convenient to discuss the scheme variation of a pQCD approximant~\cite{Boito:2016pwf, Caprini:2018agy}. It is noted that the solution of the RGE of the $C$-scheme coupling can be greatly simplified and an analytic solution can be achieved~\cite{Wu:2018cmb}. In practice, the value of the parameter $C$ can be chosen to match any conventional renormalization scheme. Ref.\cite{Wu:2018cmb} also shows that the scheme-independent RGE for the $C$-scheme coupling leads to scheme-independent pQCD predictions for any physical observables; i.e., by using the $C$-scheme coupling, a strict demonstration of the scheme-independence of PMC prediction to all-orders for any renormalization schemes can be achieved. Thus, by combining the $C$-scheme coupling together with the PMC-s approach, the resulting predictions become completely independent of the choice of the renormalization scheme and the initial renormalization scale, satisfying all of the conditions of RGI. This approach thus systematically eliminates the scheme and scale ambiguities of pQCD predictions, greatly improving the precision of tests of the Standard Model. Furthermore, since the perturbative coefficients obtained using the PMC-s are identical to those of a conformal theory, one can derive all-orders ``commensurate scale relations"~\cite{Brodsky:2013vpa, Shen:2016dnq, Brodsky:1994eh} between physical observables evaluated at specific relative scales. An example is the ``Generalized Crewther Relation"~\cite{Broadhurst:1993ru, Brodsky:1995tb, Crewther:1997ux}, which shows that the product of $R_{e^+ e^-} (s)$ times the integral over the spin-dependent structure functions $g_1(x,Q^2)$ which enters the Bjorken sum rule at a specific value of $Q^2/s$ has no leading-twist radiative QCD corrections at all orders.

The predictive power of pQCD depends on two important issues: how to eliminate the renormalization scheme-and-scale ambiguities at fixed order, and how to reliably estimate the contributions of unknown higher-order terms using information from the known pQCD series. Since the divergent renormalon series does not appear in the conformal $\beta=0$ perturbative series generated by the PMC, there is an opportunity to use some resummation procedures such as the Pad\'e method to predict higher-order terms and thus increase the precision and reliability of pQCD predictions. The Pad\'e approximation (PA) approach provides a systematic procedure for promoting a finite Taylor series to an analytic function~\cite{Basdevant:1972fe, Samuel:1995jc, Samuel:1992qg}. In particular, the PA approach can be used to estimate the $(n+1)_{\rm th}$-order coefficient by incorporating all known coefficients up to order $n$. Some applications of the PA approach, together with alternatives to the PA approach, have been discussed in the literature~\cite{Brodsky:1997vq, Gardi:1996iq, Ellis:1997sb, Burrows:1996dk, Ellis:1996zn, Jack:1997jn, Boito:2018rwt}. As stated in Ref.\cite{Brodsky:1997vq}, studies of higher-order perturbative QCD diagrams are often made by first decomposing them in a skeleton expansion, in which each term contains chains of vacuum polarization bubbles inserted in virtual-gluon propagators. Then, they can be studied in the BLM/PMC approach, which seeks the optimal scale for evaluating each term in the skeleton expansion. The last step, the sum over skeleton graphs, is then similar to the summation of perturbative contributions for a corresponding theory with a conformal theory. It has been shown that the next-to-leading order BLM/PMC procedure is equal to $[0/1]$-type PA~\cite{Gardi:1996iq}. It is helpful to see whether PA approach can achieve reliable predictions of the unknown higher-order terms by using the renormalon-free PMC scheme-and-scale-independent method, which underlies the BLM method and generalizes it to all orders.

The remaining parts of this paper are organized as follows.

In Sec.\ref{sec:1}, we review the developments of the RGE of the QCD running coupling and its solution. We shall first show the conventional RGE and the extended RGE which govern the scheme-and-scale runnings of the QCD running coupling, and then give their solutions. The analytic $\alpha_s$-running differ significantly at scales below a few GeV from the exact (numerical) solution of RGE~\cite{Olive:2016xmw}, especially if the RGE is less than five-loop level. A comparison of the exact numerical solution and the analytic solution of the RGE under the $\overline{\rm MS}$-scheme within the low and high energy scales will be presented. We then define the $C$-scheme coupling $\hat\alpha_s$, deduce its much simpler scheme-independent RGE, give its analytic solution, and provide the relation between the $C$-scheme coupling and a conventional coupling. Using this relation, one can transform the conventional coupling series into the $C$-scheme coupling series, and discuss the scheme dependence conveniently.

In Sec.\ref{sec:2}, we present an overview of the PMC multi-scale approach, and present its formalism by using the $R_\delta$-scheme. The residual scale dependence after applying the PMC is discussed. We also review some recent PMC applications, such as Higgs hadroproduction and top-quark pair production at the LHC. We also discuss the $\gamma\gamma^*\to\eta_c$ form factors up to N$^2$LO level, which show the PMC properties in detail and emphasize the importance of a correct renormalization scale setting.

In Sec.\ref{sec:3}, we shall show how scheme-and-scale independent fixed order predictions can be achieved by applying the PMC-s method with the help of the $C$-scheme coupling. We shall first give the formulae for both the dimensional-regularized ${\cal R}_\delta$-scheme and the general $C$-scheme within the PMC-s method. We then demonstrate the equivalence of the PMC predictions, which use either the $C$-scheme coupling or a conventional coupling, respectively. Furthermore, by rewriting the pQCD prediction in terms of the $C$-scheme coupling, we show how the scheme-and-scale independent all-orders predictions can be achieved by applying the PMC-s method. As an example, we present numerical results for the non-singlet Adler function to four-loop level. A practical way to achieve scheme-and-scale independent fixed order prediction by using the PMS method shall also be presented.

In Sec.\ref{sec:4}, we shall show that by using the conformal series derived using the PMC procedures, in combination with the PA approach, one can achieve quantitatively useful estimates for the unknown higher-order terms from the known perturbative series. Comparison of the PA approach using the conventional scale-dependent pQCD series and the PMC scale-independent conformal series shall be presented. We then illustrate the PMC+PA procedure via three hadronic observables $R_{e^+e^-}$, $R_{\tau}$, and $\Gamma(H \to b \bar{b})$. We show that by applying the PA approach to the renormalon-free conformal series, one can achieve quantitatively more reliable estimates for the unknown higher-order terms based on the known pQCD series. If the conformal series has been calculated up to $n_{\rm th}$-order level, the $[0/n-1]$-type PA series provides an important estimate for the higher-order terms.

In Sec.\ref{sec:5}, we summarize and present an outlook.

\section{The renormalization scheme-and-scale running of the QCD running coupling}
\label{sec:1}

The $Q^2\to 0$ limit of the Thomson cross-section provides a natural definition of the QED running coupling, $\alpha(Q^2)=e^2/4\pi=1/137.0359...$~\cite{Olive:2016xmw}, which characterizes the strength of the electromagnetic interaction among elementary charged particles and serves as an initial condition for the RGE which determines $\alpha(Q^2)$ for all $Q^2$-values. The QED running coupling $\alpha(Q^2)$ serves as an expansion parameter for the perturbative QED series. Because the UV divergences in the vertex and fermion self-energy corrections exactly cancel by the Ward identity~\cite{Ward:1950xp, Takahashi:1957}, the net UV divergences are associated with vacuum polarization. Thus only the vacuum-polarization corrections renormalize the QED running coupling, and by resuming all vacuum polarization contributions to the dressed photon propagator, there is in principal no scheme-and-scale ambiguities in QED processes --- This is the so-called GM-L scale setting method for the QED processes~\cite{GellMann:1954fq}.

The non-Abelian nature and the complicated definition of the QCD running coupling make the QCD case more involved. In fact, whether a QCD scale setting method can be reduced to the GM-L method in the QED Abelian limit could be an important criterion for its correctness; Ref.~\cite{Binger:2003by} has shown that one must use the same scale-setting procedure for QCD and QED to obtain grand unification.

As has been discussed in the Introduction, the conventional renormalization scheme-and-scale ambiguities for the fixed-order pQCD prediction are caused by the mismatching of the perturbative coefficients and the QCD running coupling at any order. It is important to obtain the correct value of the running coupling for the considered process, which can be done with the help of the RGE. Thus a more precise solution for the RGE will lead to a precise definition of the QCD running coupling.

\subsection{The conventional renormalization group equation}

The definition of the QCD running coupling $\alpha_s(\mu)$ depends on theoretical conventions such as the choice of the renormalization scheme. Its running behavior in the renormalization scale $\mu$ -- its RGE -- is governed by its logarithmic derivative, the $\beta$-function:
\begin{equation}
\mu^2 \frac{{\rm d}a_\mu}{{\rm d}\mu^2} \,= \, \beta(a_\mu) \,=\, - a_\mu^2 \sum_{i=0}^{\infty}\beta_i a_\mu^i.
\label{eq:betafun}
\end{equation}
For simplicity, we shall define $a_{\mu}= \alpha_s(\mu)/\pi$, where $\mu$ is the renormalization scale, throughout the paper. Various terms in $\beta_0$, $\beta_1$, $\cdots$, correspond to the one-loop, two-loop, $\cdots$, contributions to the RGE, respectively. The first two terms $\beta_0=(11-{2\over 3} n_f)/4$ and $\beta_1=(102-{38 \over 3}n_f)/4^2$, where $n_f$ is the number of active quarks, are universal in the mass-independent renormalization schemes due to decoupling theorem~\cite{Appelquist:1974tg}; the remaining $\{\beta_i\}$-terms are scheme-dependent. According to the decoupling theorem~\cite{Appelquist:1974tg}, a quark with mass $m_{f}^2\gg\mu^2$ can be ignored, and we can usually neglect $m_f$-terms when $m_{f}^2\ll\mu^2$. Then, for every renormalization scale $\mu$, one can divide the quarks into active ones with $m_f =0$ and inactive ones that can be ignored. The explicit form for the $\{\beta_i\}$-terms up to five-loop level in the $\overline{\rm MS}$-scheme are available in Refs.~\cite{Gross:1973id, Politzer:1973fx, Caswell:1974gg, Jones:1974mm, Tarasov:1980au, Larin:1993tp, vanRitbergen:1997va, Chetyrkin:2004mf, Czakon:2004bu, Baikov:2016tgj, Herzog:2017ohr, Luthe:2017ttg}.

If one integrates the RGE (\ref{eq:betafun}), one obtains
\begin{eqnarray}
\ln{\mu_0^2} - \frac{1}{\beta_0 a_{\mu_0}} - \frac{\beta_1}{\beta_0^2}\ln{a_{\mu_0}} - \int_0^{a_{\mu_0}} \frac{{\rm d}a}{\tilde{\beta}(a)} = \ln \mu^2 - \frac{1}{\beta_0 a_\mu} - \frac{\beta_1}{\beta_0^2}\ln{a_\mu} - \int_0^{a_\mu} \frac{{\rm d}a}{\tilde{\beta}(a)},
\label{eq:scaleInvariant}
\end{eqnarray}
where $\mu_0$ is an arbitrary reference scale. The $\tilde{\beta}$-function is defined as
\begin{equation}
\frac{1}{\tilde\beta(a)} \,\equiv\, \frac{1}{\beta(a)} + \frac{1}{\beta_0 a^2} - \frac{\beta_1}{\beta_0^2 a}.
\end{equation}
The advantage of introducing the $\tilde{\beta}$-function lies in the fact that the integral $\int_0^{a_\mu} {{\rm d}a}/{\tilde{\beta}(a)}$ is free of singularities in the limit $a_\mu \to 0$. Up to five-loop level, the integral can be expressed as a power series in $a_\mu$,
\begin{eqnarray}
\int_0^{a_\mu} \frac{{\rm d}a}{\tilde{\beta}(a)} &=& \left(\frac{\beta_2}{\beta_0^2}-\frac{\beta_1^2}{\beta_0^3}\right) a_\mu
+\left(\frac{\beta_3}{2\beta_0^2}-\frac{\beta_2 \beta_1}{\beta_0^3}+\frac{\beta_1^3}{2\beta_0^4}\right) a_\mu^2 \nonumber\\
&& +\left(\frac{\beta_4}{3\beta_0^2}-\frac{\beta_2^2}{3\beta_0^3}-\frac{2\beta_3 \beta_1}{3\beta_0^3}
+\frac{\beta_2 \beta_1^2}{\beta_0^4}-\frac{\beta_1^4}{3\beta_0^5}\right) a_\mu^3 + {\cal O}(a_\mu^4).
\label{eq:integral}
\end{eqnarray}

It is convenient to define an {\it asymptotic scale} $\Lambda$ by collecting all $\mu_0$-dependent terms on the left-hand-side of Eq.(\ref{eq:scaleInvariant}) into its definition, and then the evolution of the QCD running coupling $a_\mu$ is independent of a specific choice for $\mu_0$, i.e.
\begin{eqnarray}
\ln\frac{\mu^2}{\Lambda^2} = \frac{1}{\beta_0 a_\mu} + \frac{\beta_1}{\beta_0^2}\ln{a_\mu} + \int_0^{a_\mu} \frac{{\rm d}a}{\tilde{\beta}(a)}.
\label{eq:LambdaQCD}
\end{eqnarray}
The asymptotic scale $\Lambda$ is, by definition, scheme dependent. Given a measurement which sets the value of the running coupling at a given scale, one can fix $\Lambda$ for a given scheme by matching the measured value of the coupling to its predicted value as determined by Eq.(\ref{eq:LambdaQCD}). Notice that this new asymptotic scale $\Lambda$ differs from the generally adopted asymptotic scale $\Lambda_{\rm QCD}$ (c.f. the definition given by the PDG~\cite{Olive:2016xmw}) by an overall parameter; i.e.,
\begin{equation}
\Lambda = \beta_0^{({\beta_1}/{2\beta_0^2})}\Lambda_{\rm QCD}.
\label{eq:LambdaRelation}
\end{equation}
This difference is caused by absorbing different integration constants into the definition of the asymptotic scales. Another example of differing conventions is the 't Hooft scheme~\cite{tHooft}, where the associated asymptotic scale is $\Lambda^{\rm 'tH}=(\beta_0^2/\beta_1)^{2\beta_0^2/\beta_1} \Lambda_{\rm QCD}$~\cite{Brodsky:2011ta, Lu:1992nt}.

Using the relation Eq.(\ref{eq:LambdaRelation}) and iteratively solving Eq.(\ref{eq:LambdaQCD}) yields~\cite{Kniehl:2006bg}
\begin{eqnarray}
a_{\mu} &=&\frac{1}{\beta_0 L} - \frac{b_1 \ln L}{(\beta_0 L)^2}
+\frac{1}{(\beta_0 L)^3}\left[b_1^2(\ln^2 L- \ln L -1)+b_2\right] \nonumber\\
&& +\frac{1}{(\beta_0 L)^4} \left[b_1^3 \left(-\ln^3 L+\frac{5}{2} \ln^2 L+ 2\ln L -\frac{1}{2}\right) -3b_1b_2 \ln L+\frac{b_3}{2}\right] \nonumber \\
&&+\frac{1}{(\beta_0 L)^5} \left[3b_2 b_1^2 (2 \ln^2 L -\ln L-1) +b_1^4 \left(\ln^4 L- \frac{13}{3} \ln^3 L -\frac{3}{2} \ln^2 L+4 \ln L +\frac{7}{6} \right) \right. \nonumber \\
&& \left. -b_3b_1 \left(2\ln L+\frac{1}{6}\right) +\left(\frac{5}{3}b_2^2+\frac{1}{3}b_4\right)\right] +{\cal O}\left(\frac{1}{(\beta_0 L)^6}\right),
\label{convscalerunning}
\end{eqnarray}
where the 5-loop terms which are proportional to ${1}/{(\beta_0 L)^5}$ have been presented. Here $L=\ln(\mu^2/\Lambda_{\rm QCD}^2)$ and $b_i=\beta_i/\beta_0$.

\begin{figure}[htb]
\epsfysize=9.0cm
\begin{center}
\begin{minipage}[t]{10 cm}
\epsfig{file=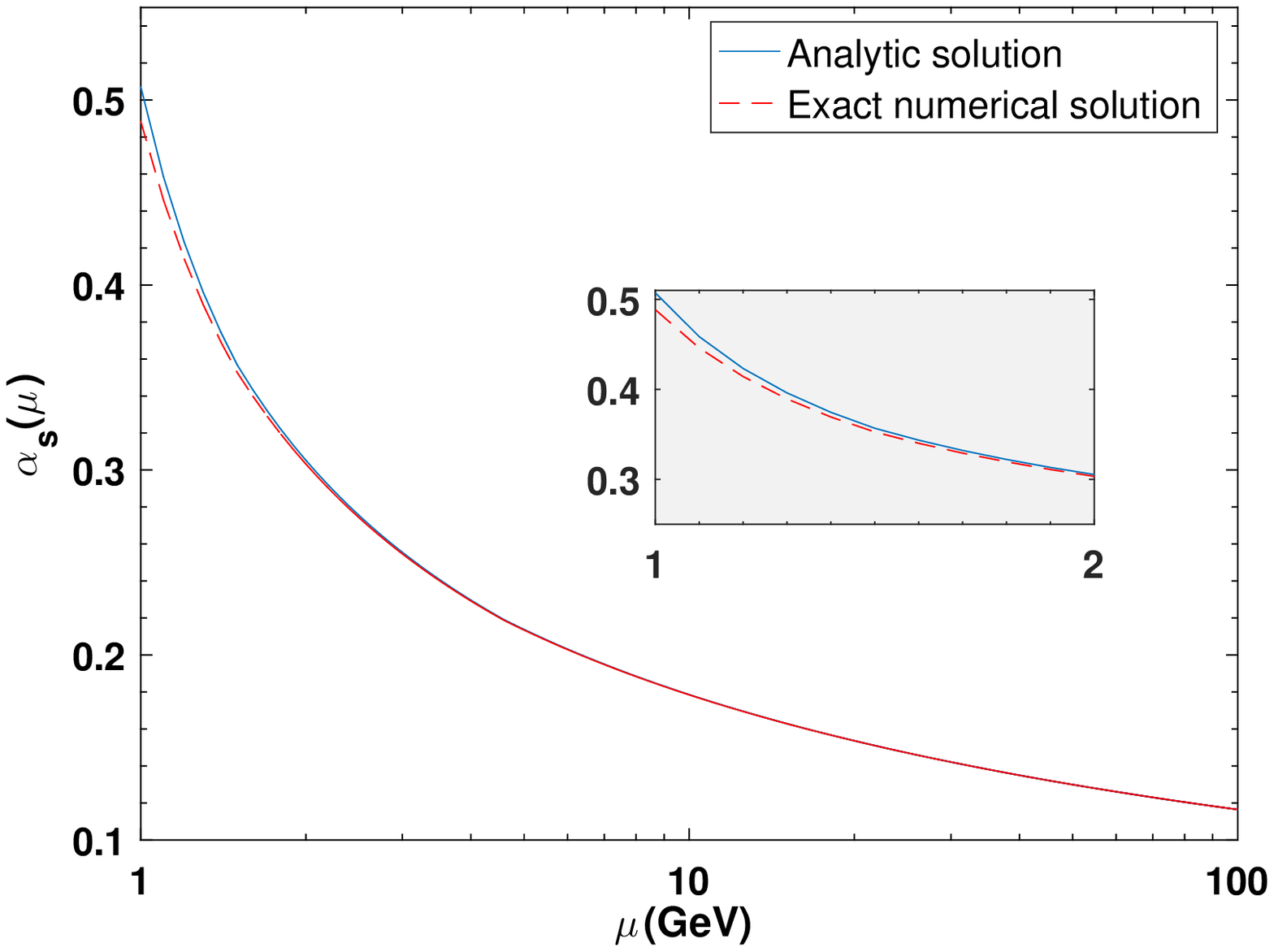,scale=0.5}
\end{minipage}
\begin{minipage}[t]{16.5 cm}
\caption{Comparison of the $\overline{\rm MS}$-scheme QCD running couplings $\alpha_s(\mu)$ at the four-loop level in low scale region, where the solid lines and dashed lines are analytic and exact numerical solutions of the RGE, respectively. The decoupling is performed at the pole mass of the respective heavy quark. The asymptotic scale is fixed by using $\alpha_s^{\overline{\rm MS}}(M_Z) =0.1181(11)$~\cite{Olive:2016xmw}. \label{fig:alphas4loop}}
\end{minipage}
\end{center}
\end{figure}

\begin{figure}[htb]
\epsfysize=9.0cm
\begin{center}
\begin{minipage}[t]{10 cm}
\epsfig{file=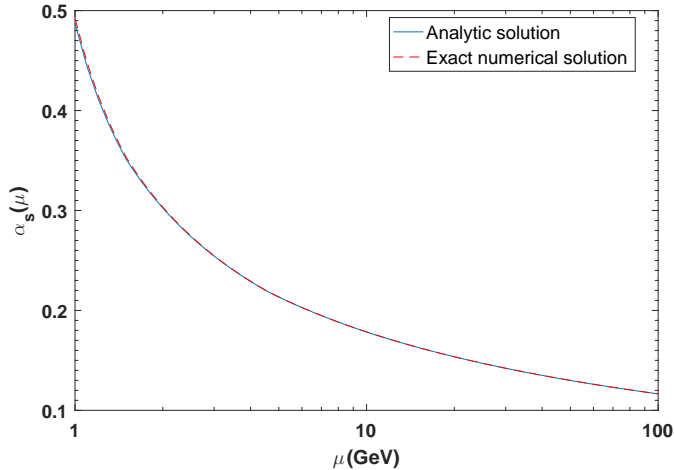,scale=0.5}
\end{minipage}
\begin{minipage}[t]{16.5 cm}
\caption{Comparison of the $\overline{\rm MS}$-scheme QCD running couplings $\alpha_s(\mu)$ at the five-loop level, where the solid lines and dashed lines are analytic and exact numerical solutions of the RGE, respectively. The decoupling is performed at the pole mass of the respective heavy quark. The asymptotic scale is fixed by using $\alpha_s^{\overline{\rm MS}}(M_Z) =0.1181(11)$~\cite{Olive:2016xmw}. \label{fig:alphas5loop}}
\end{minipage}
\end{center}
\end{figure}

In the literature, the RunDec package~\cite{Chetyrkin:2000yt, Schmidt:2012az, Herren:2017osy} is often used for computing the running of the coupling constant. We present comparisons of the $\overline{\rm MS}$-scheme QCD running couplings under the usual analytic solution and the exact numerical solution of the RGE up to four-loop level and five-loop level in Figures \ref{fig:alphas4loop} and \ref{fig:alphas5loop}, respectively.
For $n_f = 3, 4, 5$ and $6$ quark flavors, the decoupling is performed at the pole mass of the respective heavy quark~\cite{Herren:2017osy}, i.e., at the charm-, bottom- and top-quark pole masses of 1.5, 4.8 and 173.21 GeV, respectively. The decoupling constants for $\alpha_s$, which establishes the transition from $\alpha_s$ defined in the $n_f$-flavour effective theory to $\alpha_s$ in the $(n_f-1)$-flavour effective theory using the $\overline{\rm MS}$ definition of the heavy quark mass, are available from Refs.~\cite{Chetyrkin:1997sg, Weinberg:1980wa, Ovrut:1980dg, Wetzel:1981qg, Bernreuther:1981sg, Bernreuther:1983ds, Larin:1994va, Chetyrkin:1997un, Schroder:2005hy, Chetyrkin:2005ia, Grozin:2011nk, Liu:2015fxa, Marquard:2015qpa}. In large scale region, $\mu>>\Lambda_{\rm QCD}$, the analytic solution is a good approximation for $\alpha_s$-running even for lower-order predictions. Figure \ref{fig:alphas4loop} shows that at or below the four-loop level, the usually adopted analytic $\alpha_s$-running differ significantly from the exact solution of RGE at scales below a few GeV. As shown by Figure \ref{fig:alphas5loop}, since the 5-loop terms are always negative in small scale region and their magnitudes become large for smaller scale, the net difference between the analytic and exact RGE solutions becomes negligible. For momentum transfers in the $100$ GeV - TeV range, $\alpha_s \sim 0.1$, whereas the QCD theory is strongly interacting for scales around and below 1 GeV. Thus if momentum flow is close to a few GeV, one needs to adopt the $\alpha_s$-value determined from the exact numerical solution of the RGE.

It is useful to notice that by using Eq.(\ref{eq:LambdaQCD}), we can obtain a relation of the couplings at two scales such as $\mu$ and $Q$ under the same scheme:
\begin{eqnarray}\label{eq:PMCcoupling1}
\left(\frac{1}{\beta_0 a_{\mu}} + \frac{\beta_1}{\beta^2_0}\ln{a_{\mu}} + \int_0^{a_{\mu}} \frac{{\rm d}a}{\tilde{\beta}(a)}\right) -\left( \frac{1}{\beta_0 a_Q} + \frac{\beta_1}{\beta_0^2}\ln{a_Q} + \int_0^{a_Q} \frac{{\rm d}a}{\tilde{\beta}(a)} \right) = \ln\frac{\mu^2}{Q^2}.
\end{eqnarray}
If the QCD running coupling is measured at a reference scale $Q$, then we can fix its value at any other scale without determining the asymptotic scale $\Lambda$, thus avoiding any uncertainty coming from the determination of $\Lambda$.

\subsection{The extended renormalization group equation}

The RGE approach relates the running coupling at different scales in a continuous way, avoiding the large expansion coefficients of the running couplings at dissimilar scales; thus better pQCD predictions can be achieved. Stimulating by this idea, Stevenson~\cite{Stevenson:1980du, Stevenson:1981vj} has suggested the use of new scheme-running equations which incorporate both the scale and scheme running behaviors in a consistent way. This procedure is called the {\it extended RGE approach}~\cite{Lu:1992nt}. A review of the extended RGE and its solution can be found in Ref.\cite{Wu:2013ei}.

As an application of the extended RGE approach, by using the relation of the $\beta$-functions between different schemes, i.e. $\beta_{\cal S}(a_\mu^{\cal S})=\beta_{\cal R}(a_\mu^{\cal R}) {\partial a_\mu^{\cal S}}/{\partial a_\mu^{\cal R}}$, one can reproduce the Celmaster-Gonsalves relation~\cite{Celmaster:1979dm, Celmaster:1979km} for the asymptotic scales of different schemes~\cite{vonSmekal:209ae, Gracey:2013sca, Zeng:2015gha}; i.e.
\begin{eqnarray}
\frac{\Lambda^{\cal S}}{\Lambda^{\cal R}} = \exp\left(-\frac{f_2}{2\beta_0}\right).
\label{CGR}
\end{eqnarray}
Here ${\cal S}$ and ${\cal R}$ designate two arbitrary renormalization schemes, and the coefficient $f_2$ is the next-to-leading order term of the coupling $\alpha^{\cal S}_\mu$ expanded in powers of $\alpha^{\cal R}_\mu$, i.e. $a^{\cal R}_\mu = a^{\cal S}_\mu + f_2 (a^{\cal S}_\mu)^2+ f_3 (a^{\cal S}_\mu)^3 +\cdots$.

\subsection{The $C$-scheme running coupling and its scheme-invariant renormalization group equation}

The scheme-and-scale running behaviors as determined by either the conventional RGE or the extended RGE depends explicitly on the scheme parameters $\{\beta_{i\geq2}\}$. It is thus difficult to achieve an analytical solution, and one needs to use perturbative theory.

Boito, Jamin and Miravitllas~\cite{Boito:2016pwf} have suggested an unusual way to deal with the scheme dependence of QCD running couplings based on the Celmaster-Gonsalves relation. They have shown that one can introduce a new class of running couplings $\hat{a}_\mu$, characterized by a single parameter $C$, whose variation directly compensates for the usual scheme dependence of the scale parameter $\Lambda$ of the corresponding conventional coupling $a_\mu$. In the following, we shall first introduce the $C$-scheme coupling $\hat{a}_\mu$ and then demonstrate that -- in contrast to the standard RGE behavior (\ref{eq:betafun}) of $a_\mu$, the scale dependence of $\hat{a}_\mu$ is independent of the scheme-dependent $\{\beta_{i\ge2}\}$-terms, and it is thus explicitly scheme-invariant.

Eq.(\ref{eq:LambdaQCD}) implies that the conventional $a_\mu$ coupling satisfies the following scheme-dependent scale-running behavior
\begin{eqnarray}\label{eq:general}
\frac{1}{a_\mu} + \frac{\beta_1}{\beta_0}\ln{a_\mu} = \beta_0 \left( \ln\frac{\mu^2}{\Lambda^2} - \int_0^{a_\mu} \frac{{\rm d}a}{\tilde{\beta}(a)} \right).
\end{eqnarray}
One can define a new coupling $\hat{a}_\mu=\hat{\alpha}_s(\mu)/\pi$ in the following way~\cite{Boito:2016pwf}:
\begin{eqnarray}
\frac{1}{\hat a_\mu} + \frac{\beta_1}{\beta_0} \ln\hat a_\mu
&=& \beta_0 \left( \ln\frac{\mu^2}{\Lambda^2} + C \right) \,,
\label{eq:ahat}
\end{eqnarray}
where the phenomenological parameter $C$ is introduced, which incorporates the effects of all scheme-dependent $\{\beta_{i\ge2}\}$-terms; i.e. $C=- \int_0^{a_\mu} {{\rm d}a}/{\tilde{\beta}(a)}$. Different choices of $C$ thus correspond to different renormalization schemes. By choosing a specific value for $C = C_{\rm RS}$, the running coupling of the $C$-scheme will become equivalent to the coupling of any conventional renormalization scheme. A subtle point for this equivalence is that the value of $C$ implicitly depends on the renormalization scale where the $C$-scheme coupling and the conventional coupling using a specific renormalization scheme (RS) are matched. However, as will be shown below, such a price is worth it; there are many advantages in using this single parameter $C$ to characterize the scheme-dependence of the running coupling.

The solution of Eq.(\ref{eq:ahat}) can be written in terms of Lambert $W$-function, which is the solution of the equation
\begin{equation}
W(z)\exp[W(z)] = z, \label{wfunction}
\end{equation}
To this end, we multiply both sides of Eq.(\ref{eq:ahat}) by a factor $(-{\beta_0}/{\beta_1})$ and then exponentiate it as the following form
\begin{equation}
\exp \left[ -\frac{\beta_0^2}{\beta_1} \left(\ln\frac{\mu^2}{\Lambda^2}+C\right) \right]
= \exp \left[ -\frac{\beta_0}{\beta_1} \left(\frac{1}{\hat{a}_\mu}+\frac{\beta_1}{\beta_0}\ln\hat{a}_\mu\right) \right]
= \frac{1}{\hat{a}_\mu} \exp \left[-\frac{\beta_0}{\beta_1}\frac{1}{\hat{a}_\mu} \right],
\label{eq:solu0}
\end{equation}
which can be further rewritten as
\begin{equation}
-\frac{\beta_0}{\beta_1} \exp \left[ -\frac{\beta_0^2}{\beta_1} \left(\ln\frac{\mu^2}{\Lambda^2}+C\right) \right]
= -\frac{\beta_0}{\beta_1} \frac{1}{\hat{a}_\mu} \exp \left[-\frac{\beta_0}{\beta_1}\frac{1}{\hat{a}_\mu} \right].
\label{eq:solu1}
\end{equation}
Comparing Eq.(\ref{eq:solu1}) with Eq.(\ref{wfunction}), we obtain
\begin{eqnarray}
z = -\frac{\beta_0}{\beta_1} \exp \left[ -\frac{\beta_0^2}{\beta_1} \left(\ln\frac{\mu^2}{\Lambda^2}+C\right) \right]
\;\; {\rm and} \;\; W(z) = -\frac{\beta_0}{\beta_1}\frac{1}{\hat{a}_\mu}.
\label{eq:solu2}
\end{eqnarray}
The function $W(z)$ is a multi-valued function with an infinite number of branches denoted by $W_n(z)$~\cite{Corless:1996zz}. The correct physical branch can be determined by the requirement that $\hat{a}_\mu$ must be real and positive for a real positive scale $\mu$~\footnote{This conclusion is valid, at least for $\mu^2 \gg \Lambda^2 e^{-C}$.}, which inversely, indicates $W(z)$ should be real and negative. Since in practice $n_f\leq6$, we have $z<0$, and the physical branch is $W_{-1}(z)$. One also finds that $W_{-1}(z)$ monotonically decreases within the region of $z\in(-1/e,0)$, with $W_{-1}(z)\in(-\infty,-1)$. The ultraviolet limit $\mu\to\infty$ corresponds to $z \to 0^{-}$ and $W_{-1}(z) \to -\infty$, leading to $\hat{a}_\mu \to 0^{+}$, as required by asymptotic freedom. Finally, the solution of Eq.(\ref{eq:ahat}) is
\begin{equation}
\hat{a}_\mu = -\frac{\beta_0}{\beta_1 W_{-1}(z)}.
\end{equation}

Using Eq.(\ref{eq:ahat}), we obtain the RGE for the $C$-scheme coupling $\hat{a}_\mu$:
\begin{equation}
\mu^2 \frac{\partial \hat{a}_\mu}{\partial \mu^2} = \hat\beta(\hat{a}_\mu) = -
\frac{\beta_0 \hat{a}_\mu^2}{ 1 - \frac{\beta_1}{\beta_0} \hat{a}_\mu }, \label{eq:betahat}
\end{equation}
which has a much simpler form than the standard RGE (\ref{eq:betafun}). By coincidence, this RGE agrees with the one suggested by Refs.\cite{Brown:1992pk, Lee:1996yk, Brambilla:2017hcq}, in which a new strong coupling is introduced with the purpose of improving the convergence of pQCD series and whose RGE is derived by using the approximation, $\beta_j \approx \beta_0 \left(\beta_1/\beta_0\right)^{j}$ ($j\geq 0$). The introducing of the $C$-scheme coupling provides a natural explanation of how its RGE comes from without introducing any approximations. In fact the new strong coupling introduced in Ref.\cite{Brambilla:2017hcq} corresponds to the special case, e.g. $C=0$, of the $C$-scheme coupling. As shall be shown later, the introducing of $C$-scheme coupling not only improves the pQCD convergence but also provides a good basis for solving the conventional renormalization scheme-and-scale ambiguities~\footnote{If one further sets $\beta_1 \approx\beta_0^2$ (leading to $\beta_j \approx \beta^{j+1}_0$), one can achieve the simplest RGE $\beta({a}_\mu) = - {\beta_0 {a}_\mu^2}/{(1 - \beta_0 {a}_\mu) }$, which has been adopted to eliminate the divergent renormalon terms $n! a_\mu^n \beta^n_0$ in the pQCD series~\cite{Beneke:1998ui}.}.

At the same time, by using Eq.(\ref{eq:ahat}), one may also observe that
\begin{equation}
\frac{\partial \hat{a}_\mu}{\partial C} = \hat\beta(\hat{a}_\mu).
\label{eq:betahatC}
\end{equation}
Eqs.(\ref{eq:betahat}, \ref{eq:betahatC}) indicate that
\begin{itemize}
\item The $\hat\beta$-function (\ref{eq:betahat}) is by definition scheme-independent. Thus the scale-running behavior of the $C$-scheme coupling $\hat{a}_\mu$ is explicitly scheme-independent since it only depends on the scheme-independent $\beta$-coefficients $\beta_0$ and $\beta_1$. Thus even though the $C$-scheme coupling $\hat{a}_\mu$ itself is implicitly scheme-dependent, its scale-running behavior can be scheme-independent.

\item The scale-running and scheme-running behaviors of $\hat{a}_\mu$ have been explicitly separated -- each of them satisfy the same $\hat\beta$-function. As is the case of the conventional RGE (\ref{eq:betafun}), the new RGE~(\ref{eq:betahat}) for the $C$-scheme coupling can also be solved iteratively and perturbatively. By comparing the perturbative expansion
  \begin{equation}
   \hat\beta(\hat{a}_\mu)= - \beta_0 \hat{a}_\mu^2 \sum_{i=0}^{\infty}\left({\beta_1} /{\beta_0}\right)^{i} \hat{a}_\mu^{i}
  \end{equation}
   with the RGE (\ref{eq:betafun}), the solution of $\hat{a}_\mu$ up to four-loop level can be obtained from the solution of RGE (\ref{eq:betafun}), e.g. Eq.(\ref{convscalerunning}), by replacing
 \begin{displaymath}
 a_\mu \to \hat{a}_\mu, \;\; b_i= \beta_i/\beta_0 \to (\beta_1/\beta_0)^i .
 \end{displaymath}

\item Integrating RGE~(\ref{eq:betahat}) yields a relation of $\hat{a}_\mu$ for any two scales $\mu_1$ and $\mu_2$, i.e.,
 \begin{equation} \label{ahatrun}
 \frac{1}{\hat{a}_{\mu_2}} \,=\, \frac{1}{\hat{a}_{\mu_1}} + {\beta_0}\ln\frac{\mu_2^2}{\mu_1^2}
 - \frac{\beta_1}{\beta_0}\ln\frac{\hat{a}_{\mu_2}}{\hat{a}_{\mu_1}} \,.
 \end{equation}
 Thus if $\hat{a}$ at the reference scale $\mu_1$ is known, we can determine its value at any other scale $\mu_2$.

\item Given a proper choice of $C$, any conventional coupling $a_\mu$ which is defined in any renormalization scheme can be uniquely expressed by a corresponding $C$-scheme coupling $\hat{a}_\mu$. For example, following the idea of {\it effective charge} approach~\cite{Grunberg:1980ja, Grunberg:1982fw}, any pQCD calculable physical observable can be used to define an effective coupling $\hat{a}_\mu$. If the defined effective $C$-scheme coupling $\hat{a}_\mu$ for an observable is independent of $C$, Eq.(\ref{eq:betahatC}) indicates that $\hat\beta(\hat{a}_\mu)=0$, and we will then obtain a scheme-independent conformal series in the effective coupling $\hat{a}_\mu$ of the corresponding observable.

\end{itemize}

\subsection{Relation between the $C$-scheme coupling $\hat{a}_\mu$ and the conventional running coupling $a_\mu$}

The pQCD calculation is usually done by using the conventional running coupling $a_\mu$. We can transform it into a pQCD series for the $C$-scheme coupling $\hat{a}_\mu$ by using the relation between $\hat{a}_\mu$ and the conventional coupling $a_\mu$. Using Eq.(\ref{eq:general}), we transform Eq.(\ref{eq:ahat}) to the following form
\begin{eqnarray}
\frac{1}{\hat a_\mu} + \frac{\beta_1}{\beta_0} \ln\hat a_\mu = {\beta_0}\,C + \frac{1}{a_\mu} +
\frac{\beta_1}{\beta_0}\ln a_\mu + \beta_0 \!\int_0^{a_\mu}\frac{{\rm d}a}{\tilde\beta(a)} \,;
\label{eq:C-coupling}
\end{eqnarray}
solving it recursively, we obtain the required relation
\begin{eqnarray}
a_\mu &=& \hat{a}_\mu+C\beta_0\hat{a}_\mu^2+ \left(\frac{\beta_2}{\beta_0}- \frac{\beta_1^2}{\beta_0^2}+\beta_0^2 C^2+\beta_1 C\right)\hat{a}_\mu^3 \nonumber\\
& & +\left[\frac{\beta _3}{2 \beta _0}-\frac{\beta_1^3}{2 \beta _0^3}+\left(3 \beta _2-\frac{2 \beta _1^2}{\beta _0}\right) C+\frac{5}{2}\beta_0 \beta_1 C^2 +\beta_0^3 C^3\right] \hat{a}_\mu^4 + {\cal O}(\hat{a}_\mu^5) ,
\label{eq:Expandhata}
\end{eqnarray}
or inversely,
\begin{eqnarray}
\hat{a}_\mu &=& a_\mu-C\beta_0 a_\mu^2+\left(\frac{\beta _1^2}{\beta _0^2}-\frac{\beta_2}{\beta_0}+\beta_0^2 C^2-\beta_1 C\right)a_\mu^3 \nonumber\\
&& +\bigg[\frac{\beta_1^3}{2\beta_0^3}-\frac{\beta_3}{2\beta_0}+\left(2 \beta _2-\frac{3 \beta _1^2}{\beta _0}\right) C +\frac{5}{2}\beta_0\beta_1 C^2 -\beta_0^3 C^3 \bigg]a_\mu^4 + {\cal O}(a_\mu^5). \label{eq:Expandhata2}
\end{eqnarray}

As an explicit example, considering the conventional coupling $a_\mu$ with the $\{\beta_{i\ge2}\}$-functions under the $\overline{\rm MS}$-scheme, we have~\cite{Boito:2016pwf},
\begin{eqnarray}
a^{\overline{\rm MS}}_\mu &=& \hat{a}_\mu + \frac{9}{4}C\hat{a}_\mu^2 + \left( \frac{3397}{2592} + 4 C +
\frac{81}{16} C^2 \right) \hat{a}_\mu^3 \nonumber\\
&& + \bigg( \frac{741103}{186624} +\frac{18383}{1152}\,C + \frac{45}{2}\,C^2 + \frac{729}{64}\,C^3 + \frac{445}{144}\zeta(3) \bigg) \hat{a}_\mu^4 + \cdots,
\label{aofahat}
\end{eqnarray}
where we have set the active flavor number $n_f=3$, and $\zeta(i)$ is the Riemann $\zeta$-function. Eq.(\ref{aofahat}) indicates that the value of $C$ needs to be a function of the scale $\mu$ in order to ensure the equivalence of the $C$-scheme coupling $\hat{a}_\mu$ and the $\overline{\rm MS}$-scheme coupling $a^{\overline{\rm MS}}_\mu$.

\begin{figure}[htb]
\epsfysize=9.0cm
\begin{center}
\begin{minipage}[t]{10 cm}
\epsfig{file=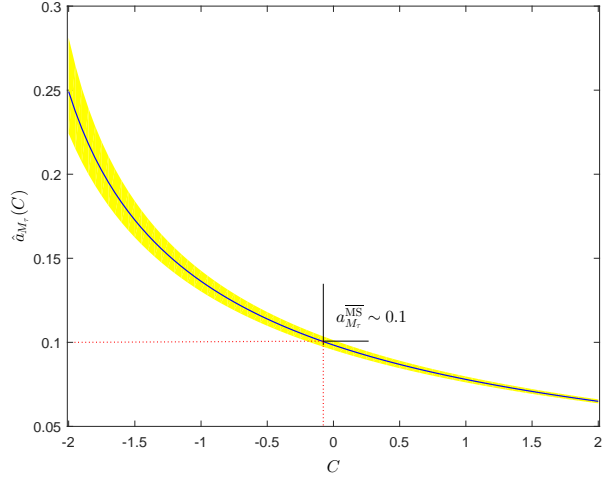,scale=0.45}
\end{minipage}
\begin{minipage}[t]{16.5 cm}
\caption{The $C$-scheme coupling $\hat{a}_{M_\tau}$ as a function of $C$ at the scale $M_\tau$, which is calculated by using the relation (\ref{eq:C-coupling}) up to four-loop level. We adopt $\alpha^{\overline{\rm MS}}_s(M_\tau)=0.3159(95)$ as the reference value. The solid line represents the center value, and the shaded band corresponds to the uncertainty $\Delta\alpha_s(M_\tau)=\pm0.0095$. The crossing point of the two dotted lines indicates $\hat{a}_{M_\tau}(C=-0.0818)=a_{M_\tau}^{\overline{\rm MS}}$. \label{fig:Ccoupling}}
\end{minipage}
\end{center}
\end{figure}

To show explicitly how the $C$-scheme coupling $\hat{a}_\mu$ depends on the parameter $C$, we present the coupling $\hat{a}_\mu$ at the scale $\mu={M_\tau}$ as a function of $C$ in Figure~\ref{fig:Ccoupling}. Here we adopt the world average $\alpha_s^{\overline{\rm MS}}(M_Z) =0.1181(11)$~\cite{Olive:2016xmw} as the reference value, which runs down to $\alpha_s^{\overline{\rm MS}}(M_\tau)=0.3159(95)$ using the four-loop RGE. Figure~\ref{fig:Ccoupling} shows that the coupling $\hat{a}_\mu$ monotonously decreases as a function of $C$. This is confirmed by the fact that the $C$-scheme $\hat\beta(\hat{a}_\mu)$-function (\ref{eq:betahat}) is generally negative -- the negative $\hat\beta(\hat{a}_\mu)$-function implies that the coupling must monotonically decrease with the increment of $C$. By choosing a suitable $C$, the new coupling $\hat{a}_\mu$ becomes equivalent to the coupling $a_\mu$ defined for any corresponding conventional scheme; i.e. $a_\mu = \hat{a}_{\mu}|_{C}$. At a different scale $\mu$, a different $C$ needs to be introduced in order to ensure the equivalence of the couplings at the same scale. For examples, we have
\begin{displaymath}
 a_{M_\tau}^{\overline{\rm MS}}=\hat{a}_{M_\tau}(C=-0.0818) \;\; {\rm and} \;\; a_{M_Z}^{\overline{\rm MS}}=\hat{a}_{M_Z}(C=0.7285).
\end{displaymath}

\section{The pQCD predictions using the PMC scale-setting approach}
\label{sec:2}

Generally, the pQCD approximant of an observable up to $n_{\rm th}$-order level can be expressed as
\begin{eqnarray}
\rho_{n}(Q) = \sum_{i=1}^{n} r_{i}(\mu/Q) a_{\mu}^{i}, \label{pQCDexp}
\end{eqnarray}
where $\mu$ is the renormalization scale and $Q$ is the kinematic scale of the process at which it is measured or the typical momentum flow of the process. Without losing generality, we have set the power of the running coupling associated with the tree-level term as $1$.

At any finite order, the renormalization scheme-and-scale dependence from $a_{\mu}$ and $r_i(\mu/Q)$ usually do not exactly cancel; this leads to the renormalization scheme-and-scale ambiguity. The PMC has been introduced to eliminate the conventional renormalization scheme-and-scale ambiguity by finding the optimal behavior of the running coupling via a systematic and process-independent way. In the following subsections, we shall first present an overview of the PMC and describe its standard formalism. Then we discuss the residual scale dependence of the predictions after applying the PMC. Finally, we give some recent PMC applications, such as the Higgs hadroproduction and the top-quark pair production at the LHC, and the $\gamma\gamma^*\to\eta_c$ form factor up to N$^2$LO level.

There are also cases in which additional momentum flows occur, whose scale uncertainties can also be eliminated by applying the PMC. For example, there are two types of log terms, $\ln(\mu/M_{Z})$ and $\ln(\mu/M_{t})$~\cite{Baikov:2008jh, Baikov:2010je, Baikov:2012zn, Baikov:2012er}, for the axial singlet $r^A_S$ of the hadronic $Z$ decays. By applying the PMC, one finds the optimal scale is $Q^{\rm AS} \simeq 100$ GeV~\cite{Wang:2014aqa}, indicating that the typical momentum flow for $r^A_S$ is closer to $M_Z$ than $M_t$. The PMC can also be systematically applied to multi-scale problems. The typical momentum flow can be distinct; thus, one should apply the PMC separately in each region. For example, two optimal scales arise at the N$^2$LO level for the production of massive quark-anti-quark pairs ($Q \bar Q$) close to threshold~\cite{Brodsky:1995ds}, with one being proportional to $\sqrt{\hat{s}}$ and the other to $v\sqrt{\hat{s}}$, where $v$ is the $Q$ and $\bar Q$ relative velocity.

\subsection{An overview of the PMC scale setting}
\label{subsec:pmcdevelop}

The PMC procedure follows these steps
\begin{itemize}
\item[-] First, we perform a pQCD calculation of an observable by using general regularization and renormalization procedures at an arbitrary initial renormalization scale $\mu$ and by taking any renormalization scheme. The initial renormalization scale can be arbitrarily chosen, which only needs to be large enough ($\mu>>\Lambda_{\rm QCD}$) to ensure the reliability of the perturbative calculation. One may choose the renormalization scheme to be the usually adopted $\overline{\rm MS}$-scheme; after applying the PMC, the final pQCD prediction will be shown to be independent to this choice, since the PMC is consistent with RGI.

\item[-] Second, we identify the non-conformal $\{\beta_i\}$-terms in the pQCD series. This can be achieved with the help of the degeneracy relations among different orders~\cite{Mojaza:2012mf, Brodsky:2013vpa}, which identify which terms in the pQCD series are associated with the RGE and which terms are not.

    By using the displacement relation for the running coupling at any two scales, e.g. Eq.(\ref{scaledis}), one can obtain the general pattern of the $\{\beta_i\}$-terms at each order, which naturally implies the wanted degeneracy relations among different terms; e.g., the coefficients for $\beta_0 a^2_{\mu}$, $\beta_1 a^3_{\mu}$, $\cdots$, $\beta_i a^{i+2}_{\mu}$ are the same. It has been demonstrated that the degeneracy relations hold using any renormalization scheme~\cite{Bi:2015wea}. The dimensional-like ${\cal R}_\delta$-scheme provides a natural explanation of the degeneracy relations which are general properties of the non-Abelian gauge theory and underly the resulting conformal features of the pQCD series.

    Alternatively, one can use the $\delta$ dependence of the series to identify the $\{\beta_i\}$-terms~\cite{Brodsky:2013vpa}. One can also rearrange all the perturbative coefficients, which are usually expressed as an $n_f$-power series, into $\{\beta_i\}$-terms or non-$\{\beta_i\}$-terms. One needs to be careful using this method to ensure that the UV-free light-quark loops are not related to the $\{\beta_i\}$-terms; they should be identified as conformal terms and should be kept unchanged when doing the $n_f\to\{\beta_i\}$ transformation. The separation of UV-divergent and UV-free terms is very important. This fact has already been shown in QED case, in which electron-loop light-by-light contribution to the sixth-order muon anomalous moment is sizable but UV-free and should be treated as conformal terms~\cite{Aldins:1969jz}. There are many examples for the QCD case. For example, by carefully dealing with the UV-free light-by-light diagrams at the N$^2$LO level, the resulting PMC prediction agrees with the BaBar measurements within errors, thus provides a solution for the $\gamma\gamma^* \to \eta_c$ form factor puzzle~\cite{Wang:2018lry}.

   In practice, one can also apply the PMC by directly dealing with the $n_f$-power series without transforming them into the $\{\beta_i\}$-terms~\cite{Brodsky:2011ta}. This procedure is based on the observation that one can rearrange all the Feynman diagrams of a process in form of a cascade; i.e., the ``new" terms emerging at each order can be equivalently regarded as a one-loop correction to all the ``old" lower-order terms. All of the $n_f$-terms can then be absorbed into the running coupling following the basic $\beta$-pattern in the scale-displacement formula, i.e. Eq.\eqref{scaledis}. More explicitly, in this treatment, the PMC scales can be derived in the following way: The LO PMC scale $Q_{1}$ is obtained by eliminating all the $n_f$-terms with the highest power at each order, and at this step, the coefficients of the lower-power $n_f$-terms are changed simultaneously to ensure that the correct LO $\alpha_s$-running is obtained; the NLO PMC scale $Q_{2}$ is obtained by eliminating the $n_f$-terms of one less power in the new series obtaining a third series with less $n_f$-terms; and so on until all $n_f$-terms are eliminated.

   If the $n_f$-terms are treated correctly, the results for both treatments shall be equivalent since they lead to the same resummed ``conformal" series up to all orders. Those two PMC approaches differ, however, at the non-conformal level, by predicting slightly different PMC scales of the running coupling. This difference arises due to different ways of resumming the non-conformal $\{\beta_i\}$-terms, but this difference decreases rapidly when additional loop corrections are included~\cite{Bi:2015wea}.

\item[-] Third, we absorb different types of $\{\beta_i\}$-terms into the running coupling via an order-by-order manner with the help of degeneracy relations. Different types of $\{\beta_i\}$-terms as determined from the RGE lead to different running behaviors of the running coupling at different orders, and hence, determine the distinct scales at each order. As a result, the PMC scales themselves are perturbative expansion series in the running coupling. Since a different scale generally appears at each order, we call this approach as the PMC multi-scale approach.

\item[-] Finally, since all the non-conformal $\{\beta_i\}$-terms have been resummed into the running coupling, the remaining terms in the perturbative series will be identical to those of the corresponding conformal theory, thus leading to a generally scheme-independent prediction. Because of the uncalculated high-order terms, there is residual scale dependence for the PMC prediction. However such residual renormalization scale dependence is generally small either due to the perturbative nature of the PMC scales or due to the fast convergence of the conformal pQCD series~\footnote{By choosing a proper scale for the highest-order terms, whose value cannot be fixed, one can achieve a scheme-independent prediction due to commensurate scale relations among the predictions under different schemes~\cite{Brodsky:1994eh}.}. This explains why one refers to the PMC method as ``principle of maximum conformality". The scheme independence of the PMC prediction is a general result, satisfying the central property of RGI.
\end{itemize}

\subsection{The PMC scale setting formulism for dimensional-like ${\cal R}_\delta$-scheme} \label{PMCRdelta}

In this subsection, we take the dimensional-like ${\cal R}_\delta$-scheme to show the how to do PMC scale setting. The ${\cal R}_\delta$-scheme introduces a generalization of the conventional dimensional regularization schemes, where a constant $(-\delta)$ is subtracted in addition to the standard subtraction $(\ln 4 \pi - \gamma_E)$ of the $\overline{\rm MS}$-scheme~\cite{Mojaza:2012mf}. Different $\delta$-values indicate different dimensional-like schemes. For examples, $\delta=0$ is the $\overline{\rm MS}$-scheme, $\delta=\ln4\pi-\gamma_E$ is the ${\rm MS}$-scheme, and $\delta=-2$ is the G-scheme~\cite{Chetyrkin:1980pr}.

The pQCD approximants among different ${\cal R}_\delta$-schemes are simply related by a scale shift~\cite{Mojaza:2012mf, Brodsky:2013vpa}. One can derive a general pQCD expression in the ${\cal R}_\delta$-schemes by using the displacement relation between the couplings at different scales,
\begin{eqnarray}\label{eq:running}
a_{\mu} = a_{\mu_\delta} + \sum_{n=1}^\infty \frac{1}{n!} { \frac{{\rm d}^n a_\mu}{({\rm d} \ln \mu^2)^n}\bigg|_{\mu=\mu_\delta} (-\delta)^n},
\end{eqnarray}
where $\delta=\ln\mu^2_\delta/\mu^2$. Thus one can rewrite the conventional pQCD series (\ref{pQCDexp}) as
\begin{eqnarray}
\rho_{n}(Q) &=& r_1 a_{\mu_\delta} + (r_2 + \beta_0 r_1 \delta) a_{\mu_\delta}^2 +[r_3 +\beta_1 r_1 \delta + 2\beta_0 r_2 \delta + \beta_0^2 r_1 \delta^2] a_{\mu_\delta}^3 \nonumber\\
&&+[r_4 +\beta_2 r_1 \delta + 2\beta_1 r_2 \delta + 3\beta_0 r_3 \delta + 3\beta_0^2 r_2 \delta^2 +\beta_0^3 r_1 \delta^3 +\frac{5}{2} \beta_1 \beta_0 r_1 \delta^2] a_{\mu_\delta}^4 +\cdots.
\label{eq:rhodelta1}
\end{eqnarray}
It is easy to confirm that,
\begin{equation}
\frac{\partial \rho_n}{\partial \delta} = -\beta(a_{\mu_\delta}) \frac{\partial \rho_n}{\partial a_{\mu_\delta}}.
\label{eq:deltaRGI}
\end{equation}
This shows that when the non-conformal $\{\beta_i\}$-terms associated with the $\beta(a_{\mu_\delta})$-function have been removed, one can achieve a scheme-independent prediction for the physical observable $\rho$; i.e. $\beta(a_{\mu_\delta})\to 0$ indicates ${\partial \rho_n}/{\partial \delta}\to 0$. The PMC scales determined by using those non-conformal terms depend on the choice of renormalization scheme, which however are compensated by scheme-dependent coefficients at each order, leading to the final scheme-independent conformal series.

The running coupling $\alpha_s$ at each perturbative order has its own $\{\beta_i\}$-series governed by the RGE. The $\beta$-pattern for the pQCD series at each order is a superposition of all of the $\{\beta_i\}$-terms which govern the lower-order $\alpha_s$ contributions at this particular order. The pQCD prediction $\rho_{n}$ with its explicit $\beta$-pattern can be rewritten as
\begin{eqnarray}
\rho_{n}(Q) &=& \sum_{i=1}^{n} r_{i}(\mu/Q) a_{\mu}^{i}  \label{rho} \\
&=& r_{1,0} a_{\mu} + \left[r_{2,0} + \beta_0 r_{2,1} \right] a^2_{\mu} + \left[r_{3,0} + \beta_1 r_{2,1} + 2 \beta_0 r_{3,1} + \beta _0^2 r_{3,2} \right] a^3_{\mu} \nonumber\\
&& +\left[r_{4,0} + \beta_2 r_{2,1} + 2\beta_1 r_{3,1} + \frac{5}{2} \beta_1 \beta_0 r_{3,2} +3\beta_0 r_{4,1} + 3 \beta_0^2 r_{4,2} + \beta_0^3 r_{4,3} \right] a^4_{\mu} + \cdots .
\label{eq:rhodelta2}
\end{eqnarray}
where we have introduced the notes $r_{i,j}$. The non-conformal coefficients $r_{i,j(\geq1)}$ are general functions of $\mu$ and $Q$, which are usually in form of $\ln\mu^2/Q^2$. For convenience, we identify the coefficients $r_{i,j(\geq1)}$ as $r_{i,j}=\sum_{k=0}^{j} C_j^k \ln^k(\mu^2/Q^2) \hat{r}_{i-k,j-k}$, in which $\hat{r}_{i,j}=r_{i,j}|_{\mu=Q}$ and the combination coefficients $C_j^k=({j!}/{k!(j-k)!})$. All the conformal coefficients are free from $\mu$-dependence, e.g., $r_{i,0}\equiv\hat{r}_{i,0}$.

Next, we rewrite the pQCD expansion (\ref{eq:rhodelta2}) into a compact form as
\begin{eqnarray}
\rho_n(Q) = \sum\limits^{n}_{i \ge 1} {r_{i,0}}{a_\mu^i} + \sum\limits^{i+j\leq n}_{i\ge1, j\ge1} (-1)^{j} \left[i\beta(a_\mu) a_\mu^{i-1}\right] r_{i+j,j} \Delta_{i}^{(j-1)}(a_\mu),
\label{eq:generalization}
\end{eqnarray}
where the summation keeps the expansion up to $a_\mu^n$-order. For a fourth-order prediction, we need to know the first three $\Delta_{i}^{(j-1)}(x)$, which are
\begin{eqnarray}
\Delta_i^{(0)}(x) &=& 1, \\
\Delta_i^{(1)}(x) &=& \frac{1}{2!}\left[\frac{\partial\beta(x)}{\partial x}+ (i-1)\frac{\beta(x)}{x}\right], \\
\Delta_i^{(2)}(x) &=& \frac{1}{3!}\left[\beta(x) \frac{\partial^2 \beta(x)}{(\partial x)^2} + \left(\frac{\partial\beta(x)}{\partial x}\right)^2 + 3(i-1)\frac{\beta(x)}{x}\frac{\partial\beta(x)}{\partial x} + (i-1)(i-2)\frac{\beta(x)^2}{x^2} \right].
\end{eqnarray}

Following the standard PMC procedures, we can obtain the following conformal series for a pQCD approximant,
\begin{equation}
\rho_{n}(Q)|_{\rm PMC}=\sum\limits^{n}_{i=1} r_{i,0} a_{Q_i}^i,
\end{equation}
where only the conformal coefficients $r_{i,0}$ remain, and the PMC scales $Q_i$ for each order are determined by recursively absorbing the $\{\beta_i\}$-terms into the coupling at the corresponding order and by resumming the known type of $\{\beta_i\}$-terms up to all orders. The PMC scales satisfy
\begin{equation}
\sum_{j\geq 0}{\Delta_i^{(j)}(a_\mu) \ln^{j+1}\frac{Q_i^2}{\mu^2}} = \sum_{j\geq 0} (-1)^{j+1} {\Delta_i^{(j)}(a_\mu) \frac{r_{i+j+1,j+1}}{r_{i,0}}}.
\label{eq:Qn}
\end{equation}
Then we have
\begin{eqnarray}
\ln\frac{Q_i^2}{Q^2} &=& \sum_{0\leq j \leq (n-1-i)} P_{i,j} a^j_\mu. \;\;\; i\in[1,(n-1)].
\label{eq:pmcscale1}
\end{eqnarray}
For an $n_{\rm th}$-order pQCD prediction, we can fix $(n-1)$ PMC scales. Solving Eq.(\ref{eq:Qn}) iteratively, we can obtain the perturbative coefficients $P_{i,j}$ needed for those $(n-1)$ PMC scales. For example, for a fourth-order pQCD prediction, we have
\begin{eqnarray}
P_{i,0} &=& -\frac{r_{i+1,1}}{r_{i,0}}, \\
P_{i,1} &=& \frac{(i+1)(r_{i+1,1}^2-r_{i,0}r_{i+2,2})}{2r_{i,0}^2} \beta_0, \\
P_{i,2} &=& \frac{(i+2)(r_{i+1,1}^2-r_{i,0}r_{i+2,2})}{2r_{i,0}^2} \beta_1 \nonumber\\
           & & -\frac{(i+1)[(2i+1)r_{i+1,1}^3-3(i+1)r_{i,0} r_{i+1,1}r_{i+2,2}+(i+2)r_{i,0}^2 r_{i+3,3}]}{6r_{i,0}^3}\beta_0^2.
\end{eqnarray}
The above expressions show that the PMC scale $Q_i$ is given as a perturbative series, e.g. $Q_1$ is at the ${\rm N^{n-1}LLO}$ level, $Q_2$ is at the ${\rm N^{n-2}LLO}$ level, and etc., for a $n_{\rm th}$-order prediction.

The PMC scale setting procedures have been successfully applied in many high-energy processes, such as the top-pair production at the ${\rm N^2LO}$ level~\cite{Brodsky:2012sz, Brodsky:2012rj, Brodsky:2012ik, Wang:2015lna, Wang:2014sua}, the electron-positron annihilation into hadrons at the ${\rm N^3LO}$ level, the decay width $\Gamma(H\to b\bar{b})$ at ${\rm N^3LO}$ level~\cite{Wang:2013bla}, the hadronic $Z$ decays at ${\rm N^3LO}$ level~\cite{Wang:2014aqa}, the decay width $\Gamma (H \to gg)$ at ${\rm N^4LO}$ level~\cite{Zeng:2018jzf}, the decay width $\Gamma(H\to \gamma\gamma)$ at ${\rm N^2LO}$ level~\cite{Wang:2013akk}, the $\rho$ parameter at ${\rm N^3LO}$ level~\cite{Wang:2014wua}, the $\Upsilon(1S)$ decays at ${\rm N^3LO}$ level~\cite{Shen:2015cta}, and etc. These applications not only show the essential features of PMC but also emphasize the importance of a proper renormalization scale setting for achieving precise fixed-order pQCD predictions.

\begin{figure}[htb]
\epsfysize=9.0cm
\begin{center}
\begin{minipage}[t]{10 cm}
\epsfig{file=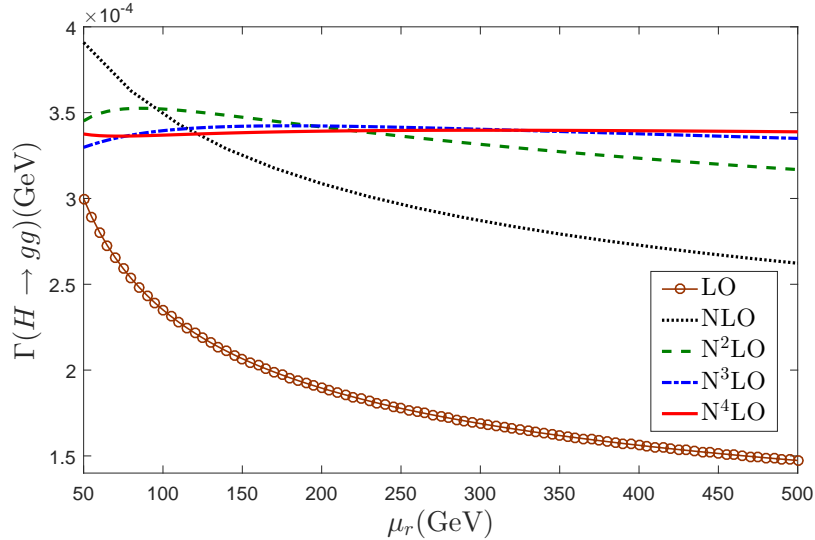,scale=0.6}
\end{minipage}
\begin{minipage}[t]{16.5 cm}
\caption{Total decay width $\Gamma (H \to gg)$ versus the initial choice of renormalization scale $\mu_r$ using conventional scale setting up to five-loop level~\cite{Zeng:2018jzf}, the scale dependence becomes smaller as more loop terms are taken into consideration. The solid line with circle symbols, the dotted line, the dashed line, the dash-dot line and the solid line are for the predictions up to $\rm{LO}$, $\rm{NLO}$, $\rm{N^2LO}$, $\rm{N^3LO}$, and $\rm{N^4LO}$ levels, respectively.}
\label{HtoggConv}
\end{minipage}
\end{center}
\end{figure}

\begin{figure}[htb]
\epsfysize=9.0cm
\begin{center}
\begin{minipage}[t]{10 cm}
\epsfig{file=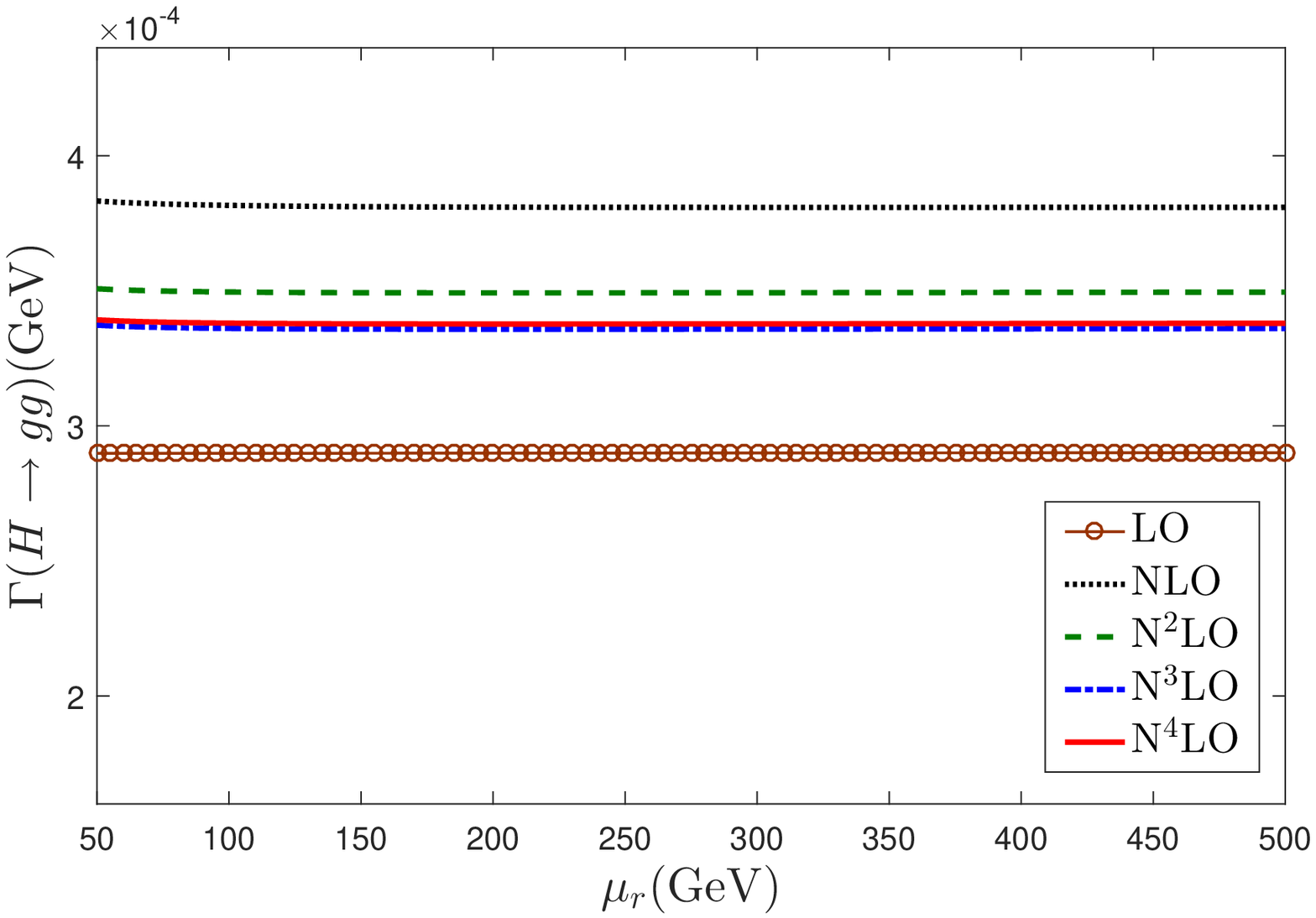,scale=0.6}
\end{minipage}
\begin{minipage}[t]{16.5 cm}
\caption{Total decay width $\Gamma (H \to gg)$ versus the initial choice of renormalization scale $\mu_r$ after applying the PMC up to five-loop level~\cite{Zeng:2018jzf}, whose value is scale-independent even for low-order predictions. The solid line with circle symbols, the dotted line, the dashed line, the dash-dot line and the solid line are for the predictions up to $\rm{LO}$, $\rm{NLO}$, $\rm{N^2LO}$, $\rm{N^3LO}$, and $\rm{N^4LO}$ levels, respectively.}
\label{mMOMPMC}
\end{minipage}
\end{center}
\end{figure}

As an example, we present the renormalization scale dependence of the total decay width $\Gamma (H \to gg)$ up to $\rm{N^4LO}$ level in Figures \ref{HtoggConv} and \ref{mMOMPMC}, which well explains how the pQCD series behaves before and after applying the PMC. Figure \ref{HtoggConv} shows that using conventional scale setting, the renormalization scale dependence becomes smaller when more loop terms have been taken into consideration. This trend agrees with the conventional wisdom that by finishing a higher-enough-order calculation, one can finally achieve desirable convergent and renormalization scale-invariant estimations. As a comparison, Figure \ref{mMOMPMC} shows that the PMC prediction for the total decay width is renormalization scale independent even for low-order predictions, and the PMC prediction quickly approaches the ``physical" value of $\Gamma (H \to gg)$ due to a much faster pQCD convergence.

\subsection{The residual renormalization scale dependence}

The PMC scale is determined by absorbing the $\{\beta_i\}$-terms of the process, where the $\beta$-pattern at each order is determined by the recursive use of the RGE (\ref{eq:betafun}). Since the determined PMC scale is independent of the choice of the initial renormalization scale, the conventional scale ambiguity is eliminated.

There are two kinds of residual scale dependence for an $n_{\rm th}$-order pQCD prediction $\rho_n$. The first one is caused by the unknown terms in the determined PMC scales, such as $Q_{1,\cdots,n-1}$, due to their perturbative nature. The second residual scale dependence is for the undetermined PMC scale $Q_n$ for the highest perturbative term of the pQCD approximant, since we have no $\{\beta_i\}$-terms to fix its value. In practice, one can set its value as the latest determined PMC scale, i.e. $Q_{n}=Q_{n-1}$; such a choice of $Q_{n}$ ensures the scheme independence of the PMC prediction. It should be pointed out that such residual scale dependence is different from the arbitrary conventional scale dependence. The first kind of residual scale dependence is also reduced by the exponential suppression, leading to negligible residual scale dependence. The magnitude of the residual scale dependence depends on perturbative nature of the PMC scale, and thus depends heavily on how well we know the $\{\beta_i\}$-terms of the pQCD series. The precision of the PMC scale for high-order terms decreases at higher-and-higher orders due to the less known $\{\beta_i\}$-terms in those higher-order terms. In practice, we have found that those two residual scale dependence are quite small even at low orders. This is due to a generally faster pQCD convergence after applying the PMC. Some PMC examples can be found in Ref.\cite{Wu:2015rga}.

\begin{figure}[htb]
\epsfysize=9.0cm
\begin{center}
\begin{minipage}[t]{10 cm}
\epsfig{file=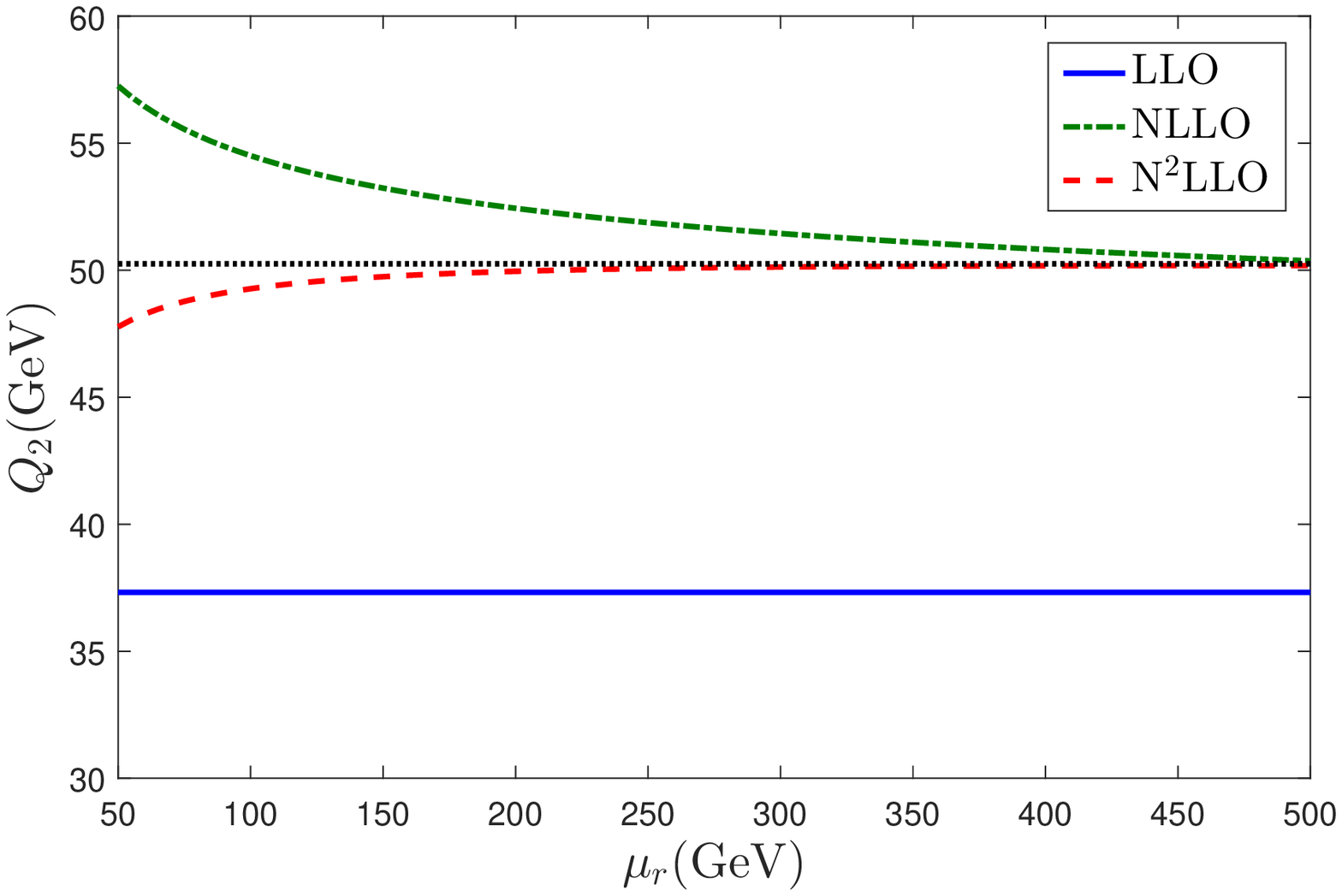,scale=0.6}
\end{minipage}
\begin{minipage}[t]{16.5 cm}
\caption{The PMC scale $Q_2$ versus the initial choice of renormalization scale $\mu_r$ for the NLO-terms of the pQCD series from the $\rm{LLO}$-level up to $\rm{N^2LLO}$-level versus the initial scale choices~\cite{Zeng:2018jzf}. The dotted line represents the approximate asymptotic limit for $Q_2$. It shows that the precision of $Q_2$ increases when more loop terms have been taken into consideration. }
\label{PMCQ2}
\end{minipage}
\end{center}
\end{figure}

\begin{figure}[htb]
\epsfysize=9.0cm
\begin{center}
\begin{minipage}[t]{10 cm}
\epsfig{file=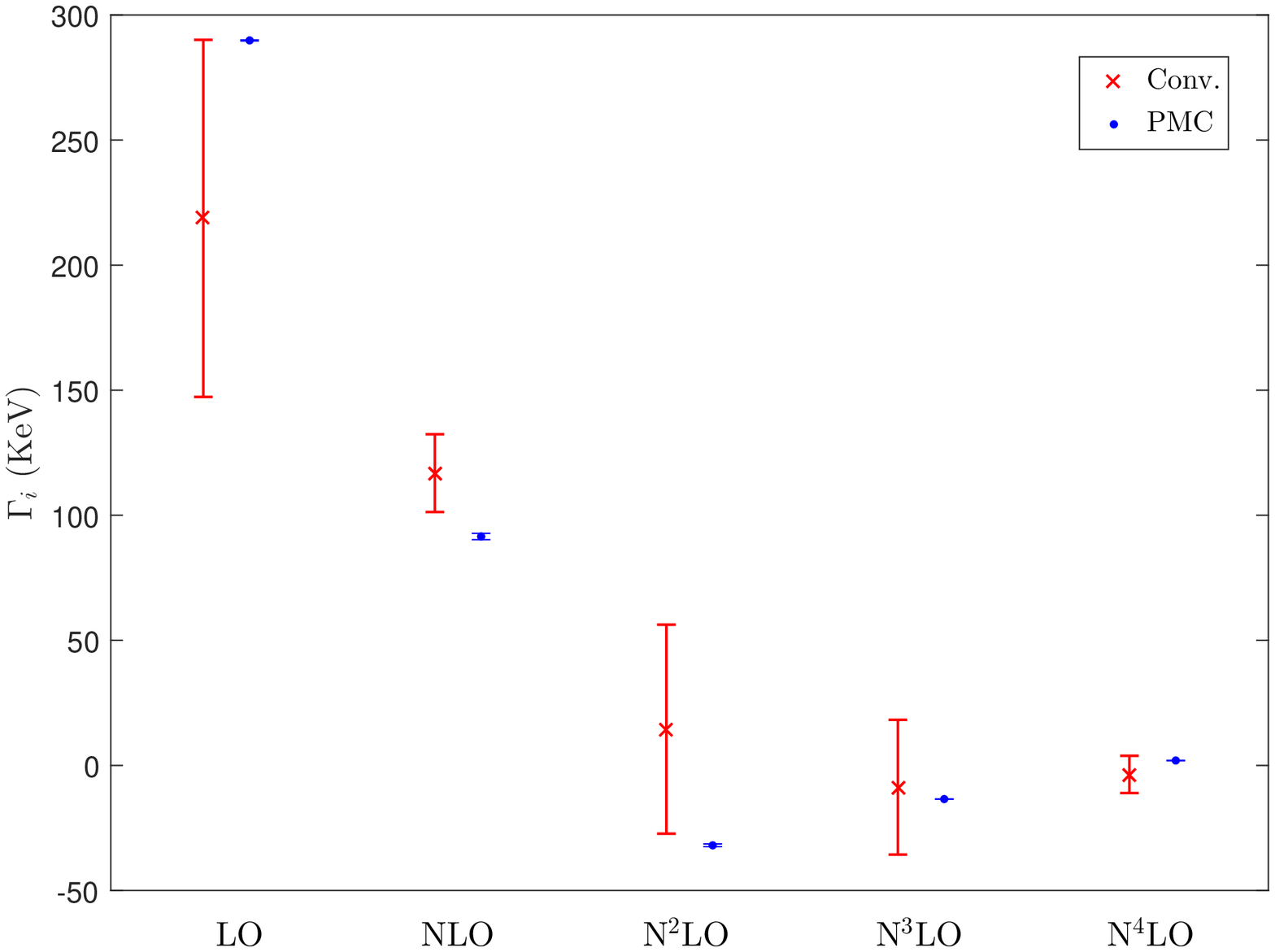,scale=0.6}
\end{minipage}
\begin{minipage}[t]{16.5 cm}
\caption{Scale uncertainties of the individual decay width $\Gamma_i$ (in unit: $\rm{KeV}$) under conventional and PMC scale setting approaches~\cite{Zeng:2018jzf}. $i=\rm{LO}$, $\rm{NLO}$, $\rm{N^2LO}$, $\rm{N^3LO}$, and $\rm{N^4LO}$, respectively. The central values are for $\mu = M_H $, and the errors are for $\mu \in \left[{M_H}/{2},4 M_H\right]$. This figure shows that the separate scale errors for each perturbative term $\Gamma_i|_{\rm Conv.}$ are quite large under conventional scale setting. And the residual scale dependence for the PMC prediction is negligibly small for each term $\Gamma_i|_{\rm PMC}$. \label{individual} }
\end{minipage}
\end{center}
\end{figure}

In some cases, a weak perturbative convergence may be exist in a PMC scale; the residual scale dependence for this particular scale may be large, leading to a comparatively larger residual scale dependence for the pQCD approximant. As an example, it has been found that there is comparatively large $\mu_r$ dependence for the NLO PMC scale $Q_2$ for $H\to gg$ process, leading to a larger residual scale dependence for $H\to gg$ decay width~\cite{Zeng:2015gha, Zeng:2018jzf}. Figure \ref{PMCQ2} shows that the precision of $Q_2$ increases as more loop terms have been taken into consideration. By varying the initial renormalization scale within the region of $[M_H/2, 4M_H]$, the scale dependence of $Q_2$ at the $\rm{NLLO}$ level is about $\Delta Q_2 \sim 7~\rm{GeV}$, which changes down to $\Delta Q_2 \sim 2~\rm{GeV}$ for $Q_2$ at the $\rm{N^2LLO}$ level. This will lead to residual scale dependence for the total decay width. When one uses the known $\rm{N^{4}LO}$ pQCD series for the $H\to gg$ decay~\cite{Herzog:2017dtz}, the PMC prediction of the total decay width is
\begin{equation}
\Gamma (H\to gg)\rm{|_{ PMC }} =337.9_{-0.1}^{+0.9}\;\;{\rm KeV},
\end{equation}
where the error is the residual scale dependence, which mainly comes from the NLO-terms.

As a comparison, it has been found that the renormalization scale uncertainty for $\rm{N^{4}LO}$ using the conventional scale setting is $\left({}_{-1.3}^{+2.1}\right)$ KeV~\cite{Zeng:2018jzf}, which is larger than that of PMC prediction but is also small. This indicates that if one knows enough higher-order terms, the conventional scale uncertainty can also be suppressed to a certain degree. However we should point out that such a small scale dependence for conventional scale setting is caused by the large cancellation of the scale dependence among different orders; the scale dependence for each perturbative term is still very large, which cannot be cured by higher-order terms. Up to $\rm{N^{4}LO}$ level, the total decay width uses $\rm{\Gamma}=\sum_{i={\rm LO}}^{\rm N^{4}LO} \Gamma_i$. We present a comparison of the scale uncertainties of the individual decay width $\Gamma_i$ under conventional and PMC scale-settings in Figure \ref{individual}, where the error bars are determined by
\begin{equation}
\Delta=\pm| \Gamma_{\rm{N^kLO}}(\mu)-\Gamma_{\rm{N^kLO}}(M_H)|_{\rm MAX}.
\end{equation}
Here the symbol `MAX' stands for the maximum value by varying $\mu$ within the range of $\left[{M_H}/{2},4 M_H\right]$. Figure \ref{individual} shows that the separate scale errors for each perturbative term $\Gamma_i$ are quite large using conventional scale setting, which can be as large as an order of magnitude. On the other hand, those uncertainties for each order are rather small using PMC scale setting; the maximum uncertainty is due to the somewhat large residual scale dependence for the NLO-terms, $\Delta|_{\rm PMC,NLO}=\pm1.3$ GeV. It should be noted that due to the conformal nature of the PMC series, the PMC predictions are scheme-independent at any fixed order. This fixed-order scheme independence is also ensured by the commensurate scale relations among different observables~\cite{Brodsky:1994eh}.

\subsection{Some recent applications of PMC scale setting}

In this subsection, we present some recent PMC applications, which show essential features of PMC and the importance of proper renormalization scale-setting. Some subtle points in using the PMC will also be explained, which can be treated as useful references for future applications.

\subsubsection{The hadroproduction of the Higgs boson}

The total cross section for the production of Higgs boson at hadron colliders can be treated as the convolution of the hard-scattering partonic cross section $\hat \sigma_{ij}$ with the corresponding parton luminosity ${\cal L}_{ij}$, i.e.
\begin{equation}
\sigma_{H_1 H_2 \to {H X}} = \sum_{i,j} \int\limits_{m^2_{H}}^{S}\, ds \,\, {\cal L}_{ij}(s, S, \mu_f) \hat \sigma_{ij}(s,L,R) , \label{basic}
\end{equation}
where the parton luminosity
\begin{equation}
{\cal L}_{ij} = {1\over S} \int\limits_s^S {d\hat{s}\over \hat{s}} f_{i/H_1}\left(x_1,\mu_f\right) f_{j/H_2}\left(x_2,\mu_f\right). \label{eq:Lij}
\end{equation}
Here the indices $i,j$ run over all possible parton flavors in proton $H_1$ or $H_2$, $x_1= {\hat{s} / S}$ and $x_2= {s / \hat{s}}$. $S$ denotes the hadronic center-of-mass energy squared, and $s=x_1 x_2 S$ is the subprocess center-of-mass energy squared. The subprocess cross section $\hat \sigma_{ij}$ depends on both the renormalization scale $\mu_r$ and the factorization scale $\mu_f$, and the parton luminosity depends on $\mu_f$. We define two ratios $L=\mu_f^2/m_H^2$ and $R=\mu_r^2/\mu_f^2$, where $m_H$ is the Higgs boson mass. The parton distribution functions (PDF) underlying the parton luminosity $f_{i/H_{\alpha}}(x_\alpha,\mu_f)$ ($\alpha=1$ or $2$) describes the probability of finding a parton of type $i$ with light-front momentum fraction between $x_\alpha$ and $x_{\alpha} +dx_{\alpha}$ in the proton $H_{\alpha}$. The two-dimensional integration over $s$ and $\hat{s}$ can be performed numerically by using the VEGAS program~\cite{Lepage:1977sw}. For this purpose, one can set $s=m_H^2 (S/m_H^2)^{y_1}$ and $\hat{s}=s(S/s)^{y_2}$, and transform the two-dimensional integration into an integration over two variables $y_{1,2}\in[0,1]$.

Analytic expressions using the $\overline{\rm MS}$-scheme for the partonic cross section $\hat{\sigma}_{ij}$ up to N$^2$LO level can be found in Refs.\cite{Anastasiou:2002yz, Ravindran:2003um}, which can be used for the PMC analysis. There are two types of large logarithmic terms $\ln(\mu_r/m_{H})$ and $\ln(\mu_r/m_{t})$ in $\hat{\sigma}_{ij}$. Thus a single guessed scale, using conventional scale-setting, such as $\mu_r=m_H$, cannot eliminate all of the large logarithmic terms. This explains why there are large $K$ factors for the high-order terms, confirming the importance of achieving exact values for each order. The PMC uses the RGE to determine the optimal running behavior of $\alpha_s$ at each order, and the large scale uncertainty for each order using conventional scale setting can be eliminated. To be specific, the PMC introduces multiple scales for physical applications which depend on multiple kinematic variables, which is caused by the fact that different typical momentum flows could exist in different kinematic regions. Similar conditions have been observed in the hadronic $Z$ decays~\cite{Wang:2014aqa} and the heavy-quark pair production via $q\bar{q}$ fusion~\cite{Brodsky:1995ds}. For example, the process $q \bar q \to Q\bar Q$ near the heavy quark ($Q$) threshold involves not only the invariant variable $\hat{s} \sim 4 M^2_Q$, but also the variable $v_{\rm rel}^2 \hat s$ with $v_{\rm rel}$ being the relative velocity, which enters the Sudakov final-state corrections.

\begin{table}[htb]
\centering
\begin{tabular}{cccccc}
\hline
 & Tevatron & \multicolumn{4}{c}{LHC } \\
\hline
$\sqrt{S}$ & 1.96 TeV & 7 TeV & 8 TeV & 13 TeV & 14 TeV \\
\hline
Conv. & $0.63^{+0.13}_{-0.11}$ & $13.92^{+2.25}_{-2.06}$ & $18.12^{+2.87}_{-2.66}$ & $44.26^{+6.61}_{-6.43}$ & $50.33^{+7.47}_{-7.31}$ \\
PMC & $0.86^{+0.13}_{-0.12}$ & $18.04^{+1.36}_{-1.32}$ & $23.37^{+1.65}_{-1.59}$ & $56.34^{+3.45}_{-3.00}$ & $63.94^{+3.88}_{-3.30}$ \\
\hline
\end{tabular}
\caption{The total hadronic cross section $\sigma_{\rm sum}$ (in unit: pb) using the conventional (Conv.) and PMC scale-settings~\cite{Wang:2016wgw}, where the uncertainties are for $\mu_r \in[m_H/2,2m_H]$ and $\mu_f \in[m_H/2,2m_H]$. } \label{HiggsproTlhc}
\end{table}

We use $\sigma_{\rm sum}$ to stand for the sum of the total hadronic production cross sections $\sigma_{(ij)}$ with $(ij) = (gg)$, $(q{\bar q})$, $(gq)$, $(g\bar{q})$ and $(qq')$, respectively. Numerical results for $\sigma_{\rm sum}$ at the Tevatron and LHC are presented in Table~\ref{HiggsproTlhc}~\cite{Wang:2016wgw}, where the uncertainties are for $\mu_r \in[m_H/2,2m_H]$ and $\mu_f \in[m_H/2,2m_H]$. As a comparison, the results using conventional scale-setting are also presented. After applying the PMC, $\sigma_{\rm sum}$ is increased by $\sim37\%$ at the Tevatron, and by $\sim30\%$ at the LHC for $\sqrt{S}=$7, 8, 13 and 14 TeV, respectively.

\begin{table}[htb]
\centering
\begin{tabular}{cccc}
\hline
Decay channel & \multicolumn{3}{c}{$\sigma_{\rm Incl}$} \\ \cline{2-4}
 & 7 TeV & 8 TeV & 13 TeV \\
\hline
$H\to\gamma\gamma$~\cite{Aad:2015lha, TOTCS:ATLAS, ATLAS:2017ovn} & $35^{+13}_{-12}$ & $30.5^{+7.5}_{-7.4}$ & $47.9^{+9.1}_{-8.6}$ \\
$H\to ZZ^{*}\to 4l$~\cite{Aad:2015lha,TOTCS:ATLAS, ATLAS:2017ovn} & $33^{+21}_{-16}$ & $37^{+9}_{-8}$ & $68.0^{+11.4}_{-10.4}$ \\
LHC-XS~\cite{deFlorian:2016spz} & $19.2\pm0.9$ & $24.5\pm1.1$ & $55.6^{+2.4}_{-3.4}$ \\
PMC & $21.21^{+1.36}_{-1.32}$ & $27.37^{+1.65}_{-1.59}$ & $65.72^{+3.46}_{-3.01}$ \\
\hline
\end{tabular}
\caption{Total inclusive cross sections (in unit: pb) for Higgs production at the LHC for the CM collision energies $\sqrt{S}=7$, 8 and 13 TeV, respectively~\cite{Wang:2016wgw}. The inclusive cross section is $\sigma_{\rm Incl}=\sigma_{\rm sum}+\sigma_{\rm xH}+\sigma_{\rm EW}$. } \label{HiggsproATLAS}
\end{table}

\begin{figure}[htb]
\epsfysize=9.0cm
\begin{center}
\begin{minipage}[t]{10 cm}
\epsfig{file=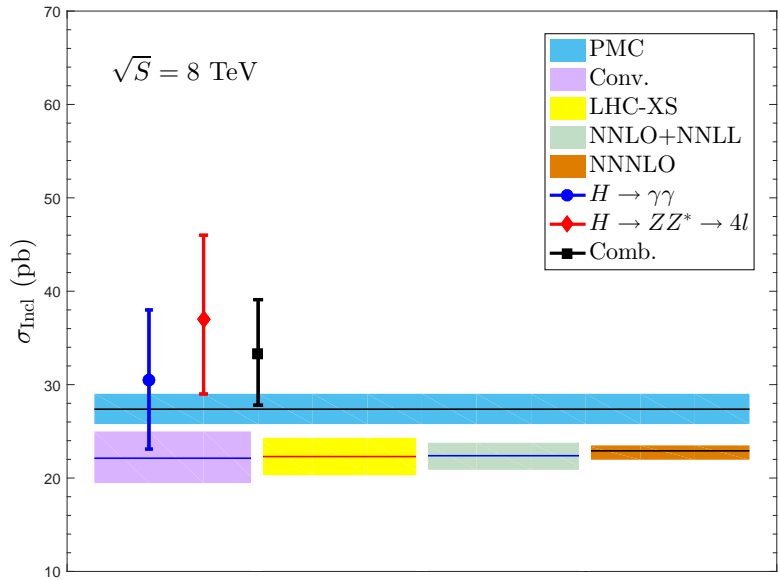,scale=0.6}
\end{minipage}
\begin{minipage}[t]{16.5 cm}
\caption{Comparison of the N$^2$LO conventional versus PMC predictions for the total inclusive cross section $\sigma_{\rm Incl}$~\cite{Wang:2016wgw} with the latest ATLAS measurements at 8 TeV~\cite{Aad:2015lha}. The LHC-XS predictions~\cite{Heinemeyer:2013tqa}, the N$^2$LO+NNLL prediction~\cite{deFlorian:2012yg}, and the N$^3$LO prediction~\cite{Anastasiou:2015ema} are presented as a comparison. The solid lines are central values. \label{ATLASConvPMC}}
\end{minipage}
\end{center}
\end{figure}

To compare with the LHC measurements for Higgs boson production cross-section~\cite{Aad:2015lha, TOTCS:ATLAS, ATLAS:2017ovn}, we need to include the contributions from other known production modes, such as the vector-boson fusion production process, the $WH/ZH$ Higgs associated production process, the Higgs production associated with heavy quarks, etc. We use $\sigma_{\rm xH}$ to stand for the sum of those production cross sections from the channels via $Z$, $W$, $t\bar{t}$, $b\bar{b}$ and $\cdots$, and use $\sigma_{\rm EW}$ to stand for those production cross sections from the channels with electroweak corrections. The values of $\sigma_{\rm xH}$ and $\sigma_{\rm EW}$ are small in comparison to the dominant gluon-fusion $\sigma_{\rm ggH}$ contribution. Taking $\sqrt{S}=8$ TeV and $m_H=125$ GeV, one predicts $\sigma_{xH}=3.08+0.10$ pb~\cite{Aad:2015lha, Heinemeyer:2013tqa}; the electro-weak correction up to two-loop level leads to a $+5.1\%$ shift with respect to the N$^2$LO-level QCD cross sections~\cite{Actis:2008ts, Actis:2008ug}. Taking those contributions into consideration, the PMC predictions for the total inclusive cross section $\sigma_{\rm Incl}$ at the LHC for several center-of-mass (CM) collision energies are presented in Table~\ref{HiggsproATLAS}; the LHC ATLAS predictions via $H\to\gamma\gamma$ and $H\to ZZ^{*}\to 4l$ decay channels~\cite{Aad:2015lha, TOTCS:ATLAS} are also given. The PMC results are larger than the central values of the LHC-XS prediction~\cite{deFlorian:2016spz} by about $10\%$, $12\%$ and $18\%$ for $\sqrt{S}=7$, 8 and 13 TeV, respectively, which shows a better agreement with the data. This is clearly shown by Figure \ref{ATLASConvPMC}, in which a comparison of our present N$^2$LO conventional and PMC predictions for $\sigma_{\rm Incl}$ with the ATLAS measurements at 8 TeV is presented. Because of the large uncertainty for the ATLAS data, we need more data to draw definite conclusion on the SM predictions. The more accurate measurements with high integrated luminosity for $\sqrt{S}$=13 TeV will be helpful to test the PMC and conventional predictions.

\begin{table}[htb]
\centering
\begin{tabular}{cccc}
\hline
$\sigma_{\rm fid}(pp\rightarrow H\rightarrow\gamma\gamma)$ & 7 TeV & 8 TeV & 13 TeV \\
\hline
ATLAS data~\cite{sigmafid:ATLAS} & $49\pm18$ & $42.5^{+10.3}_{-10.2}$ & $52^{+40}_{-37}$ \\
CMS data~\cite{CMS:2017nyv} & - & - & $84_{-12}^{+13}$ \\
ATLAS data~\cite{ATLAS:2018uso} & - & - & $60.4 \pm 8.6$ \\
LHC-XS~\cite{deFlorian:2016spz} & $24.7\pm2.6$ & $31.0\pm3.2$ & $66.1^{+6.8}_{-6.6}$ \\
PMC prediction & $30.1^{+2.3}_{-2.2}$ & $38.3^{+2.9}_{-2.8}$ & $85.8^{+5.7}_{-5.3}$ \\
\hline
\end{tabular}
\caption{The fiducial cross section $\sigma_{\rm fid}(pp\to H\to \gamma\gamma)$ (in unit: fb) at the LHC for CM collision energies $\sqrt{S}=$7, 8 and 13 TeV, respectively~\cite{Wang:2016wgw}. } \label{sigmafidATLAS}
\end{table}

It has been suggested that the fiducial cross section $\sigma_{\rm fid}$ can also be used to test the theoretical predictions, which is defined as
\begin{eqnarray}
\sigma_{\rm fid}(pp\to H\to \gamma\gamma)=\sigma_{\rm Incl}{\cal B}_{H\to \gamma\gamma}{\cal A}.
\end{eqnarray}
The ${\cal A}$ is the acceptance factor, whose values for three typical proton-proton CM collision energies are~\cite{sigmafid:ATLAS}, ${\cal A}|_{\rm 7TeV}=0.620\pm0.007$, ${\cal A}|_{\rm 8TeV}=0.611\pm0.012$ and ${\cal A}|_{\rm 13TeV}=0.570\pm0.006$. The ${\cal B}_{H\to \gamma\gamma}$ is the branching ratio of $H\to \gamma\gamma$. By using the $\Gamma(H\to \gamma\gamma)$ with conventional scale-setting, the LHC-XS group predicts ${\cal B}_{H\to \gamma \gamma} =0.00228\pm0.00011$~\cite{Heinemeyer:2013tqa}. A PMC analysis for $\Gamma(H\to \gamma\gamma)$ up to three-loop or five-loop level has been given in Refs.\cite{Wang:2013akk, Yu:2018hgw}. Using the formulae given there, we obtain $\Gamma(H\to \gamma\gamma)|_{\rm PMC}=9.34\times10^{-3}$ MeV for $m_H=125$ GeV. Using this value, together with Higgs total decay width $\Gamma_{\rm Total}=(4.07\pm0.16)\times 10^{-3}$ GeV~\cite{Heinemeyer:2013tqa}, we obtain ${\cal B}_{H\to \gamma\gamma}|_{\rm PMC}=0.00229\pm0.00009$. The PMC predictions for $\sigma_{\rm fid}(pp\to H\to \gamma\gamma)$ at the LHC are given in Table~\ref{sigmafidATLAS}, where the ATLAS and CMS measurements~\cite{sigmafid:ATLAS, CMS:2017nyv, ATLAS:2018uso} and the LHC-XS predictions~\cite{deFlorian:2016spz} are also presented. The PMC fiducial cross sections are larger than the LHC-XS ones by $\sim22\%$, $\sim24\%$ and $\sim30\%$ for $\sqrt{S}=$7 TeV, 8 TeV and 13 TeV, respectively. Table~\ref{sigmafidATLAS} shows no significant differences between the measured fiducial cross sections and the SM predictions, and the PMC predictions show better agreement with the measurements at $\sqrt{S}=7$ TeV and 8 TeV.

\subsubsection{Top-quark pair production at hadron colliders and the top-quark pole mass}

Similar to the hadronic production of the Higgs boson, the total cross section for the top-quark pair production at the hadronic colliders can also be written as the convolution of the factorized partonic cross section $\hat{\sigma}_{ij}$ with the parton luminosities ${\cal L}_{ij}$:
\begin{eqnarray}
\sigma_{H_1 H_2 \to {t\bar{t} X}} = \sum_{i,j} \int\limits_{4m_{t}^2}^{S}\, ds \,\, {\cal L}_{ij}(s, S, \mu_f) \hat \sigma_{ij}(s,\alpha_s(\mu_r),\mu_r,\mu_f),
\end{eqnarray}
where the parton luminosities ${\cal L}_{ij}$ has been defined in Eq.(\ref{eq:Lij}), and the partonic cross section $\hat{\sigma}_{ij}$ has been computed up to N$^2$LO level,
\begin{eqnarray}
\hat{\sigma}_{ij} = \frac{1}{m_{t}^2} \left[f^0_{ij}(\rho,\mu_r,\mu_f) \alpha^2_s(\mu_r)+f^1_{ij}(\rho,\mu_r,\mu_f)\alpha^3_s(\mu_r) +f^2_{ij}(\rho,\mu_r,\mu_f)\alpha^4_s(\mu_r)+{\cal O}(\alpha_s^5)\right]
\end{eqnarray}
where $\rho=4m_{t}^2/s$, $(ij)=\{(q\bar{q}),(gg),(gq),(g\bar{q})\}$ stands for the four production channels, respectively. In the literature, the perturbative coefficients up to N$^2$LO level have been calculated by various groups, e.g. Refs.\cite{Nason:1987xz, Nason:1989zy, Beenakker:1988bq, Beenakker:1990maa, Czakon:2013goa, Moch:2008qy, Czakon:2008ii, Beneke:2011mq, Kidonakis:2010dk, Baernreuther:2012ws, Czakon:2012pz}. More explicitly, the LO, NLO and N$^2$LO coefficients $f^0_{ij}$, $f^1_{ij}$ and $f^2_{ij}$ in an $n_f$-power series can be explicitly read from the HATHOR program~\cite{Aliev:2010zk} and the Top++ program~\cite{Czakon:2011xx}. By identifying the $n_f$-terms associated with the $\{\beta_i\}$-terms in the coefficients $f^0_{ij}$, $f^1_{ij}$ and $f^2_{ij}$, and by using the degeneracy relations of $\beta$-pattern at different orders, one can determine the correct arguments of the strong couplings at each order and hence the PMC scales at each order by using the RGE via a recursive way~\cite{Brodsky:2012sz, Brodsky:2012rj}. The Coulomb-type corrections near the threshold region should be treated separately, since their contributions are enhanced by factors of $\pi$ and are sizable (e.g. those terms are proportional to $(\pi/v)$ or $(\pi/v)^2$~\cite{Brodsky:1995ds}, where $v=\sqrt{1-\rho}$, the heavy quark velocity). For this purpose, the Sommerfeld re-scattering formula is useful for a reliable prediction~\cite{Hagiwara:2008df, Kiyo:2008bv}.

\begin{table}[htb]
\centering
\begin{tabular}{ccccccccc}
\hline
& \multicolumn{4}{c}{Conventional scale-setting} & \multicolumn{4}{c}{PMC scale-setting} \\ \cline{2-5} \cline{6-9}
~~~ ~~~ & ~~LO~~ & ~~NLO~~ & ~~N$^2$LO~~ & ~~{\it Total}~~ & ~~LO~~ & ~~NLO~~ & ~~N$^2$LO~~ & ~~{\it Total}~~ \\
\hline
$(q\bar{q})$ channel & 4.87 & 0.96 & 0.48 & 6.32 & 4.73 & 1.73 & $-0.063$ & 6.35 \\
$(gg)$ channel  & 0.48 & 0.41 & 0.15 & 1.04 & 0.48 & 0.48 & 0.15 & 1.14 \\
$(gq)$ channel  & 0.00 & $-0.036$ & 0.0046 & $-0.032$ & 0.00 & $-0.036$ & 0.0046 & $-0.032$ \\
$(g\bar{q})$ channel & 0.00 & $-0.036$ & 0.0047& $-0.032$ & 0.00 & $-0.036$ & 0.0047 & $-0.032$ \\
sum   & 5.35 & 1.30 & 0.64 & 7.29 & 5.21 & 2.14 & 0.096 & 7.43 \\
\hline
\end{tabular}
\caption{The top-quark pair production cross sections (in unit: pb) before and after PMC scale-setting at the Tevatron with $\sqrt{S}=1.96$ TeV. $\mu_r=\mu_f=m_t$.} \label{tab2}
\end{table}

\begin{table}[htb]
\centering
\begin{tabular}{ccccccccc}
\hline
& \multicolumn{4}{c}{Conventional scale setting} & \multicolumn{4}{c}{PMC scale setting} \\ \cline{2-5} \cline{6-9}
~~~ ~~~ & ~~LO~~ & ~~NLO~~ & ~~N$^2$LO~~ & ~~{\it Total}~~ & ~~LO~~ & ~~NLO~~ & ~~N$^2$LO~~ & ~~{\it Total}~~\\
\hline
$(q\bar{q})$ channel & 23.37 & 3.42 & 1.86 & 28.69 & 22.32 & 7.23 & $-0.78$ & 28.62 \\
$(gg)$ channel  & 80.40 & 46.87 & 10.87 & 138.15 & 80.10 & 54.70 & 8.77 & 145.54 \\
$(gq)$ channel  & 0.00 & $-0.43$ & 1.41 & 1.03 & 0.00 & $-0.43$ & 1.41 & 1.03 \\
$(g\bar{q})$ channel & 0.00 & $-0.44$ & 0.24 & $-0.20$ & 0.00 & $-0.44$ & 0.24 & $-0.20$ \\
sum   & 103.77 & 49.42 & 14.38 & 167.67 & 102.42 & 61.06 & 9.64 & 174.98 \\
\hline
\end{tabular}
\caption{The top-quark pair production cross sections (in unit: pb) before and after PMC scale-setting at the LHC with $\sqrt{S}=7$ TeV. $\mu_r=\mu_f=m_t$.} \label{tab7}
\end{table}

\begin{table}[htb]
\centering
\begin{tabular}{ccccccccc}
\hline
& \multicolumn{4}{c}{Conventional scale setting} & \multicolumn{4}{c}{PMC scale setting} \\ \cline{2-5} \cline{6-9}
~~~ ~~~ & ~~LO~~ & ~~NLO~~ & ~~N$^2$LO~~ & ~~{\it Total}~~ & ~~LO~~ & ~~NLO~~ & ~~N$^2$LO~~ & ~~{\it Total}~~\\
\hline
$(q\bar{q})$ channel & 29.88 & 4.20 & 2.31 & 36.43 & 28.46 & 9.09 & $-1.06$ & 36.29 \\
$(gg)$ channel  & 118.10 & 67.43 & 15.01 & 200.57 & 117.66 & 78.53 & 11.92 & 210.86 \\
$(gq)$ channel  & 0.00 & 0.18 & 2.02 & 2.18 & 0.00 & 0.18 & 2.02 & 2.18 \\
$(g\bar{q})$ channel & 0.00 & $-0.53$ & 0.37 & $-0.15$ & 0.00 & $-0.53$ & 0.37 & $-0.15$ \\
sum   & 147.98 & 71.28 & 19.71 & 239.03 & 146.12 & 87.27 & 13.25 & 249.18 \\
\hline
\end{tabular}
\caption{The top-quark pair production cross sections (in unit: pb) before and after PMC scale-setting at the LHC with $\sqrt{S}=8$ TeV. $\mu_r=\mu_f=m_t$.}\label{tab8}
\end{table}

\begin{table}[htb]
\centering
\begin{tabular}{ccccccccc}
\hline
& \multicolumn{4}{c}{Conventional scale setting} & \multicolumn{4}{c}{PMC scale setting} \\ \cline{2-5} \cline{6-9}
~~~ ~~~ & ~~LO~~ & ~~NLO~~ & ~~N$^2$LO~~ & ~~{\it Total}~~ & ~~LO~~ & ~~NLO~~ & ~~N$^2$LO~~ & ~~{\it Total}~~\\
\hline
$(q\bar{q})$ channel & 66.47 & 8.30 & 4.73 & 79.58 & 62.86 & 19.38 & $-2.74$ & 79.08 \\
$(gg)$ channel  & 415.06 & 224.43 & 43.36 & 682.98 & 413.52 & 259.35 & 32.56 & 713.60 \\
$(gq)$ channel  & 0.00 & 7.09 & 6.52 & 13.82 & 0.00 & 7.09 & 6.52 & 13.82 \\
$(g\bar{q})$ channel & 0.00 & $-0.25$ & 1.59 & 1.33 & 0.00 & $-0.25$ & 1.59 & 1.33 \\
sum   & 481.53 & 239.57 & 56.20 & 777.72 & 476.38 & 285.57 & 37.93 & 807.83 \\
\hline
\end{tabular}
\caption{The top-quark pair production cross sections (in unit: pb) before and after PMC scale-setting at the LHC with $\sqrt{S}=13$ TeV. $\mu_r=\mu_f=m_t$.} \label{tab13}
\end{table}

Numerical results for the total top-quark pair production cross sections at the hadronic colliders Tevatron and LHC for both conventional and PMC scale settings are presented in Tables \ref{tab2}, \ref{tab7}, \ref{tab8}, and \ref{tab13}, respectively. To do numerical calculation, we update our previous predictions by using $m_t=173.3$ GeV~\cite{ATLAS:2012coa} and the CTEQ version CT14~\cite{Dulat:2015mca} as the PDF. The cross sections for the individual production channels, i.e. $(q\bar{q})$, $(gq)$, $(g\bar{q})$ and $(gg)$ channels are presented. In these tables, the initial choice of renormalization scale and factorization scale is fixed to be $\mu_r=\mu_f=m_t$.

\begin{table}[htb]
\centering
\begin{tabular}{ccccccc}
\hline
 & \multicolumn{3}{c}{Conventional} & \multicolumn{3}{c}{PMC} \\ \cline{2-4} \cline{5-7}
 ~~~$\mu_r$~~~& ~~$m_t/2$~~ & ~~$m_t$~~ & ~~$2m_t$~~~~ & ~~~~$m_t/2$~~ & ~~$m_t$~~ & ~~$2m_t$~~ \\
\hline
~$\sigma^{1.96\rm TeV}_{\rm Tevatron}$ & 7.54 & 7.29 & 7.01 & 7.43 & 7.43 & 7.43 \\
~$\sigma^{7\rm TeV}_{\rm LHC}$ & 172.07 & 167.67 & 160.46 & 174.97 & 174.98 & 174.99 \\
~$\sigma^{8\rm TeV}_{\rm LHC}$ & 244.87 & 239.03 & 228.94 & 249.16 & 249.18 & 249.19 \\
~$\sigma^{13\rm TeV}_{\rm LHC}$ & 792.36 & 777.72 & 746.92 & 807.80 & 807.83 & 807.86 \\
\hline
\end{tabular}
\caption{The N$^2$LO top-pair production cross sections for the Tevatron and LHC (in unit of pb), comparing conventional versus PMC scale settings. Here all production channels have been summed. Three typical choices for the initial renormalization scales $\mu_{r}=m_t/2$, $m_t$ and $2m_t$ have been adopted.} \label{TevLHCtotcs}
\end{table}

We present the N$^2$LO top-quark pair production cross sections at the Tevatron and LHC for both conventional and PMC scale settings in Table~\ref{TevLHCtotcs}, where four CM collision energies $\sqrt{S}=1.96$ TeV, $7$ TeV, $8$ TeV, and $13$ TeV, and three typical choices of initial renormalization scale $\mu_r=m_t/2$, $m_t$, and $2m_t$ have been assumed. Table~\ref{TevLHCtotcs} shows the PMC predictions for the top-pair total cross section: $\sigma^{1.96\rm TeV}_{\rm Tevatron}=7.43^{+0.14}_{-0.13} ~\rm pb$ at the Tevatron, $\sigma^{7\rm TeV}_{\rm LHC}=175.0^{+3.5}_{-3.5} ~\rm pb$, $\sigma^{8\rm TeV}_{\rm LHC}=249.2^{+5.0}_{-4.9} ~\rm pb$, and $\sigma^{13\rm TeV}_{\rm LHC}=807.8^{+16.0}_{-15.8} ~\rm pb$ at the LHC. These predictions agree with the Tevatron and LHC measurements within errors~\cite{Aaltonen:2013wca, Chatrchyan:2013ual, Aad:2012vip, Chatrchyan:2013kff, Aad:2015dya, Chatrchyan:2012vs, Aad:2012qf, Chatrchyan:2016abc, Aad:2014kva, Khachatryan:2016mqs, Khachatryan:2015fwh, Khachatryan:2014loa, Aad:2015pga, Khachatryan:2016kzg, Khachatryan:2015uqb, Aaboud:2015AAAA, Aaboud:2016pbd}. Table~\ref{TevLHCtotcs} shows that using conventional scale setting, the renormalization scale dependence of the N$^2$LO-level cross section is about $6\% - 7\%$ for $\mu_r\in[m_t/2, 2m_t]$. Thus achieving the exact value for each order is important for a precise lower-order pQCD prediction, especially for those observables that are heavily dependent on the value at a particular order. By analyzing the N$^2$LO pQCD series in detail, it is found that the renormalization scale dependence of each perturbative term is rather large using conventional scale setting~\cite{Wang:2015lna}. On the other hand, by using the PMC, the cross sections at each order are almost unchanged, indicating a nearly scale-independent prediction can be achieved even at lower orders. If one sets $\mu_r=m_t/2$ for conventional scale setting, the total cross section is close to the PMC prediction, whose pQCD convergence is also better than the cases of $\mu_r=m_t$ and $\mu_r=2m_t$ as has been observed in Ref.\cite{Czakon:2016dgf}. Thus, the PMC provides support for ``guessing" the optimal choice of $\mu_r\sim m_t/2$ using conventional scale setting~\cite{Wang:2014sua}.

After applying the PMC, we obtain the optimal scale of the top-quark pair production at each perturbative order in pQCD, and the resulting theoretical predictions are essentially free of the initial choice of renormalization scale. Thus a more accurate top-quark pole mass and a reasonable explanation of top-quark pair forward-backward asymmetry at the hadronic colliders can be achieved~\cite{Brodsky:2012rj, Brodsky:2012ik, Wang:2015lna, Wang:2014sua, Wang:2017kyd}.

First, to fix the top-quark mass, one can compare the pQCD prediction on the top-quark pair production cross-section with the experimental data. For this purpose, one can define a likelihood function~\cite{Aaboud:2016000}
\begin{eqnarray}
f(m_t)=\int^{+\infty}_{-\infty} f_{\rm th}(\sigma|m_t)\cdot f_{\rm exp}(\sigma|m_t)\; d\sigma.
\label{likelifunction}
\end{eqnarray}
Here $f_{\rm th}(\sigma|m_t)$ is the normalized Gaussian distribution determined theoretically,
\begin{eqnarray}
f_{\rm th}(\sigma|m_t) = \frac{1}{\sqrt{2\pi}\Delta \sigma_{\rm th}(m_t)}
 \exp \left[-\frac{\left(\sigma-\sigma_{\rm th}(m_t)\right)^2}
 {2 \Delta\sigma^2_{\rm th}(m_t)}\right].
\end{eqnarray}
The top-quark pair production cross-section is a function of the top-quark pole mass $m_t$; its decrease with increasing $m_t$ can be parameterized as~\cite{Beneke:2011mq}
\begin{eqnarray}
\sigma_{\rm th}(m_t)&=&\left(\frac{172.5}{m_t/{\rm GeV}}\right)^4\left(c_0+c_1(\frac{m_t}{\rm GeV}-172.5) +c_2 \times(\frac{m_t}{\rm GeV}-172.5)^2+c_3(\frac{m_t}{\rm GeV}-172.5)^3\right),
\label{massdepdPMC}
\end{eqnarray}
where all masses are given in units of GeV. $\Delta\sigma_{\rm th}(m_t)$ stands for the maximum error of the cross-section for a fixed $m_t$. One can estimate its value by using the CT14 error PDF sets~\cite{Dulat:2015mca} with range of $\alpha_s(M_Z)\in [0.117, 0.119]$. The values for the coefficients $c_{0,1,2,3}$ can be determined by using a wide range of the top-quark pole mass, $m_t\in [{\rm 160\; GeV}, {\rm 190\; GeV}]$. Here $\sigma_{\rm th}(m_t)$ is defined as the cross-section at a fixed $m_t$, where all input parameters are set to be their central values, $[\sigma_{\rm th}(m_t) + \Delta\sigma^{+}_{\rm th}(m_t)]$ is the maximum cross-section within the allowable parameter range, and $[\sigma_{\rm th}(m_t)-\Delta\sigma^{-}_{\rm th}(m_t)]$ is the minimum value. The function $f_{\rm exp}(\sigma|m_t)$ is the normalized Gaussian distribution determined experimentally,
\begin{eqnarray}
f_{\rm exp}(\sigma|m_t) = \frac{1}{\sqrt{2\pi}\Delta \sigma_{\rm exp}(m_t)} \exp{ \left[-\frac{\left(\sigma-\sigma_{\rm exp}(m_t)\right)^2}
 {2\Delta\sigma^2_{\rm exp}(m_t)} \right]},
\end{eqnarray}
where $\sigma_{\rm exp}(m_t)$ is the measured cross-section, and $\Delta\sigma_{\rm exp}(m_t)$ is the uncertainty for $\sigma_{\rm exp}(m_t)$. By evaluating the likelihood function, we obtain $m_t=174.6^{+3.1}_{-3.2}$ GeV~\cite{Wang:2017kyd}, where the central value is extracted from the maximum of the likelihood function, and the error ranges are obtained from the $68\%$ area around the maximum. Because the PMC predictions have less uncertainty compared to the predictions by using conventional scale-setting, the precision of top-quark pole mass is dominated by the experimental errors. For example, the PMC determination for the pole mass via the combined dilepton and the lepton + jets channels data is about $1.8\%$, which is almost the same as that of the recent determination by the D0 collaboration, $172.8^{+3.4}_{-3.2}$ GeV~\cite{Abazov:2016ekt}, whose error is $\sim 1.9\%$.

\begin{figure}[htb]
\epsfysize=9.0cm
\begin{center}
\begin{minipage}[t]{10 cm}
\epsfig{file=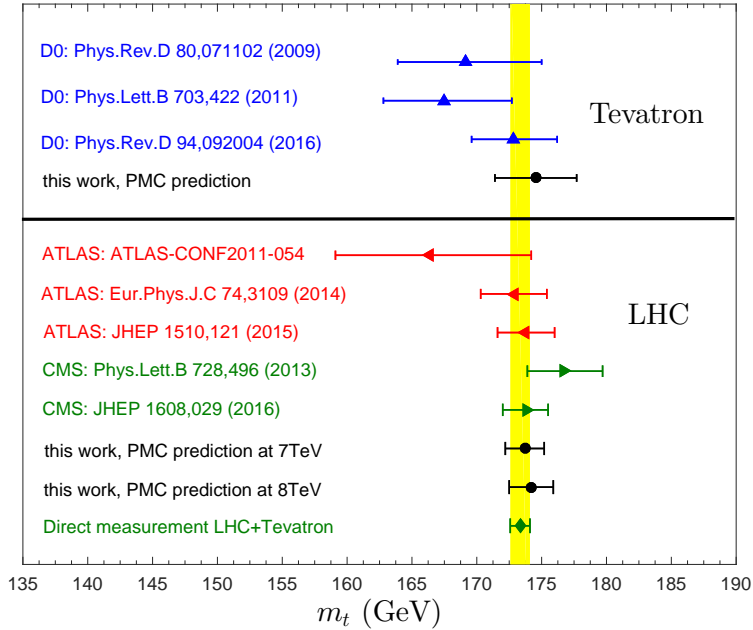,scale=0.6}
\end{minipage}
\begin{minipage}[t]{16.5 cm}
\caption{A summary of the top-quark pole mass determined indirectly from the top-quark pair production channels at the Tevatron and LHC~\cite{Wang:2017kyd}. For reference, the combination of Tevatron and LHC direct measurements of the top-quark mass is presented as a shaded band, which gives $m_t=173.34\pm0.76$ GeV~\cite{ATLAS:2014wva}. \label{toppolesummary}}
\end{minipage}
\end{center}
\end{figure}

A summary of the top-quark pole masses determined at both the Tevatron and LHC is presented in Figure \ref{toppolesummary}, where the PMC predictions and previous predictions from other collaborations~\cite{Aad:2014kva, Khachatryan:2016mqs, Aaboud:2016000, Abazov:2016ekt, Chatrchyan:2013haa, Abazov:2011pta, Aad:2015waa, Abazov:2009ae, ATLAS:2014wva} are presented.

Second, it has been found that by applying the PMC, the SM predictions for the top-quark forward-backward asymmetry at the Tevatron have only $1\sigma$ deviation from the CDF and D0 measurements~\cite{Brodsky:2012rj, Brodsky:2012ik, Wang:2015lna}. In fact, the PMC gives a scale-independent precise top-quark pair forward-backward asymmetry, $A^{\rm PMC}_{\rm FB}=9.2\%$ and $A_{\rm FB}(M_{t\bar{t}}>450\;{\rm GeV})=29.9\%$, in agreement with the corresponding CDF and D0 measurements~\cite{Aaltonen:2008hc, Aaltonen:2011kc, Aaltonen:2012it, Abazov:2007ab, Abazov:2011rq, Abazov:2014cca, Abazov:2015fna}. The large discrepancies of the top-quark forward-backward asymmetry between the SM estimate and the Tevatron data are thus greatly reduced. Moreover, the PMC prediction for $A_{\rm FB}(M_{t\bar{t}}>M_{\rm cut})$ displays an ``increasing-decreasing" behavior as $M_{\rm cut}$ is increased, consistent within errors with the measurements recently reported by D0 experiment~\cite{Abazov:2014cca}.

\begin{figure}[htb]
\epsfysize=9.0cm
\begin{center}
\begin{minipage}[t]{10 cm}
\epsfig{file=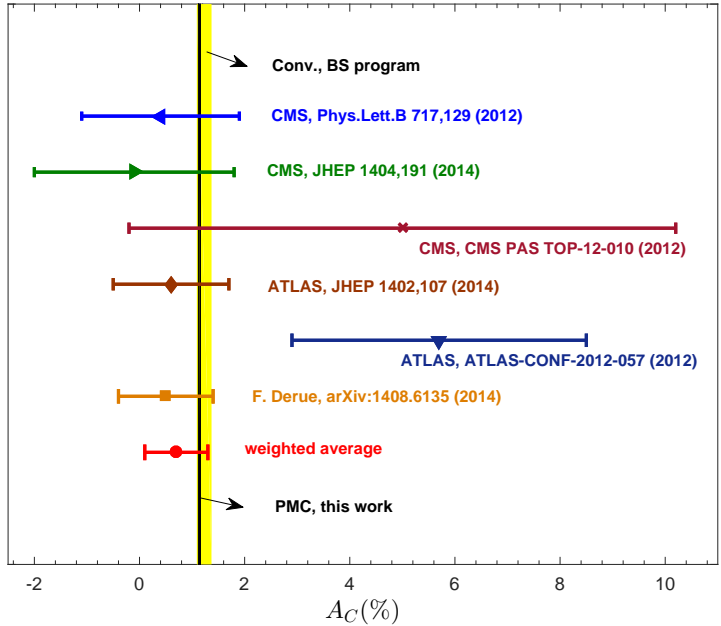,scale=0.6}
\end{minipage}
\begin{minipage}[t]{16.5 cm}
\caption{The top-quark charge asymmetry $A_{\rm C}$ assuming conventional scale setting (Conv.) and PMC scale setting for $\sqrt{S}=7$ TeV~\cite{Wang:2014sua}; the error bars are for $\mu_r \in[m_t/2,2m_t]$ and $\mu_f \in[m_t/2,2m_t]$. As a comparison, the experimental results~\cite{Chatrchyan:2012cxa, Chatrchyan:2014yta, CMS:iya, Aad:2013cea, ATLAS:2012sla, Derue:2014yja} and the prediction of Ref.\cite{Bernreuther:2012sx} are also presented. \label{cpmcasy}}
\end{minipage}
\end{center}
\end{figure}

The top-quark charge asymmetry at the LHC for the $p p \to t \bar t X$ process is defined as
\begin{equation}
A_{\rm C}=\frac{N(\Delta |y|>0)-N(\Delta |y|<0)} {N(\Delta |y|>0)+N(\Delta |y|<0)}, \label{asyc}
\end{equation}
where $\Delta |y|=|y_{t}|-|y_{\bar{t}}|$ is the difference between the absolute rapidities of the top and anti-top quarks, and $N$ is the number of events. Measurements of the top-quark charge asymmetry at the LHC have been reported in Refs.~\cite{Chatrchyan:2012cxa, Chatrchyan:2014yta, CMS:iya,Aad:2013cea, ATLAS:2012sla, Derue:2014yja}. Figure \ref{cpmcasy} gives a summary of the LHC measurements, together with the theoretical predictions. In contrast to the Tevatron $ p \bar p \to t \bar t X$ processes, the asymmetric channel $q\bar{q}\to t\bar{t}$ gives a small pQCD contribution to the top-pair production at the LHC, and the symmetric channel $gg\to t\bar{t}$ provides the dominant contribution. Thus, the predicted charge asymmetry at the LHC is smaller than the one at the Tevatron. Two typical SM predictions for the charge asymmetry at the LHC are: $A_{\rm C}|_{\rm 7 TeV} =(1.15\pm0.06)\%$ and $A_{\rm C}|_{\rm 8 TeV}=(1.02\pm0.05)\%$~\cite{Kuhn:2011ri}; $A_{\rm C}|_{\rm 7 TeV} =(1.23\pm0.05)\%$ and $A_{\rm C}|_{\rm 8 TeV}=(1.11\pm0.04)\%$~\cite{Bernreuther:2012sx}. The uncertainties of the theoretical prediction are dominated by the choice of scale. The scale errors for conventional scale setting are obtained by varying $\mu_{r} \in[m_t/2,2m_t]$, and fixing the factorization scale $\mu_{f} \equiv \mu_{r}$. As a representation, Figure \ref{cpmcasy} shows the prediction of Ref.\cite{Bernreuther:2012sx}. On the other hand, the PMC prediction is almost scale independent and a more precise comparison with the data can be achieved.

\subsubsection{The $\gamma\gamma^* \to \eta_c$ transition form factor}

The simplest exclusive charmonium production process, $\gamma^*\gamma \to \eta_c$, measured in two-photon collisions, provides another example of the importance of a proper scale-setting approach for fixed-order predictions. This is also helpful for testing Nonrelativistic QCD (NRQCD) theory~\cite{Bodwin:1994jh}. One can define a transition form factor (TFF) $F(Q^{2})$ via the following way~\cite{Lepage:1980fj}:
\begin{equation}
\langle \eta_c (p)\vert J_{\rm EM}^\mu \vert \gamma(k,\varepsilon) \rangle = i e^2 \epsilon^{\mu\nu\rho\sigma} \varepsilon_\nu q_{\rho} k_{\sigma} F(Q^2),  \label{formfactor}
\end{equation}
where $J_{\rm EM}^\mu$ is the electromagnetic current evaluated at the time-like momentum transfer squared, $Q^2=-q^2 =-(p-k)^2>0$. The BaBar collaboration has measured its value and parameterized it as $|F(Q^{2})/F(0)|=1/(1+Q^{2}/\Lambda)$~\cite{Lees:2010de}, where $\Lambda=8.5\pm0.6\pm0.7$ GeV$^{2}$. In the case of conventional scale setting, the renormalization scale is simply set as the typical momentum flow $\mu_Q=\sqrt{Q^{2}+m_c^{2}}$; the N$^2$LO NRQCD prediction cannot explain the BaBar measurements over a wide $Q^2$ range~\cite{Feng:2015uha}. Here $m_c$ is the $c$-quark mass and we set its value as $1.68$ GeV. This disagreement cannot be solved by taking higher Fock states into consideration~\cite{Guo:2011tz, Jia:2011ah}.

\begin{figure}[htb]
\epsfysize=9.0cm
\begin{center}
\begin{minipage}[t]{10 cm}
\epsfig{file=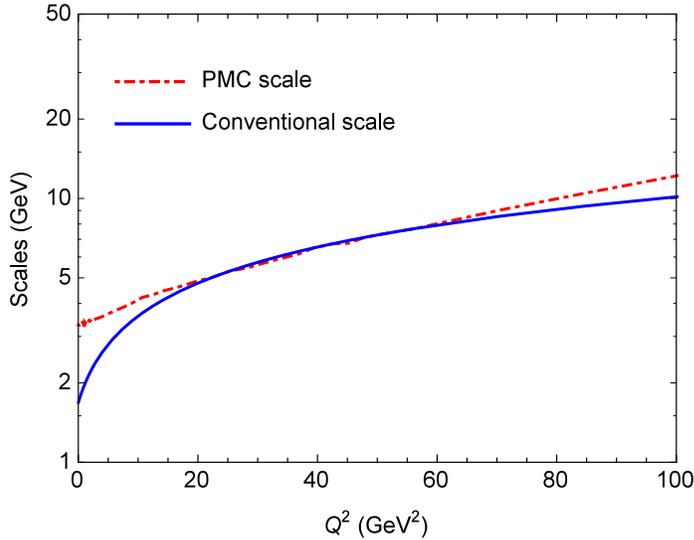,scale=1.0}
\end{minipage}
\begin{minipage}[t]{16.5 cm}
\caption{The PMC scale of the transition form factor $F(Q^2)$~\cite{Wang:2018lry}, defined in Eq.(\ref{formfactor}), versus $Q^2$. The conventional choice of scale $\mu_r=\mu_Q$ is presented as a comparison. \label{FQscalePMC}}
\end{minipage}
\end{center}
\end{figure}

Numerically, the choice of renormalization scale $\mu_r=\mu_Q$ leads to a substantially negative N$^2$LO contribution and hence a large $|F(Q^2)/F(0)|$, in disagreement with the data. Following the standard PMC scale-setting procedures, one can determine the PMC scale $\mu^{\rm PMC}_r$ of the process by carefully dealing with the light-by-light diagrams at the N$^2$LO level. The determined PMC scale varies with momentum transfer squared $Q^2$ at which the TFF is measured, and it is independent of the initial choice of $\mu_r$ (thus the conventional scale uncertainty is eliminated). We present the PMC scale $\mu^{\rm PMC}_r$ versus $Q^2$ in Figure \ref{FQscalePMC}, which is larger than the ``guessed" value $\mu_Q$ in the small and large $Q^2$-regions. In the intermediate $Q^2$-region, e.g. $Q^2\sim [20,60]$ GeV$^2$, the discrepancy between $\mu^{\rm PMC}_r$ and $\mu_Q$ is small; and the largest difference occurs at $Q^2=0$.

\begin{figure}[htb]
\epsfysize=9.0cm
\begin{center}
\begin{minipage}[t]{10 cm}
\epsfig{file=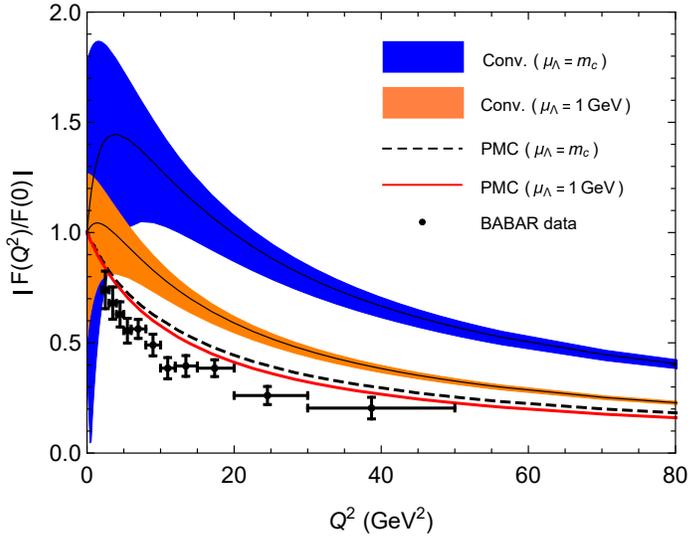,scale=1.0}
\end{minipage}
\begin{minipage}[t]{16.5 cm}
\caption{The ratio $|F(Q^2)/F(0)|$ up to N$^2$LO-level versus $Q^2$ using conventional (Conv.) and PMC scale-settings~\cite{Wang:2018lry}, where the BaBar data are presented as a comparison~\cite{Lees:2010de}. Two typical factorization scales, $\mu_\Lambda=1$ GeV and $m_c$ are adopted. The error bars are for $\mu^2_r=[\mu^2_Q/2, 2\mu^2_Q]$ with $\mu_Q=\sqrt{Q^{2}+m_c^{2}}$. \label{FQF0DataPMC168}}
\end{minipage}
\end{center}
\end{figure}

A comparison of the renormalization scale dependence for the ratio $|F(Q^2)/F(0)|$ is given in Figure \ref{FQF0DataPMC168}, which is obtained by using the same input parameters as those of Refs.\cite{Wang:2018lry, Feng:2015uha}. It shows that the PMC prediction is independent of the initial choice of scale $\mu_r$, whereas the conventional scale uncertainty is large, especially in low $Q^2$-region. The PMC prediction is close to the BaBar measurement. Thus the application of PMC supports the applicability of NRQCD to hard exclusive processes involving heavy quarkonium.

The determination of the factorization scale is a completely separate issue from the renormalization scale setting since it is present even for a conformal theory. The factorization scale can be determined by matching nonperturbative bound-state dynamics with perturbative DGLAP evolution~\cite{Gribov:1972ri, Altarelli:1977zs, Dokshitzer:1977sg}. Recently, by using the light-front holography~\cite{Brodsky:2011pw, deTeramond:2012rt}, it has been shown that the matching of high-and-low scale regimes of $\alpha_s$ can determine the scale which sets the interface between perturbative and nonperturbative hadron dynamics~\cite{Deur:2014qfa, Deur:2016cxb, Deur:2016tte,Deur:2017cvd}. Figure \ref{FQF0DataPMC168} also shows the factorization scale dependence for the ratio $|F(Q^2)/F(0)|$. In the case of conventional scale-setting, there is large factorization scale dependence. Choosing a smaller factorization scale could lower the N$^2$LO-level ratio $|F(Q^2)/F(0)|$ to a certain degree, but it cannot eliminate the large discrepancy with the data. In contrast, after applying the PMC, the prediction shows a small factorization scale dependence. This in some sense also shows the importance of a proper scale-setting approach. More explicitly, in the case $Q^2=0$, a large factorization scale uncertainty is observed using conventional scale-setting; i.e.,
\begin{equation}
F^{\rm Conv}(0)|_{\mu_r=m_c}=0.43c^{(0)},\;\; 0.22c^{(0)},\;\; -0.06c^{(0)}
\end{equation}
for factorization scale $\mu_\Lambda=1$ GeV, $m_c$ and 2$m_c$, respectively. Here the LO coefficient $c^{(0)}$ is
\begin{equation}
c^{(0)}=\frac{4e^2_c\langle\eta_c|\psi^\dag\chi(\mu_\Lambda)|0\rangle} {(Q^2+4m^2_c)\sqrt{m_c}},
\end{equation}
where $e_c=+2/3$ is the $c$-quark electric charge, and $\langle\eta_c|\psi^\dag\chi(\mu_\Lambda)|0\rangle$ represents the nonperturbative matrix-element which characterizes the probability of the $(c\bar{c})$-pair to form a $\eta_c$ bound state. The magnitude of the negative N$^2$LO term increases with increasing $\mu_\Lambda$, and the $F^{\rm Conv}(0)$ is even negative for $\mu_\Lambda =2m_c$. On the other hand, by applying the PMC, we obtain a reasonable small factorization scale dependence
\begin{equation}
F^{\rm PMC}(0)=0.61c^{(0)},\;\; 0.50c^{(0)},\;\; 0.34c^{(0)}.
\end{equation}
again for $\mu_\Lambda=1$ GeV, $m_c$ and 2$m_c$, respectively.

\section{The renormalization scheme-and-scale independent pQCD predictions}
\label{sec:3}

With the help of RGI and the RGE, the standard PMC multi-scale approach provides a rigorous method to eliminate the conventional scheme-and-scale ambiguities for pQCD predictions via a step-by-step way. As mentioned above, there are two types of residual scale dependence due to unknown higher-order terms, which are complicated by the scheme-dependent RGE of the conventional running coupling and by the complex multi-scale setting procedures. As a step forward, it is helpful to find a way to suppress or even eliminate the residual scale dependence, such that a strict scale independent prediction can be achieved at any fixed order. For this purpose, a running coupling with a simpler RGE and a simpler scale setting procedure can be helpful.

In this section, we shall show that by using the $C$-scheme coupling, together with the use of the PMC single-scale approach (PMC-s), a pQCD prediction with minimum residual scale dependence can be achieved. The RGE of the $C$-scheme coupling, as shown by Eq.(\ref{eq:betahat}), has a much simpler structure than the conventional RGE (\ref{eq:betafun}), since is only contains scheme-independent $\beta_0$- and $\beta_1$- functions. Then, we shall show that by using this simpler scheme-independent RGE, the residual scale dependence can be greatly suppressed, due to both the $\alpha_s$-power suppression and in general the exponential suppression. By applying the PMC-s procedure, the PMC predictions are exactly independent of the choice of the initial renormalization scheme-and-scale~\cite{Shen:2017pdu}. Thus, a renormalization scheme-and-scale independent fixed-order prediction can be achieved by applying the PMC-s approach.

The pQCD predictions are generally calculated using the conventional running coupling ($a_\mu$). In the following subsections, we shall first show how the pQCD predictions for conventional running coupling are transformed to those under the $C$-scheme coupling ($\hat{a}_\mu$). Next, we shall present the scale setting formulas for the PMC-s approach using both the dimensional-regularized ${\cal R}_\delta$-scheme running coupling $a_\mu$ and the general $C$-scheme coupling $\hat{a}_\mu$. Then, we shall demonstrate the equivalence of the PMC predictions under those two running couplings. Finally, we shall demonstrate that scheme-and-scale independent predictions can be achieved by applying the PMC-s approach to the pQCD series using the $C$-scheme coupling. We shall illustrate those features for the non-singlet Adler function at the four-loop level. As an addendum, we shall present a practical way to achieve a scheme-and-scale independent prediction based on the PMS scale setting approach.

\subsection{Transformation of pQCD predictions from conventional coupling to $C$-scheme coupling}

By using the relation (\ref{eq:Expandhata}) between the $C$-scheme coupling $\hat{a}_\mu$ and the conventional coupling $a_\mu$, we can transform the pQCD approximant (\ref{pQCDexp}) from the conventional coupling to the $C$-scheme coupling; i.e.,
\begin{equation}
\rho_{n}(Q) = \sum_{i=1}^{n} \hat{c}_{i}(\mu/Q) \hat{a}_{\mu}^{i}, \label{pQCDexpC}
\end{equation}
where the new perturbative coefficients $\hat{c}_{i}$ are:
\begin{eqnarray}
\hat{c}_1 &=& r_1, \label{PMC-s-coeff1}\\
\hat{c}_2 &=& r_2 + \beta_0 r_1 C, \label{PMC-s-coeff2}\\
\hat{c}_3 &=& r_3 + \left(\beta_1 r_1+2\beta_0 r_2\right)C+\beta_0^2 r_1 C^2
+ r_1\left(\frac{\beta_2}{\beta_0}-\frac{\beta_1^2}{\beta_0^2}\right), \label{PMC-s-coeff3} \\
\hat{c}_4 &=& r_4 + \left(3\beta_0 r_3+2\beta_1 r_2+3 \beta_2 r_1-\frac{2\beta_1^2 r_1}{\beta_0}\right)C
+\left(3\beta_0^2 r_2+\frac{5}{2} \beta_1 \beta_0 r_1\right)C^2 \nonumber\\
&& +r_1\beta_0^3 C^3 +r_1\left(\frac{\beta_3}{2\beta_0}-\frac{\beta_1^3}{2\beta_0^3}\right)
 +r_2\left(\frac{2\beta_2}{\beta_0}-\frac{2\beta_1^2}{\beta_0^2}\right), \label{PMC-s-coeff4} \\
&& \hspace{-6mm} \cdots \nonumber
\end{eqnarray}

The fixed-order pQCD prediction (\ref{pQCDexpC}) based on the $C$-scheme coupling is also scheme-and-scale dependent using conventional scale setting. As an attempt to achieve a more precise prediction using the $C$-scheme coupling, many authors have investigated the possibility of obtaining an ``optimized" prediction for the truncated pQCD series by exploiting its scheme dependence~\cite{Boito:2016pwf, Caprini:2018agy, Jamin:2016ihy}. In their treatment, by fixing the renormalization scale $\mu\equiv Q$ and varying $C$ within a possible domain, an optimal $C$-value, and thus an optimal scheme, is determined by requiring the absolute value of the last known term or the last non-zero term to be at its minimum.

One may observe that the idea of requiring the magnitude of the last known term of the pQCD series to be at its minimum is, in principle, similar to the postulate of the {\it Principle of Minimum Sensitivity} (PMS)~\cite{Stevenson:1980du, Stevenson:1981vj, Stevenson:1982wn, Stevenson:1982qw}, in which the optimal scheme is determined by directly requiring all unknown high-order terms to vanish, e.g. $\partial \rho_{n}(Q) /\partial {\rm (RS)}=0$, where ${\rm RS}$ stands for either the scale or scheme parameters. As shall be shown in Sec.\ref{PMSindependent} using this criteria, one can achieve scheme-and-scale independent predictions with the help of renormalization group invariants which emerge at each order~\cite{Ma:2017xef}. Even though it cannot offer correct lower-order predictions, it can be a practical and reliable way to estimate the pQCD predictions when enough higher-order terms have been included~\cite{Ma:2014oba, Wu:2014iba}. However, the PMS meets serious theoretical problems: It does not satisfy the self-consistency conditions of the renormalization group, such as reflectivity, symmetry and transitivity~\cite{Brodsky:2012ms}; its pQCD convergence is accidental and questionable; it disagrees with Gell Mann-Low scale setting when applied to QED cases; and it gives unphysical results for the jet production in $e^+ e^-$ annihilation~\cite{Kramer:1987dd, Kramer:1990zt}; etc..

The optimization to the $C$-scheme coupling approach by requiring the absolute value of the last known term to be at its minimum meets the same theoretical problems of PMS. The optimal value of $C$ is different for a different fixed-order prediction, which need to be redetermined when new perturbative terms are known. Although this approach of varying the $C$-scheme coupling could be considered as a practical way to improve pQCD precision~\cite{Boito:2016pwf, Jamin:2016ihy}, similar to the PMS approach, it cannot be considered as the solution to the conventional scheme-and-scale setting ambiguities.

In contrast, the PMC identifies all the RG-involved scheme-dependent $\{\beta_i\}$-terms in the perturbative series and eliminates them by shifting the scales of the running coupling. After applying the PMC, the coefficients match the corresponding conformal series, and thus the prediction is scheme independent. An explicit demonstration that the PMC scale setting leads to scheme-independent pQCD predictions for any dimensional-like scheme has been given in Sec.~\ref{PMCRdelta}. More explicitly, Eq.(\ref{eq:deltaRGI}) shows that after eliminating all the $\{\beta_i\}$-terms, one obtains ${\partial \rho_n|_{\rm PMC}}/{\partial \delta}=0$, proving that the PMC prediction $\rho_n|_{\rm PMC}$ is independent of $\delta $ and thus any choice of the dimensionally regulated ${\cal R}_\delta$-schemes. We will now generalize this procedure to see whether one can eliminate all scheme-dependent $C$-terms in a pQCD approximant by applying the PMC. Since the parameter $C$ identifies any choice of the renormalization scheme, we will then achieve a general demonstration of the scheme-independence of the PMC pQCD predictions for any renormalization scheme.

\subsection{The pQCD predictions using the PMC-s scale setting approach}

The PMC multi-scale approach requires considerable theoretical analysis, especially since one needs to distribute the approximal renormalization-group-involved $\{\beta_i\}$-terms into each perturbative order. To make the PMC scale setting procedures simpler and more easily to be automatized, a single-scale PMC scale setting approach (e.g. the PMC-s approach) has been suggested. The PMC-s approach~\cite{Shen:2017pdu} achieves many of the same goal as the PMC multi-scale approach. The PMC-s approach replaces the individual PMC scales at each order by an overall effective single scale, which effectively replaces the individual PMC scales derived under the PMC multi-scale approach in the sense of a mean value theorem. The PMC-s scale can be regarded as the overall effective momentum flow of the process; it shows stability and convergence with increasing order in pQCD via the pQCD approximates. Similarly, we can demonstrate the scheme-independence of the PMC-s prediction up to any fixed order. Moreover, its predictions are explicitly independent of the choice of the initial renormalization scale.

After applying the PMC-s, any perturbatively calculable physical quantity can be used to define an effective coupling. In different to the idea of FAC whose effective charge is fixed by incorporating the entire perturbative corrections into its definition~\cite{Grunberg:1982fw}~\footnote{The FAC effective coupling is fixed by comparing with the data, which reduces the predictive power of QCD theory and cannot be applied for confirming or finding new physics beyond the SM.}, the PMC-s series is still of perturbative nature up to the known perturbative order and its effective coupling is determined by using the $\{\beta_i\}$-terms of the process with the help of RGE. Thus, the PMC-s approach can be adopted as a valid substitution for the PMC multi-scale approach for setting the renormalization scale, particularly when one does not need detailed information for processes at each order.

In the following, we shall first introduce the PMC-s approach for pQCD predictions using conventional coupling and the $C$-scheme coupling. Then we demonstrate that the PMC pQCD predictions using the conventional coupling and the $C$-scheme coupling are exactly the same.

\subsubsection{The PMC-s scale setting approach for conventional coupling}

After applying the PMC-s approach to the pQCD series (\ref{eq:rhodelta2}) or (\ref{eq:generalization}) for the conventional coupling, the resulting conformal series up to $n_{\rm th}$-order level changes to
\begin{eqnarray}
\rho_{n}(Q)|_{\rm PMC-s} = \sum\limits^{n}_{i\ge1} {r_{i,0}}{a_{Q_\star}^{i}},
\label{PMC-spQCD}
\end{eqnarray}
where the effective PMC scale $Q_\star$ is determined by requiring all the renormalization group involved non-conformal terms to vanish simultaneously; i.e.,
\begin{equation}
\sum\limits^{i+j\leq n}_{i\ge1, j\ge1, 0\leq k\leq j} (-1)^{j} \ln^k\frac{Q_\star^2}{Q^2} \left[i\beta(a_{Q_\star}) a_{Q_\star}^{i-1}\right] C_j^k \Delta_{i}^{(j-1)}(a_{Q_\star}) \hat{r}_{i+j-k,j-k} = 0.
\label{eq:Qstar}
\end{equation}
Thus, similar to the PMC multi-scale approach, the single scale $Q_\star$ is also of perturbative nature. Its perturbative form takes the form
\begin{eqnarray}
\ln\frac{Q_{\star}^2}{Q^2} &=& \sum^{n-2}_{i= 0} S_i a^i_{Q_\star}.
\label{eq:pmcscale2}
\end{eqnarray}
Solving Eq.(\ref{eq:Qstar}) iteratively, we can obtain the coefficients $S_{i}$ up to any fixed-order. For example, for a fourth-order prediction, we have
\begin{eqnarray}
S_0 &=& -\frac{\hat{r}_{2,1}}{\hat{r}_{1,0}}, \label{singles1}\\
S_1 &=& \frac{2\left(\hat{r}_{2,0}\hat{r}_{2,1}-\hat{r}_{1,0}\hat{r}_{3,1}\right)}{\hat{r}_{1,0}^2}
+\frac{\hat{r}_{2,1}^2-\hat{r}_{1,0} \hat{r}_{3,2}}{\hat{r}_{1,0}^2}\beta_0, \label{singles2}\\
S_2 &=& \frac{3\hat{r}_{1,0}\left(\hat{r}_{3,0}\hat{r}_{2,1} -\hat{r}_{1,0}\hat{r}_{4,1}\right)+4\hat{r}_{2,0} \left(\hat{r}_{1,0}\hat{r}_{3,1}-\hat{r}_{2,0} \hat{r}_{2,1}\right)}{\hat{r}_{1,0}^3}
+\frac{3\hat{r}_{1,0}\hat{r}_{2,1}\hat{r}_{3,2} -\hat{r}_{1,0}^2\hat{r}_{4,3} -2\hat{r}_{2,1}^3}{\hat{r}_{1,0}^3}\beta_0^2 \nonumber\\
&& + \frac{3\left(\hat{r}_{2,1}^2-\hat{r}_{1,0}\hat{r}_{3,2}\right)}{2\hat{r}_{1,0}^2}\beta_1
+\frac{3\hat{r}_{1,0}\left(2\hat{r}_{2,1} \hat{r}_{3,1} -\hat{r}_{1,0} \hat{r}_{4,2}\right)+ \hat{r}_{2,0}\left(2 \hat{r}_{1,0} \hat{r}_{3,2}-5 \hat{r}_{2,1}^2\right)}{\hat{r}_{1,0}^3}\beta_0. \label{singles3}
\end{eqnarray}
Identical combinations in the scale expansion series emerge at different orders, consistent with the degeneracy relations among different orders of $\rho_{n}(Q)$; e.g., the coefficients of $(i+2)\beta_i a^{i+1}(Q)$ are the same. Substituting those coefficients into Eq.(\ref{eq:pmcscale2}), it is found that the scale $Q_\star$ has no dependence on the choice of the initial scale $\mu$ at any fixed order. For example, by assuming the range $\mu\in[1/2 Q, 2Q]$, the conventional approach assigns an uncertainty of $\left(^{+1.0\%}_{-3.0\%}\right)$, $\left(^{+0.3\%}_{-1.6\%}\right)$ or $\left(^{+0.4\%}_{-0.2\%}\right)$ to the two-loop, three-loop, and four-loop approximants of $R_{e^+ e^-}(Q=31.6{\rm GeV})$~\cite{Shen:2017pdu}, respectively; as a comparison, the PMC-s prediction for $R_{e^+ e^-}(Q=31.6{\rm GeV})$ is exactly unchanged within the same ranges of $\mu$. Thus the conventional renormalization scale ambiguity has been eliminated.

\begin{figure}[htb]
\epsfysize=9cm
\begin{center}
\begin{minipage}[t]{10 cm}
\epsfig{file=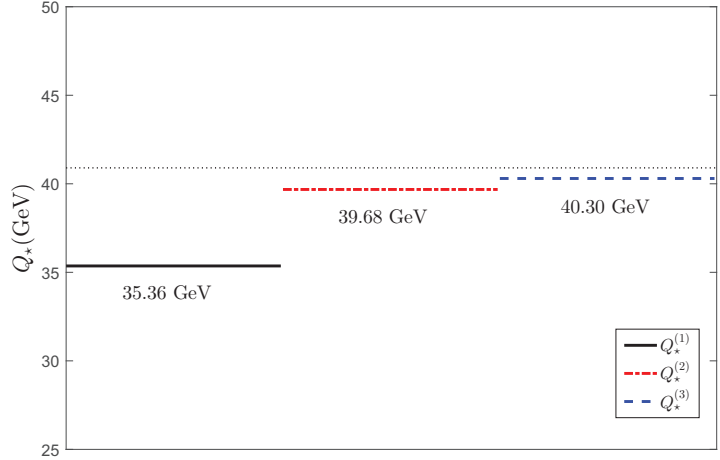,scale=0.55}
\end{minipage}
\begin{minipage}[t]{16.5 cm}
\caption{The effective PMC scale $Q_\star$ for $R(Q=31.6\;{\rm GeV})$ up to ${\rm N^{2}LLO}$ level~\cite{Shen:2017pdu}. The scale $Q^{(1)}_\star$ refers to the LLO level, $Q^{(2)}_\star$ is at the NLLO level, and $Q^{(3)}_\star$ is at the ${\rm N^{2}LLO}$ level. The value of $Q_\star$ monotonously approaches its ``exact" value (schematically shown by the dotted line) when more loop-terms are included. \label{fig:PMCs-scalex1}}
\end{minipage}
\end{center}
\end{figure}

\begin{figure}[htb]
\epsfysize=9.0cm
\begin{center}
\begin{minipage}[t]{10 cm}
\epsfig{file=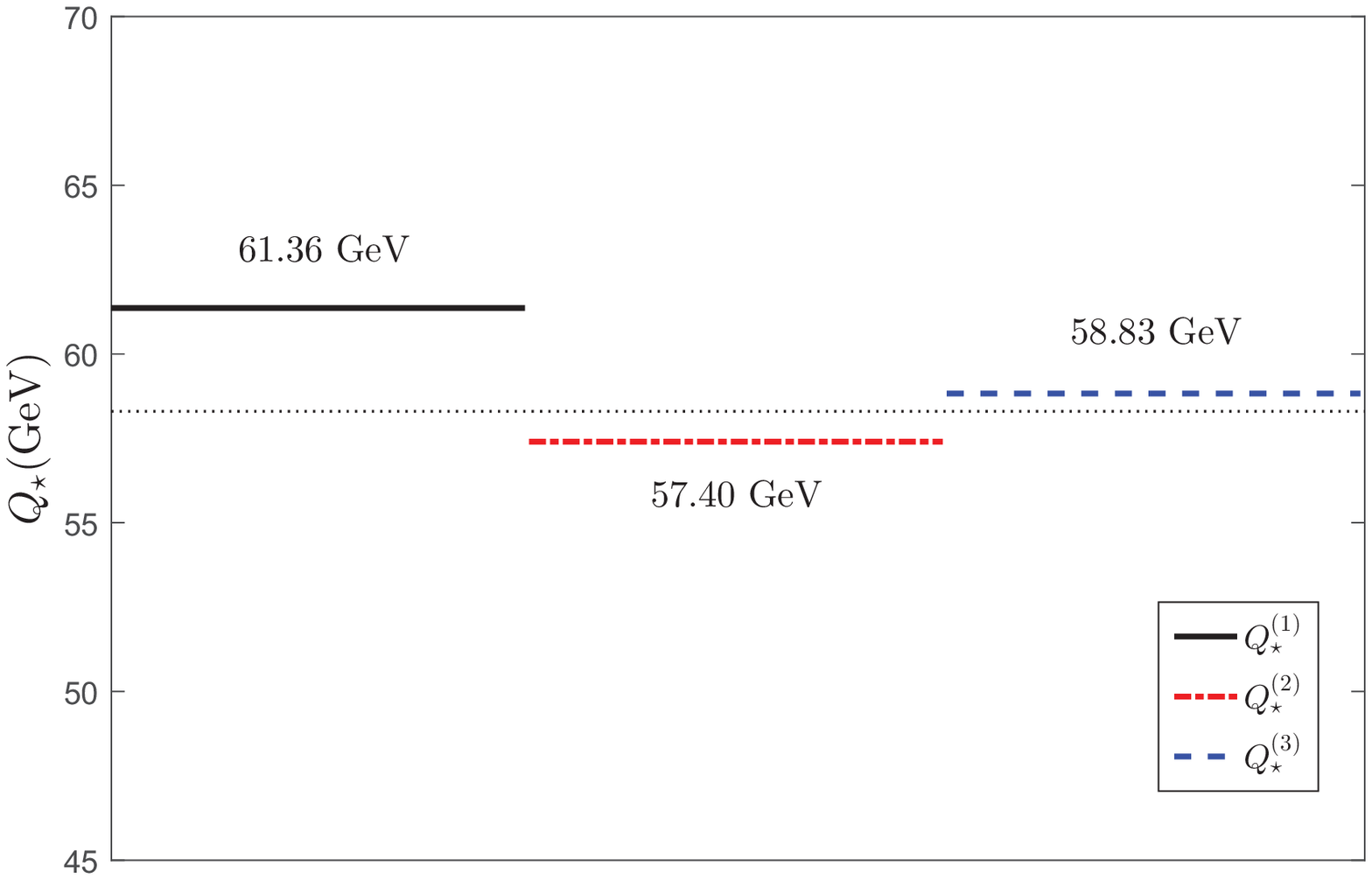,scale=0.55}
\end{minipage}
\begin{minipage}[t]{16.5 cm}
\caption{The effective PMC scale $Q_\star$ for $H\to b\bar{b}$ ($\mu=M_H$) up to ${\rm N^{2}LLO}$ level~\cite{Shen:2017pdu}. $Q^{(1)}_\star$ is at the LLO level, $Q^{(2)}_\star$ is at the NLLO level, and $Q^{(3)}_\star$ is at the ${\rm N^{2}LLO}$ level. The value of $Q_\star$ approaches its ``exact" value (schematically shown by the dotted line) via an oscillatory way when more loop-terms are included. \label{fig:PMCs-scalex2}}
\end{minipage}
\end{center}
\end{figure}

As indicated by Eq.(\ref{eq:pmcscale2}), the scale $Q_\star$ and thus the PMC-s prediction has only the first kind of residual scale dependence, which is caused by the unknown higher-order terms; its precision will be improved as more higher-order terms are included. As examples, we present the determined effective PMC scales up to ${\rm N^{2}LLO}$ level based on the four-loop prediction for observables $R(Q=31.6\;{\rm GeV})$ and $\Gamma(H\to b\bar{b})$ with $\mu=M_H$ in Figures \ref{fig:PMCs-scalex1} and \ref{fig:PMCs-scalex2}. These two figures demonstrate that the absolute scale difference between two nearby values becomes smaller as more loop corrections are included. There are two ways to approach the ``exact" value of $Q_{*}$; i.e., one is by a monotonous approach(Figure \ref{fig:PMCs-scalex1}) and the other is by an oscillating approach (Figure \ref{fig:PMCs-scalex2}). Moreover, due to the eliminating of divergent renormalon terms, the pQCD series for the observable converges rapidly; thus any residual scale dependence to the pQCD prediction is greatly suppressed. The magnitude of such residual scale dependence is generally much smaller than the case of the PMC multi-scale approach, since the precision of the PMC multi-scales are generally different due to the $\{\beta_i\}$-terms are known at different orders. More explicitly, for a four-loop prediction $\rho_4(Q)$, in the PMC multi-scale approach, the LO PMC scale $Q_1$ is at the NNLL-level, the NLO PMC scale $Q_2$ is at the NLL-level, the NNLO PMC scale is at the LL level, respectively; while in the PMC single-scale approach, the single effective PMC scale $Q_\star$ is fixed at the NNLL-level.

\subsubsection{The PMC-s scale setting approach for the general $C$-scheme coupling}

Using the relation (\ref{eq:Expandhata}) between the $C$-scheme coupling $\hat{a}_\mu$ and the conventional coupling $a_\mu$, the pQCD approximant $\rho_n(Q)$ can be transformed as
\begin{eqnarray}
\hat{\rho}_{n}(Q) = \sum_{i=1}^{n} \hat{c}_i(\mu/Q)\; \hat{a}_\mu^{i}.
\label{eq:rhoC}
\end{eqnarray}
Here $n\ge2$; i.e., we are considering at least the NLO correction to the pQCD prediction. The coefficients $\hat{c}_i$ can be related to the coefficients $r_{i,j}$ for conventional running coupling from Eqs.(\ref{eq:rhodelta2}, \ref{PMC-s-coeff1}, \ref{PMC-s-coeff2}, \ref{PMC-s-coeff3}, \ref{PMC-s-coeff4}). We shall adopt the same notation $\hat{r}_{i,j}=r_{i,j}|_{\mu=Q}$ for our following treatment, in which the conformal coefficient are labeled $r_{i,0}=\hat{r}_{i,0}$. These equations show the non-conformal part of the coefficients $\hat{c}_i$ have a much more complex $\{\beta_i\}$-structure; it can be schematically written as
\begin{eqnarray}
\hat{c}_i(\mu/Q) = \hat{r}_{i,0}+ g_i\left(\mu/Q, \left\{\beta^{m}_{j}\right\}\right) + h_i\left(\mu/Q, \left\{\beta_{k}^{l} /\beta_0^{n}\right\}\right),
\end{eqnarray}
where $j\geq 0$, $l,m,n,k \geq 1$, the functions $g_i$ and $h_i$ can be read from the known coefficients $\hat{c}_i$. Up to four-loop level, the functions $g_i$ and $h_i$ are
\begin{eqnarray}
g_1 &=& 0, \nonumber\\
g_2 &=& \beta_0 \left[\hat{r}_{1,0}(C+L)+\hat{r}_{2,1}\right], \nonumber\\
g_3 &=& 2\beta_0 \left[r_{2,0}(C+L)+\hat{r}_{3,1}\right] +\beta_1 \left[\hat{r}_{1,0}(C+L)+\hat{r}_{2,1}\right]+ \beta_0^2 \left[\hat{r}_{1,0} (C+L)^2+2 \hat{r}_{2,1}(C+L)+\hat{r}_{3,2}\right], \nonumber\\
g_4 &=& 3\beta_0 \left[\hat{r}_{3,0}(C+L)+\hat{r}_{4,1}\right]+2\beta_1 \left[\hat{r}_{2,0} (C+L)+\hat{r}_{3,1}\right]
+3\beta_2 \left[\hat{r}_{1,0} (C+L)+\hat{r}_{2,1}\right] \nonumber\\
&& +3 \beta _0^2 \left[\hat{r}_{2,0} (C+L)^2+2 \hat{r}_{3,1} (C+L)+\hat{r}_{4,2}\right]
 +\frac{5}{2} \beta _1 \beta _0 \left[\hat{r}_{1,0} (C+L)^2+2 \hat{r}_{2,1} (C+L)+\hat{r}_{3,2}\right] \nonumber\\
&& +\beta _0^3 \left[\hat{r}_{1,0} (C+L)^3+3 \hat{r}_{2,1} (C+L)^2+3\hat{r}_{3,2}(C+L)+\hat{r}_{4,3}\right], \nonumber\\
h_1 &=& h_2=0, \nonumber\\
h_3 &=& \frac{\beta_2}{\beta_0} \hat{r}_{1,0}-\frac{\beta_1^2}{\beta_0^2} \hat{r}_{1,0}, \nonumber\\
h_4 &=& \frac{\beta_3}{2\beta_0}\hat{r}_{1,0}+\frac{2\beta_2}{\beta_0}\hat{r}_{2,0} -\frac{2\beta_1^2}{\beta_0}\left[\hat{r}_{1,0}(C+L)+\hat{r}_{2,1}\right]
-\frac{\beta_1^3}{2\beta_0^3}\hat{r}_{1,0}-\frac{2\beta_1^2}{\beta_0^2}\hat{r}_{2,0}, \nonumber
\end{eqnarray}
where $L=\ln\mu^2/Q^2$. Due to this complex $\{\beta_i\}$-structure using the $C$-scheme coupling, it is hard to distribute the $g_i$ and $h_i$ functions into running couplings at different orders, which is however important for determining the correct running behavior of the coupling constant at each order. Thus to avoid this difficulty, we will treat $(g_i+h_i)$ as a whole and adopt the PMC-s scale setting approach to eliminate those renormalization group involved $\{\beta_i\}$-terms~\footnote{As a comparison, the non-conformal coefficients $r_{i,j(\ge1)}$ for the pQCD series using conventional running coupling, as shown by Eq.(\ref{eq:rhodelta2}), are superposition of RGEs for the running couplings at each order; thus they can be conveniently adopted for determining the correct PMC scale at each order.}.

By using Eq.(\ref{eq:rhoC}), we obtain
\begin{eqnarray}
\frac{\partial \hat{\rho}_n}{\partial C} = -\frac{\partial \hat{a}_\mu}{\partial C} \frac{\partial \hat{\rho}_n}{\partial \hat{a}_\mu}
= -\mu^2 \frac{\partial \hat{a}_\mu}{\partial \mu^2}\frac{\partial \hat{\rho}_n}{\partial \hat{a}_\mu}
= -\hat{\beta}(\hat{a}_\mu) \frac{\partial \hat{\rho}_n}{\partial \hat{a}_\mu},
\label{eq:Cdependence}
\end{eqnarray}
where we have used the fact that the scale-running and scheme-running of the $C$-scheme coupling $\hat{a}_\mu$ satisfy the same $\hat\beta$-function.

Eq.(\ref{eq:Cdependence}) shows that when the non-conformal terms associated with the $\hat\beta(\hat{a}_{\mu})$-function have been removed, one can achieve a scheme-independent prediction at any fixed order; i.e., $\hat\beta(\hat{a}_{\mu})\to 0$ indicates ${\partial \hat{\rho}_n}/{\partial C}\to 0$. This conclusion agrees with that of Eq.(\ref{eq:deltaRGI}) which is derived using the dimensional-like ${\cal R}_\delta$-scheme. The present conclusion however is much more general, since the value of $C$ is arbitrary and could be referred to as a general renormalization scheme. In the following, we present an additional explanation of this scheme independence.

Following the PMC-s procedures, an effective scale $Q_\star$ is introduced to eliminate all nonconformal terms. The scale $Q_\star$ is thus determined by requiring
\begin{eqnarray}
\sum_{i=1}^{n} \left[ g_i\left(Q_{\star}/Q, \left\{\beta^{m}_{j}\right\}\right) + h_{i}\left(Q_{\star}/Q, \left\{\beta_{k(\ge1)}^{l} /\beta_0^{n(\ge1)}\right\}\right)\right] \hat{a}_{Q_\star}^{i}=0.
\end{eqnarray}
This equation can be solved recursively, and similar to Eq.(\ref{eq:pmcscale2}), we can express its solution as a power series in $\hat{a}_{Q_\star}$, i.e.,
\begin{eqnarray}
\ln\frac{Q_{\star}^2}{Q^2} = \sum^{n-2}_{i= 0} \hat{S}_i \hat{a}^i_{Q_\star},
\label{eq:C-PMC-scale}
\end{eqnarray}
whose first three coefficients for the fourth-order prediction are
\begin{eqnarray}
\hat{S}_0 &=& -\frac{\hat{r}_{2,1}}{\hat{r}_{1,0}}-C, \\
\hat{S}_1 &=& \frac{2\left(\hat{r}_{2,0}\hat{r}_{2,1}-\hat{r}_{1,0}\hat{r}_{3,1}\right)}{\hat{r}_{1,0}^2}
+\frac{\hat{r}_{2,1}^2-\hat{r}_{1,0} \hat{r}_{3,2}}{\hat{r}_{1,0}^2}\beta_0 +\frac{\beta_1^2}{\beta_0^3}-\frac{\beta_2}{\beta_0^2}, \\
\hat{S}_2 &=& \frac{3\hat{r}_{1,0}\left(\hat{r}_{3,0}\hat{r}_{2,1} -\hat{r}_{1,0}\hat{r}_{4,1}\right)+4\hat{r}_{2,0} \left(\hat{r}_{1,0}\hat{r}_{3,1}-\hat{r}_{2,0} \hat{r}_{2,1}\right)}{\hat{r}_{1,0}^3}
+\frac{3\hat{r}_{1,0}\hat{r}_{2,1}\hat{r}_{3,2} -\hat{r}_{1,0}^2\hat{r}_{4,3} -2\hat{r}_{2,1}^3}{\hat{r}_{1,0}^3}\beta_0^2 \nonumber\\
&& + \frac{3\left(\hat{r}_{2,1}^2-\hat{r}_{1,0}\hat{r}_{3,2}\right)}{2\hat{r}_{1,0}^2}\beta_1
+\frac{3\hat{r}_{1,0}\left(2\hat{r}_{2,1} \hat{r}_{3,1} -\hat{r}_{1,0} \hat{r}_{4,2}\right)+ \hat{r}_{2,0}\left(2 \hat{r}_{1,0} \hat{r}_{3,2}-5 \hat{r}_{2,1}^2\right)}{\hat{r}_{1,0}^3}\beta_0 \nonumber\\
&& - \frac{\beta_1^3}{2\beta_0^4}+\frac{\beta_2 \beta_1}{\beta_0^3}-\frac{\beta_3}{2\beta_0^2}.
\end{eqnarray}
Only the first coefficient $\hat{S}_0$ depends on the scheme parameter $C$, and all the other higher-order coefficients $\hat{S}_{i}$ ($i\geq1$) are independent of $C$.

The $C$-scheme coupling $\hat{a}_{Q_\star}$ satisfies Eq.(\ref{eq:ahat}), and we obtain
\begin{eqnarray}
\frac{1}{\hat{a}_{Q_\star}}+\frac{\beta_1}{\beta_0} \ln\hat{a}_{Q_\star} &=& \beta_0\left(\ln\frac{Q_\star^2}{\Lambda^2}+C\right)\nonumber\\
&=& \beta_0 \left( \ln\frac{Q^2}{\Lambda^2}-\frac{\hat{r}_{2,1}}{\hat{r}_{1,0}} +\sum^{n-2}_{i\geq 1} \hat{S}_i \hat{a}^i_{Q_\star} \right). \label{Csas2}
\end{eqnarray}
The second equation shows that even though the effective scale $Q_{\star}$ depends on the choice of $C$, the coupling $\hat{a}_{Q_\star}$ is independent of the choice of $C$ at any fixed order. Thus, after fixing the scale $Q_\star$, we achieve a $C$-scheme independent pQCD series
\begin{eqnarray}
\hat{\rho}_n(Q)|_{\rm PMC} &=& \sum^{n}_{i\geq1} r_{i,0} \hat{a}_{Q_\star}^i .
\label{eq:PMCprediction}
\end{eqnarray}
The conventional pQCD series $\hat{\rho}_{n}(Q)$ depends on the initial choice of scheme via the coefficients $r_{i,j}$ and the $\{\beta_{i\geq2}\}$-functions. After applying the PMC-s procedures, Eq.(\ref{eq:PMCprediction}) indicates the pQCD predictions are scheme independent for any choice of the renormalization scheme.

\subsubsection{Equivalence of the pQCD predictions for conventional and $C$-scheme couplings}
\label{subsec:CequalRdelta}

The PMC-s approach replaces the individual PMC scales of the PMC multi-scale approach by an overall effective scale, whose precision and thus the precision of the PMC-s pQCD prediction is greatly improved as more higher-order terms are included. The PMC-s predictions are explicitly independent of the choice of the initial renormalization scale. Given a measurement of the running coupling at a reference scale Q, $a_Q$, one can determine the value of the asymptotic scale $\Lambda$ for a specific scheme by using its $\{\beta_i\}$-functions. We then obtain the pQCD predictions independent of any choice of scheme (represented by any choice of $C$). This demonstrates to any orders the scheme-independent of the PMC predictions -- Given one measurement which sets the value of the coupling at one kinematic point, the resulting PMC predictions are independent of the choice of the renormalization scheme. The PMC-s approach can be adopted as a valid substitution for the PMC multi-scale approach for setting the renormalization scale for high-energy processes, particularly when one does not need detailed information at each order.

The pQCD predictions under the PMC single scale setting approach for the conventional coupling (Eq.(\ref{PMC-spQCD})) and the $C$-scheme coupling (Eq.(\ref{eq:PMCprediction})) are exactly the same. This equivalence is due to the fact that
\begin{itemize}
\item By eliminating the non-conformal terms, the pQCD approximant becomes the conformal series. As shown by Eqs.(\ref{eq:Expandhata}, \ref{eq:Expandhata2}), the $C$-scheme coupling $\hat{a}_\mu$ and the conventional coupling $a_\mu$ are mutually related by the RG-involved $\{\beta_i\}$-terms. Thus, after applying the PMC-s procedures, the conformal coefficients $r_{i,0}$ at every order are identical for both cases.

\item For an $n_{\rm th}$-order prediction, the effective conventional coupling $a_{Q_\star}$ satisfies the RGE (\ref{eq:betafun}), which can be rewritten in the following form with the help of Eq.(\ref{eq:pmcscale2}), i.e.
 \begin{eqnarray}
 \frac{1}{a_{Q_\star}}+\frac{\beta_1}{\beta_0} \ln a_{Q_\star}
   = \beta_0\left[\ln\frac{Q^2}{\Lambda^2}+\sum_{i\geq 0}^{\rm n-2} S_i a_{Q_\star}^i - \left(\int_0^{a_{Q_\star}} \frac{{\rm d}a}{\tilde{\beta}(a)}\right)_{\rm n-2} \right], \label{Cs1}
 \end{eqnarray}
 where the subscript $(n-2)$ indicates the perturbative expansion is up to $a_{Q_\star}^{n-2}$-order. On the other hand, the effective $C$-scheme coupling $\hat{a}_{Q_\star}$ satisfies Eq.(\ref{Csas2}), and by using the relation
 \begin{eqnarray}
 \left( \int_0^{a_\mu} \frac{{\rm d}a}{\tilde{\beta}(a)} \right)_{n-2} = \sum_{i=1}^{\rm n-2}(S_i-\hat{S}_i) a_\mu^i,
 \end{eqnarray}
 it can be further written as
 \begin{eqnarray}
 \frac{1}{\hat{a}_{Q_\star}}+\frac{\beta_1}{\beta_0} \ln \hat{a}_{Q_\star}
   = \beta_0\left[\ln\frac{Q^2}{\Lambda^2}+\sum_{i\geq 0}^{\rm n-2} S_i \hat{a}_{Q_\star}^i - \left(\int_0^{\hat{a}_{Q_\star}} \frac{{\rm d}a}{\tilde{\beta}(a)}\right)_{\rm n-2} \right]. \label{Cs2}
 \end{eqnarray}

\item Eq.(\ref{Cs1}) or Eq.(\ref{Cs2}) indicate that both the effective couplings $a_{Q_\star}$ and $\hat{a}_{Q_\star}$ are solutions of the same equation, which can be solved iteratively. These two equations are alternatives to the same RGE, whose solutions will be identical for the choice of same scale $Q$, indicating $a_{Q_\star} \equiv \hat{a}_{Q_\star}$ for any fixed-order prediction. Thus after applying the PMC-s scale setting procedures, the resultant pQCD series for conventional and $C$-scheme couplings are the same.
\end{itemize}

\subsection{An example without renormalization scheme-and-scale dependence}

We take the non-singlet Adler function as an explicit example to explain how the scheme-and-scale independent predictions can be achieved by applying the PMC-s scale setting approach together with the $C$-scheme coupling. For this purpose, we shall transform the pQCD series derived using the conventional $\overline{\rm MS}$-scheme into the series using the $C$-scheme coupling. In doing the numerical calculations below, we adopt the world average $\alpha_s^{\overline{\rm MS}}(M_Z=91.1876\; {\rm GeV}) =0.1181(11)$~\cite{Olive:2016xmw} as the reference value for fixing the running coupling, which runs down to $\alpha_s^{\overline{\rm MS}}(M_\tau=1.777 \;{\rm GeV})=0.3159(95)$ by using the RGE.

The non-singlet Adler function is defined as~\cite{Adler:1974gd},
\begin{equation}
D^{\rm ns}(Q^2,\mu) = -12{\pi^2}{Q^2}\frac{d}{d{Q^2}}\Pi^{\rm ns}(L,{a_\mu}),  \label{eq:Adler}
\end{equation}
where $Q$ is the mass scale of the observable -- the kinematic value at which it is measured, $\mu$ encodes the renormalization scale, $a_\mu=\alpha_s(\mu)/\pi$, $\Pi^{\rm ns}(L,a_\mu)=\sum_{i \ge 0} {\Pi_i^{\rm ns}} {a_\mu^i}/{16\pi^2}$ is the non-singlet part of the polarization function for a flavor-singlet vector current, and $L=\ln{\mu^2}/{Q^2}$. The scale-running behavior of $\Pi^{\rm ns}(L, a_\mu)$ is controlled by
\begin{equation}
\left( {\mu^2}\frac{\partial}{\partial{\mu^2}} + \beta(a_\mu)\frac{\partial}{\partial{a_\mu}} \right)\Pi^{\rm ns}(L,a_\mu) = \gamma^{\rm ns}(a_\mu) ,  \label{eq:gamma}
\end{equation}
where $\gamma^{\rm ns}(a_\mu) = \sum_{i \ge 0} {\gamma^{\rm ns}_i} {a_\mu^i}/{16\pi^2}$ is the anomalous dimension for the non-singlet part of the photon-field. Then we obtain
\begin{eqnarray}
D^{\rm ns}(Q^2,\mu) = 12\pi^2 \left[\gamma^{\rm ns}(a_\mu) - \beta(a_\mu)\frac{\partial} {\partial{a_\mu}}{\Pi}^{\rm ns}(L,{a_\mu})\right]
= \frac{3}{4}\gamma^{\rm ns}_0 + \bar{D}^{\rm ns}(Q^2,\mu).
\label{eq:nsAdler}
\end{eqnarray}

Using Eqs.(\ref{eq:Adler}, \ref{eq:gamma}, \ref{eq:nsAdler}), we observe $d D^{\rm ns}(Q^2,\mu) /d\mu^2\equiv0$ at any fixed order. This shows that the pQCD approximant $D^{\rm ns}(Q^2,\mu)$ is a local RGI quantity, since it indicates the pQCD prediction to be scale-independent at any fixed order. The introduced anomalous dimension $ \gamma^{\rm ns}(a_\mu)$ is associated with the renormalization of the QED coupling, which not only determines the correct scale-running behavior of $\Pi^{\rm ns}(L,a_\mu)$ but also ensures that $D^{\rm ns}(Q^2,\mu)$ satisfies the local RGI~\cite{Wu:2014iba}. This explains why the $\gamma^{\rm ns}$-terms, which appear in the Adler function $D^{\rm ns}(Q^2,\mu)$, should be treated as conformal terms during the PMC scale setting and cannot be used to set the pQCD renormalization scales for $D^{\rm ns}(Q^2,\mu)$ -- since only those $\{\beta_i\}$-terms which are associated with the $\alpha_s$-running should be adopted for PMC scale setting.

The pQCD series of $\bar{D}^{\rm ns}(Q^2,\mu)$ up to $n_{\rm th}$-loop level can be written in the following form
\begin{eqnarray}
\bar{D}^{\rm ns}_n(Q^2,\mu) = \sum_{i=1}^n r_i(\mu/Q) a_\mu^i.
\end{eqnarray}
At present, the coefficients $\gamma^{\rm ns}_i$ and $\Pi^{\rm ns}_i$ within the $\overline{\rm MS}$-scheme have been given up to four-loop level~\cite{Baikov:2012zm}, and the coefficients $r_i$ within the $\overline{\rm MS}$-scheme up to four-loop level can be read from Refs.\cite{Baikov:2010je, Chetyrkin:1996ia}. For example, if setting $\mu=Q$ and $n_f=3$, the first four $\overline{\rm MS}$-coefficients are
\begin{displaymath}
r_1=1, \,\,\, r_2=1.6398, \,\,\, r_3=6.3710, \,\,\, r_4=49.0757.
\end{displaymath}
The coefficients at any other choices of the renormalization scale ($\mu\neq Q$) can be obtained via RGE.

\begin{itemize}
\item {\it Predictions using conventional scale setting.}
\end{itemize}

By setting $Q=M_\tau$, we can obtain a $\overline{\rm MS}$-scheme prediction for $\bar{D}^{\rm ns}(Q^2,\mu)$ up to four-loop level by using conventional scale setting (Conv.), i.e.
\begin{equation}
\bar{D}^{\rm ns}_4(M_\tau^2,\mu=M_\tau)|_{\rm Conv.} = 0.1286 \pm 0.0053 \pm 0.0094,   \label{eq:Adler-MSbar-4loop}
\end{equation}
where the first error is about $\pm 4\%$ of the central value which is caused by the $\alpha_s^{\overline{\rm MS}}(M_Z)$ uncertainty; i.e., $\Delta\alpha_s^{\overline{\rm MS}}(M_Z)=\pm0.0011$~\cite{Olive:2016xmw}, and the second error is about $\pm 7\%$ of the central value which represents a conservative estimate of the ``unknown'' high-order contribution.

As shall be shown in Sec.\ref{sec:4}, it is important to reliably estimate the contributions of unknown higher-order terms using information from the known pQCD series. In the literature, two naive ways have been adopted for estimating the second error: one is to take the error as the maximum value of the last known term of the perturbative series within reasonable choices of initial scale, and the other is to set the error directly as the difference of the predictions caused by directly varying the initial scale within reasonable region. The second way is not reliable, since it only partly estimates the non-conformal contribution but not the conformal one. Thus here we have implicitly adopted the first way to do the estimate, and for the present four-loop prediction, we take the maximum value of $|r_4(\mu/M_{\tau}) a_\mu^4|$ with $\mu\in[M_{\tau},4M_{\tau}]$ as the estimated ``unknown'' high-order contribution~\cite{Wu:2018cmb}. For a comparison, we also discuss the errors for the second way, which are obtained by varying the initial scale within the region of $[M_{\tau},4M_{\tau}]$, leading to $\bar{D}^{\rm ns}_4|_{\rm Conv.}\in[0.1083,0.1286]$. It indicates that the conventional scale error is still about $16\%$ at the four-loop level. By taking a narrower interval, e.g. $[0.61 M_{\tau}, 1.28 M_{\tau}]$~\cite{Davier:2008sk, Pich:2013lsa}, one can achieve a smaller scale error ($\sim 13\%$), e.g. $\bar{D}^{\rm ns}_4|_{\rm Conv.}\in[0.1245,0.1408]$. Thus, a five-loop or even higher loop calculation is needed to suppress the scale uncertainty using the conventional scale setting approach. At present, the unknown fifth-order coefficient has been roughly estimated by several groups, e.g. $r_5\simeq 283$~\cite{Beneke:2008ad} or $r_5\simeq 275$~\cite{Baikov:2008jh}. If using $r_5\simeq 283$, Eq.(\ref{eq:Adler-MSbar-4loop}) changes to
\begin{equation}
\bar{D}^{\rm ns}_5(M_\tau^2,\mu=M_\tau)|_{\rm Conv.} = 0.1315 \pm 0.0057 \pm 0.0065,
\label{eq:Adler-MSbar-5loop}
\end{equation}
where the second error is reduced to $\pm 5\%$, and the conventional renormalization scale uncertainty is largely reduced to $6\%$.

In addition to the renormalization scale dependence, the predictions using conventional scale setting is also scheme dependent at any fixed order. As an explanation, we adopt the $C$-scheme coupling to illustrate this scheme dependence. By using the relation (\ref{eq:Expandhata}), we rewrite $\bar{D}^{\rm ns}_n(Q^2,\mu)$ in terms of the $C$-scheme coupling $\hat{a}_\mu$ as
\begin{eqnarray}
\bar{D}_n^{\rm ns}(Q^2,C) = \sum_{i=1}^n \hat{c}_i(\mu/Q) \hat{a}_\mu^i,
\end{eqnarray}
where the coefficients $\hat{c}_i(\mu/Q)$ can be derived by using Eqs.(\ref{PMC-s-coeff1}, \ref{PMC-s-coeff2}, \ref{PMC-s-coeff3}, \ref{PMC-s-coeff4}). For example, if setting $\mu=Q$ and $n_f=3$, the $C$-dependent coefficients $\hat{c}_i$ in terms of $r_i$ up to five-loop level are
\begin{eqnarray}
\hat{c}_1(Q/Q) &=& 1, \nonumber\\
\hat{c}_2(Q/Q) &=& 1.6398+2.25C, \nonumber\\
\hat{c}_3(Q/Q) &=& 7.6816+11.3792C+5.0625C^2, \nonumber\\
\hat{c}_4(Q/Q) &=& 61.0597+72.0804C+47.4048C^2+11.3906C^3, \nonumber\\
\hat{c}_5(Q/Q) &=& r_5 + 65.4774+677.68C+408.637C^2+162.464C^3+25.6289C^4. \nonumber
\end{eqnarray}
The coefficients at any other choices of the renormalization scale ($\mu\neq Q$) can be obtained via RGE. These coefficients at NLO and higher orders explicitly depend on $C$.

\begin{figure}[htb]
\epsfysize=9.0cm
\begin{center}
\begin{minipage}[t]{10 cm}
\epsfig{file=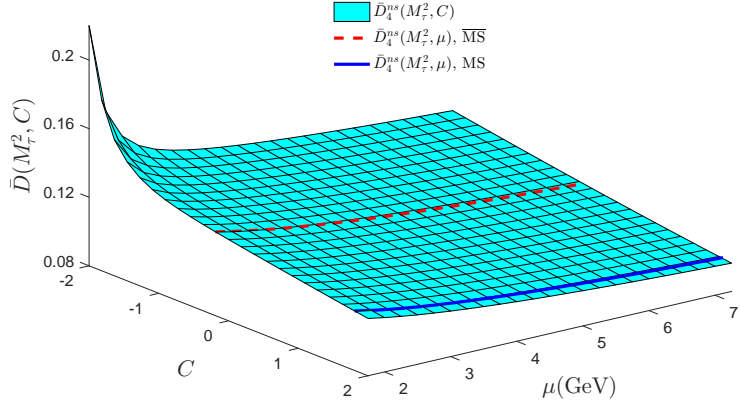,scale=0.6}
\end{minipage}
\begin{minipage}[t]{16.5 cm}
\caption{The four-loop prediction of the non-singlet Adler function $\bar{D}^{\rm ns}_{4}(M_\tau^2,C)$ using conventional scale setting as a function of parameters $C$ and $\mu$~\cite{Wu:2018cmb}, which is shown by a shaded band. Here the dashed line is for the $\overline{\rm MS}$-scheme, and the solid line is for the MS-scheme. \label{fig:Adler4l3D}}
\end{minipage}
\end{center}
\end{figure}

The parameter $C$ characterizes the scheme-dependence of the pQCD prediction. A graphical representation of the four-loop prediction on $\bar{D}^{\rm ns}(M_\tau^2,C)$ as a function of the parameters $C$ and $\mu$ is presented in Figure~\ref{fig:Adler4l3D}, where we have chosen $C\in[-2,+2]$ and $\mu \in \left[M_\tau, 4M_\tau\right]$. The relation between the $C$-scheme coupling $\hat{a}_{M_\tau}$ and the $\overline{\rm MS}$-scheme coupling $a_{M_\tau}$ ceases to be perturbative and breaks down below $C\sim -2$. Thus we adopt $C\ge -2$ in our discussions. In Figure~\ref{fig:Adler4l3D}, the shaded band shows the scheme-and-scale dependence of $\bar{D}^{\rm ns}_{4}(M_\tau^2,C)$, which still shows a large error at the four-loop prediction. Using an appropriate choice of $C$, the pQCD prediction using $C$-scheme coupling $\hat{a}_\mu$ are equivalent to predictions using some of the familiar schemes; e.g. the dashed line in Figure \ref{fig:Adler4l3D} is for the $\overline{\rm MS}$-scheme and the solid line is for the MS-scheme. To ensure equivalence, the value of $C$ should be changed for different scales. For example, by taking $C=-0.188$ one obtains the conventional $\overline{\rm MS}$ prediction for $\mu=M_\tau$, which changes to $C=-0.004$ for $\mu=4M_\tau$.

The scheme-dependence is unavoidable for the conventional scale-setting approach. If one requires the estimated ``unknown'' high-order contribution, $|\hat{c}_n(\mu/M_\tau) \hat{a}_{\mu}^n|_{\rm MAX}$, to be at its minimum, we can obtain an ``optimal" $C$-scheme for $\bar{D}_n^{\rm ns}(Q^2,C)$. For example, the ``optimal" $C$-value for a four-loop prediction with $n=4$ is, $C_{\rm Opt.}=-0.972$, leading to
\begin{equation}
\bar{D}^{\rm ns}_4(M_\tau^2,C_{\rm Opt.}=-0.972)|_{\rm Conv.} = 0.1365 \pm 0.0069 \pm 0.0083,
\label{eq:Adler-C-4loop}
\end{equation}
where the central value is for $\mu=M_\tau$, the first error is for $\Delta\alpha_s^{\overline{\rm MS}}(M_Z)=\pm0.0011$, and the second error is an estimate of the ``unknown'' high-order contribution. As for a five-loop prediction with $n=5$, if using the approximation $r_5\simeq 283$, the ``optimal" $C$-value changes to $-1.129$, and we obtain
\begin{equation}
\bar{D}^{\rm ns}_5(M_\tau^2,C_{\rm Opt.}=-1.129)|_{\rm Conv.} = 0.1338 \pm 0.0062 \pm 0.0054.
\label{eq:Adler-C-5loop}
\end{equation}

\begin{itemize}
\item {\it Predictions using PMC scale setting.}
\end{itemize}

\begin{figure}[htb]
\epsfysize=9.0cm
\begin{center}
\begin{minipage}[t]{10 cm}
\epsfig{file=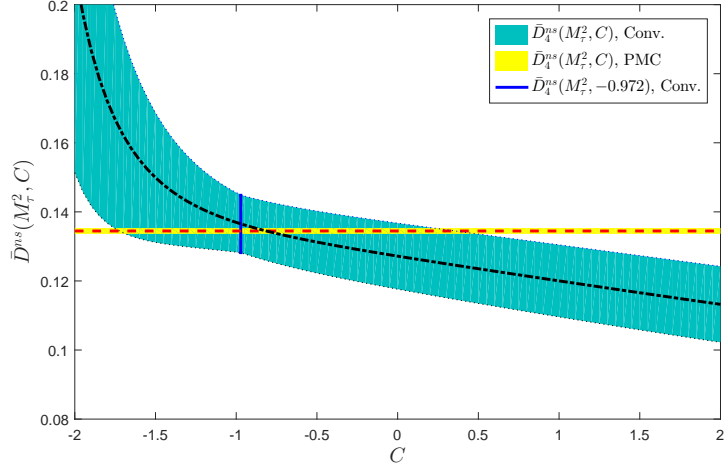,scale=0.55}
\end{minipage}
\begin{minipage}[t]{16.5 cm}
\caption{The non-singlet Adler function $\bar{D}^{\rm ns}(M_\tau^2,C)$ as a function of the parameter $C$~\cite{Wu:2018cmb}. The dash-dot line is the prediction using conventional scale setting; the darker-shaded band is the uncertainty for the four-loop prediction $\Delta = \pm|\hat{c}_4(\mu/M_\tau) \hat{a}_{\mu}^4|_{\rm MAX}$, where MAX is the maximum value for $\mu\in[M_\tau,4M_{\tau}]$. When $C=-0.972$, the error bar as shown by a vertical solid line is at its minimum. The dash line represents the four-loop PMC prediction, and the lighter-shaded band is for $\Delta = \pm |\hat{r}_{4,0}\hat{a}^4_{Q_\star}|$. The independence of the PMC prediction on the parameter $C$ demonstrates its scheme-independence. \label{fig:Adler4l2Dimproved}}
\end{minipage}
\end{center}
\end{figure}

\begin{figure}[htb]
\epsfysize=9.0cm
\begin{center}
\begin{minipage}[t]{10 cm}
\epsfig{file=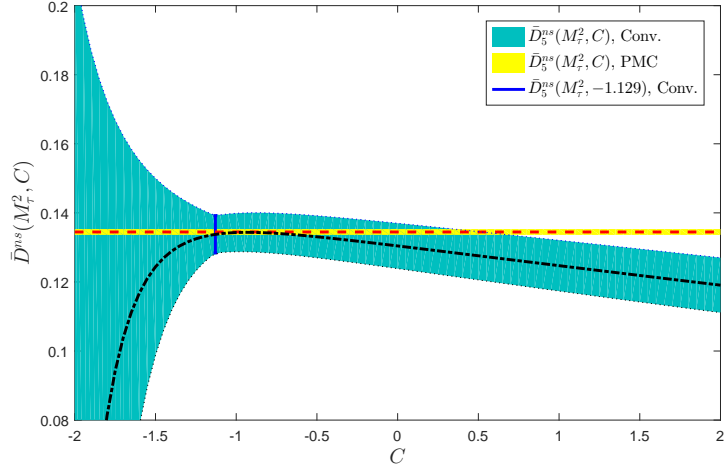,scale=0.55}
\end{minipage}
\begin{minipage}[t]{16.5 cm}
\caption{The non-singlet Adler function $\bar{D}^{\rm ns}(M_\tau^2,C)$ as a function of the parameter $C$~\cite{Wu:2018cmb}. The dash-dot line is the prediction using conventional scale setting; the darker-shaded band is the uncertainty for an approximate five-loop prediction $\Delta = \pm|\hat{c}_5(\mu/M_\tau)\hat{a}_{\mu}^5|_{\rm MAX}$, where MAX is the maximum value for $\mu\in[M_\tau,4M_{\tau}]$. When $C=-1.129$, the error bar as shown by a vertical solid line is the minimum. The dash line represents the four-loop PMC prediction, and lighter-shaded band is for $\Delta = \pm |\hat{r}_{4,0}\hat{a}^4_{Q_\star}|$ . The independence of the PMC prediction on the parameter $C$ demonstrates its scheme-independence. \label{fig:Adler4l2Dimproved2}}
\end{minipage}
\end{center}
\end{figure}

In distinct to the strong scheme dependence using conventional scale setting, which will be shown by Figures \ref{fig:Adler4l2Dimproved} and \ref{fig:Adler4l2Dimproved2}, scheme and scale independent predictions can be achieved at any fixed order after applying PMC-s scale setting approach.

To apply PMC scale setting, we need to distribute the perturbative coefficients $r_i$ into conformal $(r_{i,0})$ and non-conformal $(r_{i,j(\neq0)})$ coefficients. This can be done by using the $\beta$-pattern determined by recursively using the RGE. The general $\beta$-pattern for each perturbative order up to four-loop level is shown by Eq.(\ref{eq:rhodelta2}). Up to four-loop level, the known coefficients for conventional coupling are~\cite{Shen:2016dnq}
\begin{eqnarray}
r_{i(\geq 1),0} =\frac{3}{4} \gamma^{\rm ns}_i, \,\,\, r_{i(\geq 2),1} =\frac{3}{4} \Pi^{\rm ns}_{i-1}, \,\,\, r_{i(\geq 3),2}=0, \,\,\, r_{i(\geq 4),3}=0.
\end{eqnarray}

Following the standard PMC single-scale approach, by resumming all the RG-involved non-conformal $\{\beta_i\}$-terms into the running coupling, we obtain the PMC prediction for $\bar{D}^{\rm ns}_n$, i.e.
\begin{eqnarray}
\bar{D}^{\rm ns}_n(Q^2,C)|_{\rm PMC} = \sum_{i=1}^n r_{i,0} \hat{a}_{Q_\star}^{i} = \frac{3}{4} \sum_{i=1}^n \gamma^{\rm ns}_i \hat{a}_{Q_\star}^{i}.
\end{eqnarray}
Using the known four-loop pQCD prediction $\bar{D}^{\rm ns}_4$, the PMC scale $Q_\star$ can be determined up to next-to-next-to-leading log (N$^2$LL) level, i.e.
\begin{eqnarray}
\ln\frac{Q_{\star}^2}{Q^2} = (0.2249-C) -3.1382\hat{a}_{Q_\star} -13.3954\hat{a}_{Q_\star}^2 + {\cal O}(\hat{a}_{Q_\star}^3),
\label{eq:AdlerPMCscale}
\end{eqnarray}
where the value of the $C$-scheme coupling $\hat{a}_{Q_\star}$ can be determined by using Eq.(\ref{Csas2}).

\begin{figure}[htb]
\epsfysize=9.0cm
\begin{center}
\begin{minipage}[t]{10 cm}
\epsfig{file=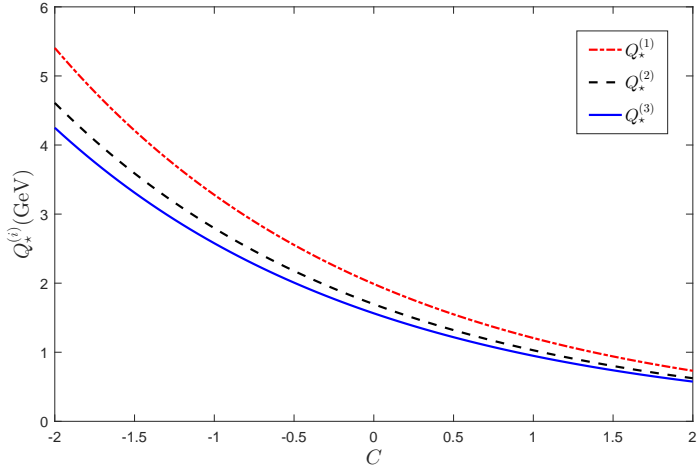,scale=0.55}
\end{minipage}
\begin{minipage}[t]{16.5 cm}
\caption{The effective PMC scale $Q^{(n-1)}_\star$ as a function of parameter $C$ for the non-singlet Adler function $\bar{D}_{n}^{\rm ns}(M_\tau^2,C)$~\cite{Wu:2018cmb}, where $n$ indicates a $n_{\rm th}$-loop prediction and the PMC scale is at the N$^{(n-1)}$LL level. \label{fig:AdlerS}}
\end{minipage}
\end{center}
\end{figure}

Eq.(\ref{eq:AdlerPMCscale}) shows that the PMC scale $Q_\star$ is independent of the choice of the initial scale, being consistent with the observation of Eq.(\ref{eq:pmcscale2}); it is, however scheme-dependent, since it depends on the parameter $C$. The PMC scale $Q_\star$ is of perturbative nature: when more loop terms are included, it becomes more accurate. We present $Q_\star$ as a function of $C$ in Figure \ref{fig:AdlerS}, in which $Q^{(1,2,3)}_\star$ are at the LL, NLL and ${\rm N^{2}LL}$ level, respectively.

Figure \ref{fig:AdlerS} shows that the scales $Q^{(1,2,3)}_\star$ decrease with the increment of the parameter $C$, and the magnitudes of these scales satisfy $Q^{(1)}_\star > Q^{(2)}_\star > Q^{(3)}_\star$ for $C\in[-2,+2]$. By taking $Q=M_\tau$, we obtain $\hat{a}_{Q_\star} \equiv 0.1056(41)$ for any choice of $C$, where the error is for $\Delta\alpha_s^{\overline{\rm MS}}(M_Z)=\pm0.0011$. This result also confirms the observation of Eq.(\ref{Csas2}) that the $C$-scheme coupling at the scale $Q_\star$ is independent of the choice of $C$. We then obtain the scheme-independent PMC prediction on $\bar{D}_4^{\rm ns}$,
\begin{eqnarray}
\bar{D}_4^{\rm ns}(M_\tau^2,C)|_{\rm PMC} = 0.1345 \pm 0.0066 \pm 0.0008,
\label{eq:Adler-PMC}
\end{eqnarray}
where the first error is for $\Delta\alpha_s^{\overline{\rm MS}}(M_Z)=\pm0.0011$, and the second error is an estimate of the ``unknown'' high-order contribution, which equals to $\pm \left|\frac{3}{4}\gamma^{\rm ns}_4 \hat{a}_{Q_\star}^{4}\right|$, since the PMC prediction is independent of the choice of initial scale $\mu$.

\begin{itemize}
\item {\it Comparison of predictions using conventional and PMC scale settings.}
\end{itemize}

For definiteness, we set the initial scale $\mu$ as $M_\tau$ to compare the scheme dependence of the non-singlet Adler function $\bar{D}^{\rm ns}_{n}(M_\tau^2,C)$ before and after applying PMC scale setting.

We present various predictions for the four-loop prediction $\bar{D}^{\rm ns}_4(M_\tau^2,C)$ in Figure~\ref{fig:Adler4l2Dimproved}. The dash-dot line stands for the prediction using conventional scale setting, which shows a rather large scheme-dependence of $\bar{D}^{\rm ns}_4(M_\tau^2,C)|_{\rm Conv.}$. The darker-shaded band stands for the conventional uncertainty for a four-loop prediction $\Delta = \pm|\hat{c}_4(\mu/M_\tau) \hat{a}_{\mu}^4|_{\rm MAX}$, where MAX is the maximum value for $\mu\in[M_\tau,4M_{\tau}]$. For small $C$ values, the error band is large; for larger $C$ values, the error band becomes slightly larger. When $C=-0.972$, the error bar is the minimum, corresponding to the optimal scheme using the conventional scale setting approach. The dash line represents the four-loop PMC prediction $\bar{D}^{\rm ns}_4(M_\tau^2,C)|_{\rm PMC}$, whose flatness indicates the scheme-independence of the PMC prediction. The lighter-shaded band is for $\Delta = \pm |\hat{r}_{4,0}\hat{a}^4_{Q_\star}|$, which is much narrower than the conventional error band due to a much faster pQCD convergence of the PMC conformal series and the elimination of scale dependence.

\begin{table}[htb]
\centering
\caption{The value of each loop-term, LO, NLO, N$^2$LO, or N$^3$LO, for the four-loop prediction $\bar{D}^{\rm ns}_4$ using conventional (Conv.) and PMC scale settings~\cite{Wu:2018cmb}, respectively. $\mu=Q=M_\tau$. The results for the $\overline{\rm MS}$-scheme, the optimal $C$-scheme with $C=-0.972$, and the $C$-scheme with $C=-0.783$~\cite{Boito:2016pwf} are presented accordingly. The PMC prediction is unchanged for any choice of $C$-scheme. $\kappa_i$ represents the relative importance among different orders. }
\begin{tabular}{ c c c c c c c c c c c c}
\hline
~~  & LO & NLO & N$^2$LO & N$^3$LO & Total & $\kappa_1$ & $\kappa_2$ & $\kappa_3$ & $\kappa_4$ \\
\hline
Conv., $\overline{\rm MS}$-scheme & 0.1006 & 0.0166 & 0.0064 & 0.0050 & 0.1286 & $78\%$ & $13\%$ & $5\%$ & $4\%$ \\
Conv., $C=-0.783$ & 0.1254 & $-0.0019$ & 0.0037 & 0.0070 & 0.1342 & $93\%$ & $-1\%$ & $3\%$ & $5\%$ \\
Conv., optimal $C$-scheme & 0.1347 & $-0.0099$ & 0.0034 & 0.0083 & 0.1365 & $99\%$ & $-7\%$ & $2\%$ & $6\%$ \\
\hline
PMC, any $C$-scheme & 0.1056 & 0.0240 & 0.0041 & 0.0008 & 0.1345 & $79\%$ & $18\%$ & $3\%$ & $<1\%$ \\
\hline
\end{tabular}
\label{tab:AdlerOrderNew}
\end{table}

Next, we discuss how the pQCD series varies according to the change of $C$-scheme before and after PMC scale setting. We present the value of each loop-term, LO, NLO, N$^2$LO, or N$^3$LO, for the four-loop prediction $\bar{D}^{\rm ns}_4$ using conventional (Conv.) and PMC scale settings in Table~\ref{tab:AdlerOrderNew}. Here the parameter $\kappa_i$ stands for the ratio of the $i_{\rm th}$-order term over the total contributions to $\bar{D}^{\rm ns}_4$, e.g. $\kappa_i=\bar{D}^{{\rm ns}, i}_{4}/\sum\limits_{i=1}^{4}\bar{D}^{{\rm ns}, i}_{4}$, where $i=1$ indicates the LO-order term, $i=2$ indicates the NLO-order term, and etc.. The pQCD convergence for the conventional $\overline{\rm MS}$-scheme is moderate. The pQCD convergence for the optimal $C$-scheme ($C=-0.972$) does not suffer from the usual $\alpha_s$-suppression, the relative size of the related high-loop terms show, $|\bar{D}^{\rm ns, LO}_{4}| \gg |\bar{D}^{\rm ns, NLO}_{4}| \sim |\bar{D}^{\rm ns, N^2LO}_{4}| \sim |\bar{D}^{\rm ns, N^3LO}_{4}|$. On the other hand, by applying the PMC, a much better pQCD convergence is naturally achieved due to the elimination of the divergent renormalon-like terms.

By using the approximate five-loop term $r_5\simeq 283$, we give the results for the five-loop prediction $\bar{D}^{\rm ns}_5(M_\tau^2,C)$ in Figure~\ref{fig:Adler4l2Dimproved2}. Figure~\ref{fig:Adler4l2Dimproved2} shows that a smaller error bar for $\bar{D}^{\rm ns}_5(M_\tau^2,C)|_{\rm Conv.}$ is achieved with a five-loop term, which first increases and then decreases with the increment of $C$, and the optimal $C$-value is slightly shifted to $C=-1.129$. The flat dash line in Figure~\ref{fig:Adler4l2Dimproved} also shows that the scheme dependence can be eliminated by applying the PMC. Due to the much faster pQCD convergence after applying PMC scale setting, the PMC prediction indicates that unknown high-order contributions could be quite small in comparison to the four-loop prediction.

\begin{figure}[htb]
\epsfysize=9.0cm
\begin{center}
\begin{minipage}[t]{10 cm}
\epsfig{file=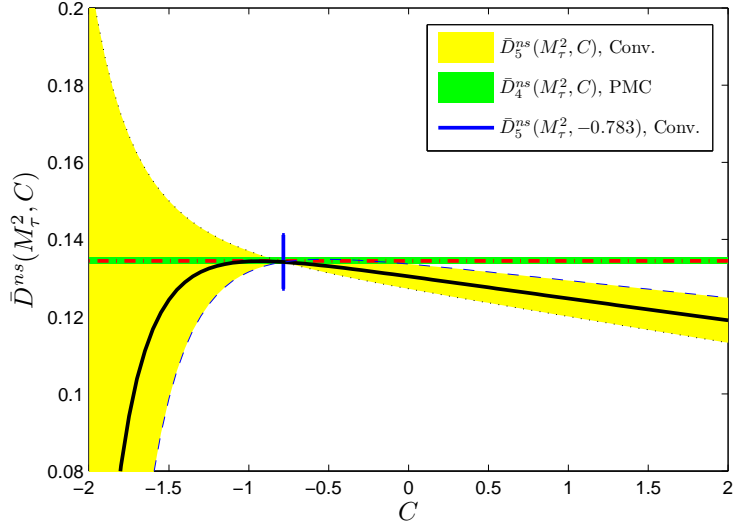,scale=0.7}
\end{minipage}
\begin{minipage}[t]{16.5 cm}
\caption{The non-singlet Adler function $\bar{D}^{\rm ns}(M_\tau^2,C)$ as a function of $C$ using conventional scale setting~\cite{Wu:2018cmb}, which agrees with that of Ref.\cite{Boito:2016pwf}. $\mu=Q=M_\tau$. The solid line is the approximate five-loop prediction with $r_5\simeq 283$ using conventional scale setting; the lighter-shaded band is its uncertainty $\Delta=\pm\hat{c}_5(M_\tau/M_\tau)\hat{a}_{M_\tau}^5$. The optimal scheme corresponds to $C=-0.783$, which leads to a vanishing $\hat{c}_5(M_\tau/M_\tau)\hat{a}_{M_\tau}^5$, and $\pm|\hat{c}_4(M_\tau/M_\tau)\hat{a}_{M_\tau}^4|$ is taken as its uncertainty. As a comparison, the dash-dot line represents the scheme-independent four-loop PMC prediction, whose darker-shaded band is for $\Delta=\pm |\hat{r}_{4,0}\hat{a}^4_{Q_\star}|$. \label{fig:Adler5loriginal}}
\end{minipage}
\end{center}
\end{figure}

An approximate method to determine the optimal $C$-scheme is suggested in Ref.\cite{Boito:2016pwf} by fixing the renormalization scale $\mu=Q$ and requiring the magnitude of the last known-term $\hat{c}_n(Q/Q)\hat{a}_Q^n$ to be at its minimum. Using this suggestion, the uncertainty is assumed to be given by the magnitude of $\hat{c}_n(Q/Q)\hat{a}_Q^n$, and if specifically $\hat{c}_n(Q/Q)\hat{a}_Q^n$ equals to zero for an optimal $C$, one sets the one-order lower term $\hat{c}_{n-1}(Q/Q)\hat{a}_Q^{n-1}$ as the uncertainty. Figure \ref{fig:Adler5loriginal} shows $\bar{D}^{\rm ns}_{5}(M_\tau^2,C)$ as a function of $C$ for $\mu=Q=M_\tau$ by using the approximate five-loop term $r_5\simeq 283$; its predicted optimal $C$ is $-0.783$, which leads to $|\hat{c}_5(M_\tau/M_\tau)\hat{a}_{M_\tau}^5|=0$ and $\bar{D}^{\rm ns}_5(M_\tau^2,C=-0.783)|_{\rm Conv.} = 0.1342 \pm 0.0063 \pm 0.0070$,
where the first error is for $\Delta\alpha_s^{\overline{\rm MS}}(M_Z)=\pm0.0011$ and the second error is equals to $\pm|\hat{c}_4(M_\tau/M_\tau)\hat{a}_{M_\tau}^4|$.

\subsection{Another way to a achieve scheme-and-scale independent predictions}
\label{PMSindependent}

It has been suggested in the literature that one can achieve the optimal scheme and scale of the pQCD approximate at any fixed order by directly requiring it to be independent of the ``unphysical" theoretical conventions such as the renormalization scheme and renormalization scale. This is the key idea of the PMS scale-setting approach~\cite{Stevenson:1980du, Stevenson:1981vj, Stevenson:1982wn, Stevenson:1982qw}. The PMS suggests that all the scheme-and-scale dependence of a fixed-order prediction can be treated as a negligible higher-order effect, and for the pQCD prediction, $\rho_n =\sum_{i=0}^{n} r_{i}(\mu) a_{\mu}^{i+1}$, we have
\begin{equation}
\partial \rho_n /\partial {\rm (RS)}={\cal O}(a_\mu^{n+2})\sim 0,
\end{equation}
where ${\rm RS}$ stands for the scheme or scale parameters~\footnote{Thus the accuracy of PMS prediction depends heavily on the perturbative convergence of the known pQCD series. It also explains why the NLO PMS predictions are generally unreliable, as is the case of the three-jet production fractions in $e^+e^-$ annihilation~\cite{Kramer:1987dd, Kramer:1990zt}.}. Equivalently, this indicates that the fixed-order approximant $\rho_n$ should satisfy the local RG invariance~\cite{Ma:2014oba, Wu:2014iba}
\begin{eqnarray}
\frac{\partial \rho_n}{\partial \tau} &=& 0, \label{eq.PMSscale}\\
\frac{\partial \rho_n}{\partial \beta_m} &=& 0, \;\; (2\leq m \leq n) \label{eq.PMSscheme}
\end{eqnarray}
where $\tau=\ln(\mu^2/\tilde\Lambda^2_{\rm QCD})$ with the asymptotic QCD scale $\tilde\Lambda_{\rm QCD}= \left({\beta_1}/ {\beta_0^2}\right)^{-\beta_1/2\beta_0^2} \Lambda_{\rm QCD}$.

The integration constants of those differential equations are scheme-and-scale independent RG invariants. For example, up to ${\rm N^3LO}$ level, there are three RG invariants
\begin{eqnarray}
\varrho_1 &=& \beta_0 \tau-{\cal C}_{1}, \label{eq.rho1} \\
\varrho_2 &=& {\cal C}_2-{\cal C}^2_1 -\frac{\beta _1 {\cal C}_1}{\beta _0}+\frac{\beta_2}{\beta _0} , \label{eq.rho2}\\
\varrho_3 &=& 2 {\cal C}_3+\frac{{\cal C}^2_1 \beta _1}{\beta _0}-2\frac{{\cal C}_1 \beta_2}{\beta _0}+\frac{\beta_3}{\beta_0} + 4 {\cal C}^3_1 -6 {\cal C}_1 {\cal C}_2, \label{eq.rho3}
\end{eqnarray}
where ${\cal C}_i=r_{i}/r_0$. Those RG invariants are helpful for transforming the pQCD approximant $\rho_n$ using the ${\cal R}$-scheme to the one using any other scheme (labeled as the ${\cal S}$-scheme)~\cite{Ma:2017xef}. More explicitly, this transformation can be achieved by applying the following two transformations simultaneously,
\begin{equation}
a^{\cal R}_s \to a^{\cal S}_s\;\; {\rm and}\;\; r_i^{\cal R} \to r_i^{\cal S}.
\end{equation}
The coupling constant $a^{\cal S}_s$ can be derived from $a^{\cal R}_s$ by using the extended RGEs, and the scheme-dependent $\beta^{\cal S}_{i\geq2}$-terms which determine $a^{\cal S}_s$ scale running behavior can be achieved by using the relation,
\begin{equation}
\beta^{\cal S}(a^{\cal S}_s) = \left({{\partial a^{\cal S}_s}}/{{\partial a^{\cal R}_s}}\right) {\beta^{\cal R}}({a^{\cal R}_s}).
\end{equation}
The perturbative coefficients $r_i^{\cal S}$ can be obtained from the coefficients $r_i^{\cal R}$ by using the renormalization group invariants $\varrho_i$, e.g. up to N$^3$LO level, we have
\begin{eqnarray}
r_1^{\cal S} &=&r_1^{\cal R}, \label{eq.coetrans1}\\
r_2^{\cal S} &=&r_2^{\cal R}+\frac{1}{\beta _0}(\beta _2^{\cal R}-\beta_2^{\cal S}), \label{eq.coetrans2}\\
r_3^{\cal S} &=&r_3^{\cal R}+\frac{2}{\beta _0} r_1^{\cal R}(\beta _2^{\cal R}-\beta_2^{\cal S}) +2 (\beta _3^{\cal R}-\beta_3^{\cal S}). \label{eq.coetrans3}
\end{eqnarray}
In combination with the known renormalization group invariants $\varrho_i$, the local RGEs (\ref{eq.PMSscale}, \ref{eq.PMSscheme}), and the solution of the RGE (\ref{eq:betafun}) up to the same order of the pQCD approximant, one can derive all of the required optimal parameters for PMS scale-setting. In practice, to apply the PMS to higher-orders, one can use the ``spiraling" numerical method~\cite{Mattingly:1992ud, Mattingly:1993ej}. This completes the description of the PMS calculation technology. The PMS predictions are independent to the initial choice of scale~\cite{Ma:2014oba}. In the following, we take $R_{e^+e^-}(Q)$ and $R_{\tau}(M_{\tau})$ to show that the scheme dependence of a pQCD approximant can also be eliminated by using the PMS.

\begin{figure}[htb]
\epsfysize=9.0cm
\begin{center}
\begin{minipage}[t]{13.5 cm}
\epsfig{file=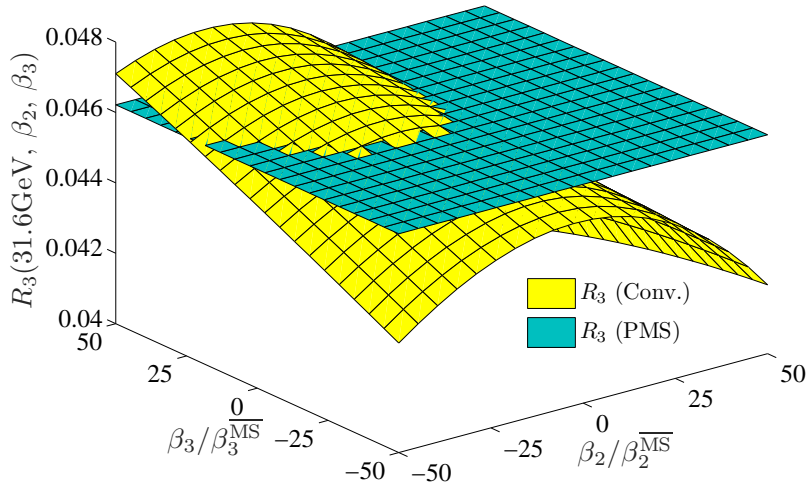,scale=0.6}
\end{minipage}
\begin{minipage}[t]{16.5 cm}
\caption{Comparison of the combined $\{\beta_{2},\beta_{3}\}$-dependence for the ${\rm N}^{3}$LO prediction on the ratio $R_{3}$ using the conventional (Conv.) scale setting and PMS~\cite{Ma:2017xef}, respectively. Q=31.6 {\rm GeV}. \label{fig:PMCs-scaley1}}
\end{minipage}
\end{center}
\end{figure}

\begin{figure}[htb]
\epsfysize=9.0cm
\begin{center}
\begin{minipage}[t]{13.5 cm}
\epsfig{file=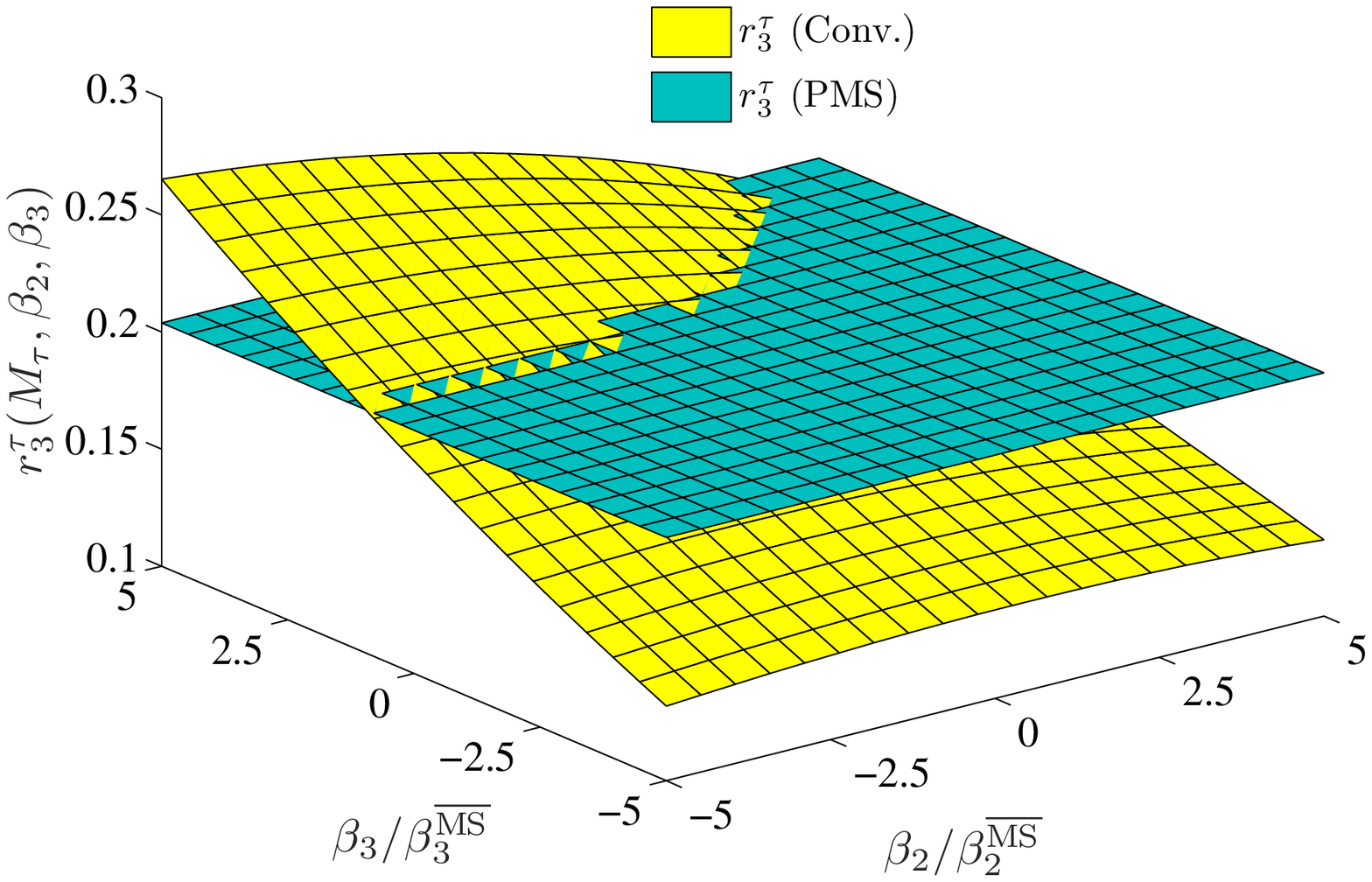,scale=0.6}
\end{minipage}
\begin{minipage}[t]{16.5 cm}
\caption{Comparison of the combined $\{\beta_{2},\beta_{3}\}$-dependence for the ${\rm N}^{3}$LO prediction on the ratio $r^{\tau}_3$ under the conventional (Conv.) scale setting and PMS~\cite{Ma:2017xef}, respectively. \label{fig:PMCs-scaley2}}
\end{minipage}
\end{center}
\end{figure}

The $R$-ratio is the characteristic parameter for the annihilation of an electron and positron into hadrons, which is defined as
\begin{eqnarray}
R_{e^+e^-}(Q)&=&\frac{\sigma\left(e^+e^-\rightarrow {\rm hadrons} \right)}{\sigma\left(e^+e^-\rightarrow \mu^+\mu^-\right)}= 3\sum_q e_q^2\left[1+R(Q)\right], \label{eq.Re+e-}
\end{eqnarray}
where $Q=\sqrt{S}$ stands for the $e^+e^-$ collision energy at which the $R$-ratio is measured. The pQCD approximant of $R(Q)$ up to N$^{n}$LO level under the $\overline{\rm MS}$-scheme reads
\begin{equation}
R_n(Q,\mu)=\sum_{i=0}^{n} {\cal C}^{\overline{\rm MS}}_{i}(Q,\mu) (a^{\overline{\rm MS}}_\mu)^{i+1},
\end{equation}
where $\mu$ stands for an arbitrary initial renormalization scale. If setting $\mu=Q$, the coefficients ${\cal C}^{\overline{\rm MS}}_{i}(Q,Q)$ up to fourth order can be obtain from Ref.\cite{Baikov:2012zn}. For any other choice of $\mu$, we will use the RGE to obtain the coefficients from ${\cal C}^{\overline{\rm MS}}_{i}(Q,Q)$.

The ratio $R_{\tau}(M_{\tau})$ for the $\tau$ decays into hadrons is defined as
\begin{eqnarray}
R_{\tau}(M_{\tau}) &=&\frac{\Gamma(\tau\rightarrow\nu_\tau+\rm{hadrons})} {\Gamma(\tau\rightarrow\nu_\tau+e^-\bar\nu_e)} = 3(|V_{ud}|^2+|V_{us}|^2) \left[1+r^{\tau}(M_{\tau})\right],
\end{eqnarray}
where the $\tau$-lepton mass $M_{\tau}=1.777$ GeV~\cite{Olive:2016xmw} and the Cabbibo-Kobayashi-Maskawa matrix elements $V_{ud}$ and $V_{us}$ satisfy the approximation, $3(|V_{ud}|^2+|V_{us}|^2)\approx 3$. The pQCD approximant of $r^{\tau}(M_{\tau})$ up to N$^{n}$LO level using the $\overline{\rm MS}$-scheme reads
\begin{equation}
r^{\tau}_n(M_{\tau},\mu)=\sum_{i=0}^{n} {\cal C}^{'\overline{\rm MS}}_{i}(M_{\tau},\mu) (a^{\overline{\rm MS}}_\mu)^{i+1}.
\end{equation}
The perturbative coefficients up to fourth order at any scale $\mu$ can be derived from the ones given in Ref.\cite{Baikov:2008jh}.

Different values of $\{\beta_{i}\}$-functions characterize different renormalization schemes. Figures \ref{fig:PMCs-scaley1} and \ref{fig:PMCs-scaley2} show the combined $\{\beta_{2},\beta_{3}\}$-dependence for the ${\rm N}^{3}$LO prediction $R_{3}$ and $r^{\tau}_3$. In these two figures, for the case of $R_3$, the $\beta_{2}$ and $\beta_{3}$ terms change simultaneously within the region of $[-50\beta_{2}^{\over {\rm MS}},+50\beta_{2}^{\over {\rm MS}}]$ and $[-50\beta_{3}^{\over {\rm MS}},+50\beta_{3}^{\over {\rm MS}}]$; and for the case of $r^{\tau}_3$, since the magnitude of the conventional scheme dependence is large, we adopt a smaller region, $\beta_m\in [-5\beta_{m}^{\over {\rm MS}},+5\beta_{m}^{\over {\rm MS}}]$. The flat planes confirm the scheme-independence of the PMS predictions over the changes of $\{\beta_2,\beta_3\}$. Thus by using the scheme equations (\ref{eq.PMSscheme}), one can not only achieve the most stable pQCD prediction around the optimal point (determined by the optimal scheme and the optimal scale), but also achieve the scheme-independent prediction for different choices of the initial renormalization scheme.

The PMS is a mathematical treatment with the purpose of finding the optimal renormalization scheme and renormalization scale for a pQCD fixed-order series. As shown above, by applying the PMS, one can achieve scheme-and-scale independent predictions with the help of renormalization group invariants such as those of Eqs.(\ref{eq.rho1}, \ref{eq.rho2}, \ref{eq.rho3}). And in certain cases when the pQCD series has good convergence and is known up to NNLO level or even higher orders, the PMS could be treated as a practical approach to soften the renormalization scheme and scale ambiguities for high-order pQCD predictions, especially for the global quantities such as total cross-section and total decay width~\cite{Ma:2014oba}. However, we should use PMS with care: In distinction to the PMC, which agrees with the standard RGI, the PMS treats the fixed-order pQCD prediction as the exact prediction of the physical observable, satisfying the local RGI~\cite{Wu:2014iba}; It breaks the standard RGI, and it does not satisfy the self-consistency conditions of the renormalization group, such as reflectivity, symmetry and transitivity~\cite{Brodsky:2012ms}.

\section{Extending the Predictive Power of Perturbative QCD}
\label{sec:4}

Due to the asymptotic freedom of the QCD theory, the QCD running coupling becomes numerically small at short distances, allowing perturbative calculations of physical observables at large momentum transfer. The complexity of the perturbative calculation greatly increases with the increment of the loop terms, and the pQCD prediction is only known to a fixed order. Thus the predictive power of pQCD depends on two important issues: how to eliminate the renormalization scheme-and-scale ambiguities at fixed order, and how to reliably estimate the contributions of unknown higher-order terms using information from the known pQCD series. At present, there is still no reliable way to estimate the unknown terms. The conventional treatment of guessing the ``typical momentum flow" as the renormalization scale not only introduces the renormalization scheme-and-scale ambiguities but also leads to a misleading pQCD prediction, especially if the conformal terms in the higher-order series are more important than the $\beta$-dependent terms. The error estimate obtained by varying the scale within an ``ad hoc" range can only obtain information from the $\beta$-dependent terms, but not from the conformal terms at higher-orders. One may hope to achieve a scheme-and-scale independent prediction by systematically computing higher-order enough QCD corrections; however, this hope is in direct conflict with the presence of the divergent $n! \alpha_s^n \beta_0^n$ renormalon series.

The Pad\'e approximation (PA) approach provides a practical procedure for promoting a finite Taylor series to an analytic function~\cite{Basdevant:1972fe, Samuel:1995jc, Samuel:1992qg}. In particular, the PA approach could be used to estimate the $(n+1)_{\rm th}$-order coefficient by incorporating all known coefficients up to order $n$. Some applications, together with alternatives to the original PA approach, have been discussed in the literature~\cite{Brodsky:1997vq, Gardi:1996iq, Ellis:1997sb, Burrows:1996dk, Ellis:1996zn, Jack:1997jn, Boito:2018rwt}. Due to the large cancelation among the coefficients at different orders, one may achieve some useful bounds on the unknown higher-order terms. In this section we will introduce a way of using the PA approach to achieve reliable predictions for the unknown higher-order terms for a pQCD series with the help of the renormalon-free conformal series determined by the PMC. The PA approach together with the usual pQCD series will also be given as a comparison.

\subsection{Pad\'e Resummation}

Generally, a perturbative series of a pQCD approximant $\rho_{n}$ up to $n_{\rm th}$-order level can be written as
\begin{equation}
\rho_{n} = \sum^{n}_{i=1} r_i a^{p+i-1},  \label{rhoest}
\end{equation}
where $a=\alpha_s/\pi$ and the index $p(\ge1)$ indicates the $\alpha_s$-order of the LO contribution. Assuming the idea of PA approach, the above perturbative series can be rewritten as a fractional $[N/M]$-type form
\begin{equation}
\rho^{[N/M]}_n = a^p \times \frac{b_0+b_1 a + \cdots + b_N a^N}{1 + c_1 a + \cdots + c_M a^M},  \label{PAAseries0}
\end{equation}
where $M\geq 1$ and $N+M+1 = n$. The coefficients $b_{j\in[0,N]}$ and $c_{k\in[1,M]}$ can be fixed by requiring the coefficients $C_{i\in[1,n]}$ defined in the following expansion series
\begin{equation}
\rho^{[N/M]}_n = \sum_{i=1}^{n} C_{i} a^{p+i-1} + C_{n+1}\; a^{p+n}+\cdots \label{PAAseries}
\end{equation}
to be the same as those of $r_{i\in[1,n]}$, e.g. $C_{i}=r_{i}$ for $i\leq n$. Then, if the coefficients $r_{i\in[1,n]}$ have been calculated, e.g. $\rho_n$ is known at the ${\rm N^{n}LO}$-order level, the fractional form (\ref{PAAseries0}) can be used to predict at least the next order higher term $C_{n+1}\; a^{p+n}$. Sometimes, the full PA expression (\ref{PAAseries0}) has also been adopted as an estimation of the whole perturbative series. The effectiveness of such all-orders PA prediction depends heavily on the pQCD convergence of the perturbative series, e.g. the precision of the known terms. So we usually use PA approach to predict only one higher order coefficient $C_{n+1}$. For example, if $[N/M]=[n-2/1]$, we have
\begin{equation}
C_{n+1}=\frac{C_n^2}{C_{n-1}}; \label{n-2/1}
\end{equation}
if $[N/M]=[n-3/2]$, we have
\begin{eqnarray}
C_{n+1}=\frac{-C_{n-1}^3+2C_{n-2}C_{n-1}C_{n}-C_{n-3}C_{n}^2}{C_{n-2}^2-C_{n-3}C_{n-1}}; \label{n-3/2}
\end{eqnarray}
if $[N/M]=[n-4/3]$, we have
\begin{eqnarray}
 C_{n+1} &=& \{C_{n-2}^4-(3 C_{n-3} C_{n-1}+2 C_{n-4} C_{n}) C_{n-2}^2 +2 [C_{n-4} C_{n-1}^2+(C_{n-3}^2+C_{n-5} C_{n-1}) C_{n}] C_{n-2} \nonumber \\
&& -C_{n-5} C_{n-1}^3+C_{n-3}^2 C_{n-1}^2+C_{n-4}^2 C_{n}^2 -C_{n-3} C_{n} (2 C_{n-4} C_{n-1}+C_{n-5} C_{n})\} \nonumber \\
&& / \{C_{n-3}^3-\left(2 C_{n-4} C_{n-2}+C_{n-5} C_{n-1}\right) C_{n-3} +C_{n-5} C_{n-2}^2+C_{n-4}^2 C_{n-1}\}; \label{n-4/3} {\rm etc.}
\end{eqnarray}
In each case, $C_{i<1}\equiv 0$. We need to know at least two $C_i$ in order to predict the unknown higher-order coefficients; thus the PA approach is applicable when we have calculated at least the NLO terms ($n=p+1$). In practice, the optimal $[N/M]$-type changes with the convergence and the precision of the perturbative series.

As shall be shown below, for the conventional pQCD series, which suffers from the renormalon divergence and whose coefficients are highly scale-dependent, its optimal type is diagonal; and for the PMC conformal series, which is free of renormalon divergence and whose conformal coefficients are generally scale-independent, its optimal choice is $[0/n-1]$. The PA approach is applicable to the PMC conformal series, and we shall show that it is applicable even for lower-order predictions.

\subsection{Estimating unknown high-order terms using conventional scale-dependent pQCD series}
\label{subsec:5.1}

It has been suggested that the diagonal Pad\'e approximation (dPA) is optimal for conventional pQCD series, e.g. $[N/N+1]$-type for $N\geq M$ or $[M-1/M]$-type for $N<M-1$, is optimal choice~\cite{Gardi:1996iq, Cvetic:1997qm}. In Ref.\cite{Gardi:1996iq} such a conclusion has been drawn by using one-loop RGE for $\alpha_s$-running, which has lately been improved by using the general RGE~\cite{Cvetic:1997qm}. In this subsection, we adopt the formalism given by Ref.\cite{Gardi:1996iq} to show the relation between the PA series and the conventional pQCD series, and then show why the dPA is preferable for the conventional pQCD series.

Ref.\cite{Gardi:1996iq} suggests a criterion for the optimal PA-type is that the PA transformation and the scale transformation should be commutative. The criterion states that by using a pQCD approximant $S(x)$ as a starting point, one may first do the PA transformation to $P(x)$, and then do the scale transformation to $P^*(y)$, or first do the scale transformation to $S(y)$ and then do the PA transformation to $P(y)$; then the preferable PA-type should ensure that $P^*(y)=P(y)$. Here $x=\alpha_s(\mu_1)/\pi$ and $y=\alpha_s(\mu_2)/\pi$. More explicitly, for a pQCD approximant $S(x)$ of given order $n+1$, we have
\begin{equation}\label{SDef}
S(x)=x(1+r_1 x+r_2 x^2+\cdots+r_n x^n).
\end{equation}
The scale-displacement relation (\ref{scaledis}) can be written as
\begin{equation}
x=y+\beta _0 \delta y^2+\left(\beta _0^2 \delta ^2+\beta _1 \delta \right)y^3 + \left(\beta _0^3 \delta ^3+\frac{5}{2} \beta _1 \beta _0 \delta ^2+\beta _2 \delta \right)y^4 +\cdots \label{scale},
\end{equation}
where $x=\alpha_s(\mu_1)/\pi$, $y=\alpha_s(\mu_2)/\pi$, and $\delta=\ln({\mu_2^2}/{\mu_1^2})$. Keeping only the $\beta_0$-series, it reduces to~\cite{Gardi:1996iq}
\begin{equation} \label{transformation}
x=y(1+\beta _0 \delta y+\beta _0^2 \delta ^2y^2 +\beta _0^3 \delta ^3y^3 +\cdots)=\frac{y}{1-\beta _0 \delta y},
\end{equation}
where the second equation is the result for an all-orders summation \footnote{It should be noted that the treatment of neglecting all $\beta_{i\ge1}$-terms is different from the large $\beta_0$ approximation, in which the $\beta_i$-terms are kept by assuming $\beta_i \simeq \beta_0^{i+1}$~\cite{Beneke:1994qe, Neubert:1994vb}. In this case, the conclusion that the dPA is preferable for conventional pQCD series is still unchanged~\cite{Cvetic:1997qm}.}.

Substituting the scale transformation (\ref{transformation}) into Eq.(\ref{SDef}), we can transform $S(x)$ to $S(y)$, i.e.
\begin{equation}
S(y)=y(1+(r_1+\beta_0\delta) y+(r_2+\beta_0^2\delta^2+2\beta_0\delta r_1) y^2 + (r_3+\beta_0^3\delta^3 +3\beta_0^2 \delta^2 r_1 +3\beta_0\delta r_2) y^3) +\cdots
\end{equation}

The $[N/M]$-type PA $P(x)$ for $S(x)$ is
\begin{equation}
P(x)=x\frac{1+a_1 x+\cdots+a_N x^N}{1+b_1 x+\cdots+b_M x^M},
\end{equation}
where the coefficients $a_i$ and $b_i$ can be determined following the same matching method described in Sec.5.1. Following the same procedure, we can obtain the $[N/M]$-type PA $P(y)$ for $S(y)$. The $[N/M]$-type PA $P^*(y)$ can be achieved by applying the scale transformation (\ref{transformation}) for $P(x)$.

It is found that the diagonal type PA, such as $[0/1]$,$[1/2]$,$[2/3]$,$[3/4]$ and etc., leads to $P^*(y)=P(y)$. For example, for a perturbative series with $n=3$, we have
\begin{eqnarray}
S(x)=x(1+r_1x+r_2x^2+r_3x^3).
\end{eqnarray}
The diagonal $[1/2]$-type PA indicates
\begin{eqnarray}
P(x)=x\left[\frac{r_1^2-r_2+(r_1^3-2 r_2 r_1+r_3)x}{r_1^2-r_2+(r_3-r_1r_2 )x+(r_2^2-r_1r_3)x^2}\right].
\end{eqnarray}
Applying the scale transformation $x=\frac{y}{1-\beta _0 \delta y}$ to $P(x)$, we obtain
\begin{eqnarray}
P^*(y)=y\left[\frac{r_1^2-r_2+(r_1^3-2 r_2 r_1+r_3-\beta _0 r_1^2 \delta+\beta _0 r_2 \delta)y}{r_1^2-r_2+(r_3-r_1 r_2-2 \beta _0 r_1^2 \delta+2 \beta _0 r_2 \delta)y+ z_1}\right],
\end{eqnarray}
where $z_1=(r_2^2-r_1 r_3-\beta _0^2 r_2 \delta^2+\beta _0^2 r_1^2 \delta^2+\beta _0 r_1 r_2 \delta-\beta _0 r_3 \delta)y^2$. Using $S(y)$ up to $y^4$-order level, we can obtain the diagonal $[1/2]$-type PA $P(y)$, i.e.
\begin{eqnarray}
P(y)=y\left[\frac{r_1^2-r_2+(r_1^3-2 r_2 r_1+r_3-\beta _0 r_1^2 \delta+\beta _0 r_2 \delta)y}{r_1^2-r_2+(r_3-r_1 r_2-2 \beta _0 r_1^2 \delta+2 \beta _0 r_2 \delta)y+z_1}\right].
\end{eqnarray}
Comparing with those two equations, we obtain $P^*(y)=P(y)$.

For the allowable non-diagonal PA types, $[2/1]$ and $[0/3]$, we have $P^*(y)\neq P(y)$. More explicitly, the $P(y)$ and $P^*(y)$ for the $[2/1]$-type PA are
\begin{eqnarray}
P(y)&=& y\left[\frac{r_2+2 \beta _0 \delta r_1+\beta _0^2 \delta ^2+(r_1 r_2-r_3+2 \beta _0 \delta r_1^2-2 \beta _0 \delta r_2)y+z_2}{r_2+2 \beta _0 \delta r_1+\beta _0^2 \delta ^2-(r_3+3 \beta _0 \delta r_2+3 \beta _0^2 \delta ^2 r_1+\beta _0^3 \delta ^3)y} \right], \\
P^*(y)&=& y\left[\frac{r_2+(r_1 r_2-r_3-2 \beta _0 \delta r_2)y+(r_2^2-r_1 r_3-\beta _0 \delta r_1 r_2+\beta _0 \delta r_3+\beta _0^2 \delta ^2 r_2)y^2}{r_2-(3 \beta _0 \delta r_2+r_3)y+(2 \beta _0 \delta r_3+3 \beta _0^2 \delta ^2 r_2)y^2-(\beta _0^2 \delta ^2 r_3+\beta _0^3 \delta ^3 r_2)y^3}\right],
\end{eqnarray}
where $z_2=(r_2^2-r_1 r_3+\beta _0 \delta r_1 r_2-\beta _0 \delta r_3+\beta _0^2 \delta ^2 r_1^2-\beta _0^2 \delta ^2 r_2)y^2$. The $P(y)$ and $P^*(y)$ for the $[0/3]$-type PA are
\begin{eqnarray}
P(y)&=& y\left[\frac{1}{1-(r_1+\beta _0 \delta)y+(r_1^2-r_2)y^2+(2 r_1 r_2-r_1^3-r_3+\beta _0 \delta r_1^2-\beta _0 \delta r_2)y^3} \right], \\
P^*(y)&=& y\left[\frac{1-2 \beta _0 \delta y+\beta _0^2 \delta ^2 y^2}{1-(r_1+3 \beta _0 \delta)y+(r_1^2-r_2+2 \beta _0 \delta r_1+3 \beta _0^2 \delta ^2)y^2+ z_3}\right],
\end{eqnarray}
where $z_3=(2 r_1 r_2-r_1^3-r_3-\beta _0 \delta r_1^2+\beta _0 \delta r_2-\beta _0^2 \delta ^2 r_1-\beta _0^3 \delta ^3)y^3$.

Because of the above transformation invariance, $P(y)=P^*(y)$, it is suggested that the diagonal PA type is the preferable one for conventional pQCD series for estimating the unknown high-order contributions. As a byproduct, it has been found that such diagonal PA series leads to scale-independent predictions~\cite{Gardi:1996iq}. This is reasonable, since it only resums one type of diagrams which involve the bubble-dressed gluon propagator~\cite{Brodsky:1997vq}, and thus is consistent with the BLM scale-setting approach.

As suggested in Ref.~\cite{Brodsky:1997vq}, the higher-order pQCD diagrams can be studied by first decomposing them in a skeleton expansion, in which each term contains different chains of vacuum polarization bubbles inserted in virtual-gluon propagators. The QCD running coupling is singular in the large-$\beta_0$ limit, and the integration over the gluon momentum yields the general renormalon singularities. By using the large-$\beta_0$ approximation, the leading-order BLM procedure equals to $[0/1]$-type PA series~\cite{Gardi:1996iq}. Recently, using Pad\'e approximant and its variant, the authors suggest a way to provide a large number of perturbative coefficients which are uncalculated so far~\cite{Boito:2018rwt}. However these predictions are based on the conventional divergent renormalon series, which is scale-dependent and scheme-dependent.

The precision of the PA approach depends heavily on the accuracy of the known perturbative coefficients. If due to large cancelations among the coefficients at different orders, the scale-dependence of the PA series could be greatly suppressed, one may then achieve some useful bounds on the unknown higher-order terms. However the scale uncertainty for each perturbative coefficient cannot be eliminated by computing higher order terms; the cancelation of the scale dependence for a higher-order pQCD approximant can be accidental. In fact, if there is large scale uncertainty for the pQCD approximant, it is unclear whether it is the nature of the pQCD series or it is due to wrong choice of scale. Basing on the largely uncertain (scale-dependent) coefficients, one also cannot draw definite conclusion whether the predicted unknown terms are reliable or not. Moreover, the large scale uncertainty for each coefficient may lead to a serious mathematical problem for the PA fractional expression; e.g., its denominator tends to zero for certain choice of scale, leading to unreasonable large prediction. Thus even though the PA approach provides a practical way to estimate unknown high-order terms, its defaults constraint the applicability of PA approach itself.

\subsection{Estimating unknown high-order terms using the renormalon-free conformal pQCD series} \label{subsec:5.2}

In this subsection, we extend the applicability of the PA approach by using the scale-independent conformal series; this will provide a more reliable prediction for the magnitude of uncalculated terms.

QCD theory is non-conformal due to the emergence of the RG-involved $\beta$-terms in perturbative series. Eq.(\ref{eq:deltaRGI}) and Eq.(\ref{eq:Cdependence}) shows that when all of $\beta$-terms are eliminated by applying the PMC; the resultant conformal pQCD series will be scale-independent and scheme-independent. Furthermore, because the divergent renormalon series does not appear in the PMC conformal series, one can thus improve the PA resummation procedures to predict higher-order terms and increase the precision and reliability of pQCD predictions~\cite{Du:2018dma}.

As is the case of Eq.(\ref{eq:rhodelta2}), we rewrite the perturbative series (\ref{rhoest}) of the pQCD approximant $\rho_{n}$ as
\begin{eqnarray}
\rho_{n}(Q)|_{\rm Conv.}
&=& r_{1,0}{a^p_\mu} + \left[ r_{2,0} + p \beta_0 r_{2,1} \right]{a^{p+1}_\mu} + \big[ r_{3,0} + p \beta_1 r_{2,1} + (p+1){\beta _0}r_{3,1} +\nonumber\\
&& \frac{p(p+1)}{2} \beta_0^2 r_{3,2} \big]{a^{p+2}_\mu} + \big[ r_{4,0} + p{\beta_2}{r_{2,1}}  + (p+1){\beta_1}{r_{3,1}} + \frac{p(3+2p)}{2}{\beta_1}{\beta_0}{r_{3,2}} \nonumber\\
&& + (p+2){\beta_0}{r_{4,1}}+ \frac{(p+1)(p+2)}{2}\beta_0^2{r_{4,2}} + \frac{p(p+1)(p+2)}{3!}\beta_0^3{r_{4,3}} \big]{a^{p+3}_\mu} + \cdots.
\label{degeneracyrho}
\end{eqnarray}
Following the standard PMC-s procedure, we obtain
\begin{eqnarray}
\rho_n(Q)|_{\rm PMC}=\sum_{i=1}^n r_{i,0}a^{p+i-1}_{Q_{*}},
\label{pmcs}
\end{eqnarray}
where $Q_{*}$ is the determined single PMC scale, whose analytical form is similar to Eqs.(\ref{eq:pmcscale2}, \ref{singles1}, \ref{singles2}, \ref{singles3}). In the following, we shall apply the PMC conformal series (\ref{pmcs}) to three physical observables $\ree$, $\rtau$ and $\hbb$ which are known up to four-loop level and show how the ``unknown" terms predicted by the PA approach varies when one inputs more-and-more known higher-order terms.

The ratio $\ree$ is defined as
\begin{equation}
R_{e^+ e^-}(Q) = \frac{\sigma\left(e^+e^-\to {\rm hadrons} \right)}{\sigma\left(e^+e^-\to \mu^+ \mu^-\right)} = 3\sum_q e_q^2\left[1+R(Q)\right], \label{ree}
\end{equation}
where $Q=\sqrt{s}$ is the $e^+e^-$ collision energy, and we take $Q=31.6 \;{\rm GeV}$~\cite{Marshall:1988ri} as an example where it is well measured. The pQCD approximant of $R(Q)$ is, $R_n(Q)= \sum_{i=1}^{n} r_i(\mu/Q) a^{i}_\mu$. The perturbative coefficients $r_i$ at $\mu=Q$ have been calculated using the $\overline{\rm MS}$-scheme, whose analytical form can be found in Refs.~\cite{Baikov:2008jh, Baikov:2010je, Baikov:2012zn, Baikov:2012zm}.

The ratio $\rtau$ is defined as
\begin{equation}
R_{\tau}(M_{\tau})=\frac{\sigma(\tau\rightarrow\nu_{\tau}+\rm{hadrons)}}{\sigma(\tau\rightarrow\nu_{\tau}+\bar{\nu}_e+e^-)} = 3\sum\left|V_{ff'}\right|^2\left(1+\tilde{R}(M_{\tau})\right),
\end{equation}
where $V_{ff'}$ are the Cabbibo-Kobayashi-Maskawa matrix elements, $\sum\left|V_{ff'}\right|^2 =\left(\left|V_{ud}\right|^2+\left|V_{us}\right|^2\right)\approx 1$ and $M_{\tau}= 1.777$ GeV. The pQCD approximant of $\tilde{R}(M_{\tau})$ is, $\tilde{R}_{n}(M_{\tau})= \sum_{i=1}^{n}r_i(\mu_r/M_{\tau})a^{i}_\mu$; the coefficients can be obtained by using the relation between $R_{\tau}(M_{\tau})$ and $R(\sqrt{s})$~\cite{Lam:1977cu}.

The decay width $\hbb$ is defined as
\begin{equation}
\Gamma(H\to b\bar{b})=\frac{3G_{F} M_{H} m_{b}^{2}(M_{H})} {4\sqrt{2}\pi} [1+\hat{R}(M_{H})],  \label{hbb}
\end{equation}
where the Fermi constant $G_{F}=1.16638\times10^{-5}\;\rm{GeV}^{-2}$, the Higgs mass $M_H=126$ GeV, and the $b$-quark $\overline{\rm{MS}}$-running mass is $m_b(M_H)=2.78$ GeV~\cite{Wang:2013bla}. The pQCD approximant of $\hat{R}(M_{H})$, $\hat{R}_n(M_H)= \sum_{i=1}^{n}r_i(\mu_r/M_{H}) a^{i}_\mu$, where the coefficients at $\mu=M_H$ can be found in Ref.\cite{Baikov:2005rw}.

The perturbative coefficients for each pQCD approximant at any other renormalization scales can be obtained via QCD evolution. In doing the numerical evaluation, we have assumed the running of $\alpha_s$ at the four-loop level. The asymptotic QCD scale is set by using the conventional fixed-point, $\alpha_s(M_z)=0.1181$~\cite{Olive:2016xmw}, which gives $\Lambda_{\rm{QCD}}^{n_f=5}=0.210$ GeV.

The optimal scale $Q_{*}$ for each process can be determined by applying the PMC-s approach, whose perturbative series up to ${\rm N^2LL}$-level are
\begin{eqnarray}
\ln\left.\frac{Q^2_{*}}{Q^2}\right|_{e^+e^-} &=& +0.22+0.23+0.03+{\cal O}(\alpha_s^3), \\
\ln\left.\frac{Q^2_{*}}{M_\tau^2}\right|_{\tau} &=& -1.36+0.23+0.08+{\cal O}(\alpha_s^3), \\
\ln\left.\frac{Q^2_{*}}{M_H^2}\right|_{H\to b\bar{b}} &=& -1.44-0.13+0.05+{\cal O}(\alpha_s^3).
\end{eqnarray}
If the pQCD approximants are known up to two-loop, three-loop, and four-loop level, the optimal scales are $Q_{*}|_{e^+e^-}=[35.36$, $39.68$, $40.30]$ GeV, $Q_{*}|_{\tau}=[0.90$, $1.01$, $1.05]$ GeV and $Q_{*}|_{H\to b\bar{b}}=[61.38$, $57.41$, $58.84]$ GeV, accordingly. These PMC scales $Q^*$ are independent of the initial choice of the renormalization scale $\mu$. The scale $Q_{*}|_{\tau}$ is not much larger than the asymptotic scale $\Lambda_{\rm QCD}$. Numerically, as shown by Figure \ref{fig:alphas4loop}, the usually adopted analytic $\alpha_s$-running differ significantly at scales below a few GeV from the exact solution of RGE at or below the four-loop level, we will use the exact numerical solution of the RGE throughout to evaluate $\rtau$.

\begin{table}[htb]
\centering
\begin{tabular}{ |c| c| c| c |}
\hline
~$r_{n+1,0}$ ~&~~~$n+1=3$~~~ & ~~~$n+1=4$~~~ & ~~~$n+1=5$~~~\\
 \hline
~EC~ & $-1.0 $ & $-11.0 $ & - \\
\hline
~\multirow{3}{*}{PAA}~ & [0/1]$+3.4$ & [0/2]$-9.9$ &[0/3]$-17.8$ \\
\cline{2-4}
- & - &[1/1]$+0.55$ & [1/2]$-18.0$ \\
\cline{2-4}
 & - & - & [2/1]$-120.$ \\
\hline
\end{tabular}
\caption{Comparison of the exact (``EC") $(n+1)_{\rm th}$-order conformal coefficients of $R_{n+1}$ with the predicted (``$[N/M]$-type PA") $(n+1)_{\rm th}$-order ones based on the known $n_{\rm th}$-order approximate $R_{n}$~\cite{Du:2018dma}. $Q=31.6$ GeV. }
\label{estimate-ree}
\end{table}

\begin{table}[htb]
\centering
\begin{tabular}{ |c| c| c| c |}
\hline
~$r_{n+1,0}$ ~&~~~$n+1=3$~~~ & ~~~$n+1=4$~~~ & ~~~$n+1=5$~~~\\
\hline
~EC~ & $+3.4$ & $+6.8$ & - \\
\hline
~\multirow{3}{*}{PAA}~ & [0/1]$+4.6$ & [0/2]$+4.9$ &[0/3]$+14.7$ \\
\cline{2-4}
  &- &[1/1]$+5.5$ & [1/2]$+11.5$ \\
\cline{2-4}
& - & - &[2/1]$+13.5$ \\
\hline
\end{tabular}
\caption{Comparison of the exact (``EC") $(n+1)_{\rm th}$-order conformal coefficients of $\tilde{R}_{n+1}(M_{\tau})$ with the predicted (``$[N/M]$-type PA") $(n+1)_{\rm th}$-order ones based on the known $n_{\rm th}$-order approximate $\tilde{R}_{n}(M_{\tau})$~\cite{Du:2018dma}, respectively.}
\label{estimate-rtau}
\end{table}

\begin{table}[htb]
\centering
\begin{tabular}{|c| c| c| c |}
\hline
~$r_{n+1,0}$ ~&~~~$n+1=3$~~~ & ~~~$n+1=4$~~~ & ~~~$n+1=5$~~~\\
\hline
~EC~ & $-1.36\times10^2$ & $-4.32\times10^2$ & -  \\
\hline
~\multirow{3}{*}{PAA}~ & [0/1]$+3.23\times10^1$ & [0/2]$-7.26\times10^2$ &[0/3]$+3.72\times10^3$ \\
\cline{2-4}
& - & [1/1]$+1.37\times10^3$ &[1/2]$+3.20\times10^3$ \\
\cline{2-4}
& - & - &[2/1]$-1.37\times10^3$ \\
\hline
\end{tabular}
\caption{Comparison of the exact (``EC") $(n+1)_{\rm th}$-order conformal coefficients of $\hat{R}_{n+1}(M_H)$ with the predicted (``$[N/M]$-type PA") $(n+1)_{\rm th}$-order ones based on the known $n_{\rm th}$-order approximate $\hat{R}_{n}(M_H)$~\cite{Du:2018dma}, respectively. }
\label{estimate-hbb}
\end{table}

As an explanation of how different PA series affects the predicted values and how the predicted conformal coefficients change with increasing perturbative orders, we present a comparison of the exact $(n+1)_{\rm th}$-order conformal coefficients with the PA approach predicted ones based on the known $n_{\rm th}$-order approximates $R_{n}(Q=31.6~ {\rm GeV})$, $\tilde{R}_{n}(M_{\tau})$ and $\hat{R}_{n}(M_H)$ in Tables~\ref{estimate-ree}, ~\ref{estimate-rtau} and~\ref{estimate-hbb}, respectively. Here the $[N/M]$-type PA series is for $N+M=n-1$ with $N\geq0$ and $M\geq1$. In some special cases, the diagonal-type PA series seemingly behaves better than that of the $[0/n-1]$-type due to accidentally larger cancellation among the coefficients at different orders, e.g. the diagonal $[1/1]$-type PA works better than the $[0/2]$-type one for $\tilde{R}_{4}(M_{\tau})$, whose normalized differences are $19\%$ and $28\%$, respectively. In general cases, these tables show that the $[N/M]=[0/n-1]$-type PA series, which corresponds to a geometric series, provides predictions closest to the known pQCD results.

At present, there is no strong theoretical support why $[0/n-1]$-type PA series is most preferable. However, it is interesting to note that the $[0/n-1]$-type PA series is consistent with the ``Generalized Crewther Relations" (GSICRs)~\cite{Shen:2016dnq}. The GSICR, which provides a remarkable all-orders connection between the pQCD predictions for deep inelastic neutrino-nucleon scattering and hadronic $e^+e^-$ annihilation, shows that the conformal coefficients are all equal to $1$; e.g. $\widehat{\alpha}_d(Q)=\sum_{i}\widehat{\alpha}^{i}_{g_1}(Q_*)$ or equivalently, $(1+\widehat{\alpha}_d(Q))(1-\widehat{\alpha}_{g_1}(Q_*))=1$, where $Q_*$ satisfies
\begin{equation}
\ln\left.\frac{Q_*^2}{Q^2}\right|_{g_1} = 1.308 + [-0.802 + 0.039 n_f] \widehat{\alpha}_{g_1}(Q_*) + [16.100 - 2.584 n_f + 0.102 n_f^2] \widehat{\alpha}_{g_1}^2(Q_*) + \cdots.
\end{equation}
By using the $[0/n-1]$-type PA series -- the geometric series -- all of the predicted conformal coefficients are also equal to $1$.

The $[0/n-1]$-type PA series also agrees with the GM-L scale-setting procedure to obtain scale-independent perturbative QED predictions; e.g., the renormalization scale for the electron-muon elastic scattering through one-photon exchange is set as the virtuality of the exchanged photon, $\mu_r^2 = q^2 = t$. By taking an arbitrary initial renormalization scale $t_0$, we have
\begin{equation}
\alpha_{em}(t) = \frac{\alpha_{em}(t_0)}{1 - \Pi(t,t_0)} \;\;,
\end{equation}
where $\Pi(t,t_0) = \frac{\Pi(t,0) -\Pi(t_0,0)}{1-\Pi(t_0,0)}$, which sums all vacuum polarization contributions, both proper and improper, to the dressed photon propagator. The PMC reduces in the $N_C \to 0$ Abelian limit to the GM-L method~\cite{Brodsky:1997jk}, and the preferable $[0/n-1]$-type makes the PA geometric series self-consistent with the GM-L/PMC prediction.

We also find that the $[0/n-1]$-type PA series works well for the conformal series of the ${\cal N}=4$ supersymmetric Yang-Mills theory. As an example, we present a PA prediction on the $\rm{N^2LO}$ and $\rm{N^3LO}$ Balitsky-Fadin-Kuraev-Lipatov (BFKL) Pomeron eigenvalues. By using the PA approach together with the known LO and NLO coefficients given in Ref.~\cite{Costa:2012cb}, we find that the $\rm{N^2LO}$ BFKL coefficient is $0.86\times10^4$ for $\Delta=0.45$, where $\Delta$ is the full conformal dimension of the twisted-two operator. On the other hand, the $\rm{N^2LO}$ BFKL coefficient has been calculated in planar ${\cal N}=4$ supersymmetric Yang-Mills theory~\cite{Gromov:2015vua} by using the quantum spectral curve integrability-based approach~\cite{Gromov:2013pga, Gromov:2014caa}, which gives $1.08\times10^4$~\cite{Gromov:2015vua}. Thus the normalized difference between those two $\rm{N^2LO}$ values is only about $20\%$. Moreover, the $\rm{N^3LO}$ coefficient by using the $[0/2]$-PA type is $-3.07\times10^5$. Within the framework of ${\cal N}=4$ supersymmetric Yang-Mills, the exact value of $\rm{N^3LO}$ coefficient is uncalculated at present, which however can be approximated by using a data-fitting process suggested in Ref.~\cite{Gromov:2015wca}, giving $-3.66\times10^5$. The normalized difference between those two $\rm{N^3LO}$ values is also only about $20\%$.

Tables~\ref{estimate-ree}, ~\ref{estimate-rtau} and~\ref{estimate-hbb} show that as more loop terms are introduced, the predicted conformal coefficients become closer to their exact value. To show this clearly, we define the normalized difference between the exact conformal coefficient and the predicted one as
\begin{equation}
\Delta_{n}=\left|\frac{r_{n,0}|_{\rm PAA}-r_{n,0}|_{\rm EC}}{r_{n,0}|_{\rm EC}}\right| ,
\end{equation}
where ``EC" and ``PAA" stand for exact and predicted conformal coefficients, respectively. By using the exact terms, known up to two-loop and three-loop levels accordingly, the normalized differences for the $3_{\rm th}$-order and the $4_{\rm th}$-order conformal coefficients become suppressed from $440\%$ to $10\%$ for $R(Q=31.6~{\rm GeV})$, from $35\%$ to $28\%$ for $\tilde{R}(M_{\tau})$, and from $124\%$ to $68\%$ for $\hat{R}(M_H)$. There are large differences for the conformal coefficients if we only know the QCD corrections at the two-loop level; however this decreases rapidly when we know more loop terms. Following this trend, the normalized differences for the $5_{\rm th}$-order conformal coefficients should be much smaller than the $4_{\rm th}$-order ones. Conservatively, if we set the normalized difference ($\Delta_5$) of the $5_{\rm th}$-loop as the same one of the $4_{\rm}$-loop ($\Delta_4$), we can inversely predict the ``exact value" of the uncalculated $5_{\rm th}$-loop conformal coefficients (labelled as ``${\rm EC'}$"):
\begin{eqnarray}
r^{e^+ e^-}_{5,0}|_{\rm EC'} &=& -18.0\pm1.8, \\
r^{\tau}_{5,0}|_{\rm EC'} &=& 16.0\pm 4.5, \\
r^{H\to b\bar{b}}_{5,0}|_{\rm EC'} &=& (6.92\pm4.71)\times10^3,
\end{eqnarray}
where the central values are obtained by averaging the two ``${\rm EC'}$" values determined by $\frac{r_{5,0}|_{\rm PAA}}{(1\pm\Delta_4)}$.

The difference between the exact and predicted conformal coefficients is reduced by the power of $\alpha_s/\pi$. Due to the fast pQCD convergence of conformal series, a precise prediction of uncalculated contributions to the pQCD approximant could be achieved even at lower orders. The precision of the predictive power of the PA series should become most useful for physical observables, such as the cross-section and the decay width, or the measurable ratios constructed from those observables.

\begin{table}[htb]
\centering
\resizebox{\textwidth}{15mm}
{
\begin{tabular}{ccccccc}
\hline
 &~~${\rm EC}$, $n=2$~~&~~${\rm PAA}$, $n=3$~~&~~${\rm EC}$, $n=3$~~&~~${\rm PAA}$, $n=4$~~&~~ ${\rm EC}$, $n=4$~~&~~${\rm PAA}$, $n=5$~~ \\
 \hline
$R_n(Q)|_{\rm PMC-s}$ & 0.04745 & 0.04772(0.04777) & 0.04635 & 0.04631(0.04631) & 0.04619 & 0.04619(0.04619) \\
$\tilde{R}_{n}(M_{\tau})|_{\rm PMC-s}$ & 0.1879 & 0.2035(0.2394) & 0.2103 & 0.2128(0.2134) & 0.2089  & 0.2100(0.2104)  \\
$\hat{R}_{n}(M_H)|_{\rm PMC-s}$ & 0.2482 & 0.2503(0.2505) & 0.2422 &  0.2402(0.2406) & 0.2401  & 0.2405(0.2405)  \\
$R_n(Q)|_{\rm Conv.}$ & $0.04763^{+0.00045}_{-0.00139}$ & $0.04781^{+0.00043}_{-0.00053}$ & $0.04648_{-0.00071}^{+0.00012}$ & $0.04632_{-0.00025}^{+0.00018}$ &$0.04617_{-0.00009}^{+0.00015}$ &$0.04617_{-0.00001}^{+0.00007}$ \\
$\tilde{R}_{n}(M_{\tau})|_{\rm Conv.}$ & $0.1527^{+0.0610}_{-0.0323}$ & $0.1800^{+0.0515}_{-0.0330}$ & $0.1832_{-0.0334}^{+0.0385}$ & $0.1975_{-0.0296}^{+0.0140}$ & $0.1988_{-0.0299}^{+0.0140}$ & $0.2056_{-0.0247}^{+0.0029}$ \\
$\hat{R}_{n}(M_H)|_{\rm Conv.}$  & $0.2406^{+0.0074}_{-0.0104}$ & $0.2475^{+0.0027}_{-0.0066}$ & $0.2425_{-0.0053}^{+0.0002}$ & $0.2419_{-0.0040}^{+0.0002}$ & $0.2411_{-0.0040}^{+0.0001}$ & $0.2407_{-0.0040}^{+0.0002}$ \\
\hline
\end{tabular}
}
\caption{Comparison of the exact (``EC") and the predicted (``PAA") pQCD approximants $R_n(Q=31.6\;{\rm GeV})$, $\tilde{R}_{n}(M_\tau)$ and $\hat{R}_n(M_H)$ using conventional (Conv.) and PMC-s approaches up to $n_{\rm th}$-order level~\cite{Du:2018dma}. The $n_{\rm th}$-order PA prediction equals the $(n-1)_{\rm th}$-order known prediction plus the predicted $n_{\rm th}$-order terms using the [0/n-2]-type PA prediction. The PMC predictions are scale independent and the errors for conventional scale-setting are estimated by varying $\mu$ within the region of $[1/2\mu, 2\mu]$, where $\mu=Q$, $M_\tau$ and $M_H$, respectively. }
\label{Total-observable}
\end{table}

We present the comparison of the exact results for the pQCD approximants $R_{n}(Q=31.6~{\rm GeV})$, $\tilde{R}_{n}(M_{\tau})$ and $\hat{R}_{n}(M_H)$ with the [0/n-1]-type PA series predicted ones in Table~\ref{Total-observable}. The values in the parentheses are results for the corresponding full PA series, which are calculated by using Eq.(\ref{PAAseries0}). Due to the fast pQCD convergence, the differences between the truncated and full PA predictions are small, which are less than $1\%$ for $n\ge 4$ \footnote{However, we do not suggest to directly use the ratio, Eq.(\ref{PAAseries0}), for obtaining the PA prediction, especially for lower-order predictions, since it only partly resums the known-type terms from lower-orders, and the higher-order terms whose contributions, though $\alpha_s/\pi$-power suppressed, are uncontrollable are unknown.}. Similarly, we define the precision of the predictive power as the normalized difference between the exact approximant ($\rho_n$) and the prediction; i.e.
\begin{equation}
\Delta_{\rho_n}=\left|\frac{\rho_{n}|_{\rm PAA}- \rho_{n}|_{\rm EC}}{\rho_{n}|_{\rm EC}}\right|.
\end{equation}

The PMC predictions are renormalization scheme-and-scale independent, and the pQCD convergence is greatly improved due to the elimination of renormalon contributions. Highly precise values at each order can thus be achieved~\cite{Shen:2017pdu}. In contrast, predictions using conventional pQCD series are scale dependent even for higher-order predictions. We also present results using conventional scale-setting in Table~\ref{Total-observable}; it confirms the conclusion that the conformal PMC-s series is much more suitable for application of the PA approach.

By using the known (exact) approximants predicted by PMC-s scale-setting up to two-loop and three-loop levels accordingly, the differences between the exact and predicted three-loop and four-loop approximants are observed to decrease from $3.0\%$ to $0.3\%$ for $\rho_n=R_n(Q=31.6~{\rm GeV})$, from $3\%$ to $2\%$ for $\rho_n=\tilde{R}_n(M_{\tau})$, and from $3.0\%$ to $<0.1\%$ for $\rho_n=\hat{R}_n(M_H)$, respectively. The normalized differences for $R_4(Q=31.6~{\rm GeV})$, $\tilde{R}_4(M_{\tau})$ and $\hat{R}_4(M_H)$ are small, e.g. already smaller than the normalized difference for $\alpha_s/\pi$, thus the PA prediction could be a good prediction of exact value at the fourth order. If we conservatively set the normalized difference of the $5_{\rm th}$-loop to match that of the $4_{\rm}$-loop predictions, then the predicted $5_{\rm th}$-loop ``${\rm EC'}$" predictions are
\begin{eqnarray}
R_5(Q=31.6~{\rm GeV})|_{\rm EC'} &=& 0.04619\pm0.00014, \label{five-ee}\\
\tilde{R}_5(M_{\tau})|_{\rm EC'} &=& 0.2100\pm0.0042, \label{five-tau} \\
\hat{R}_5(M_H)|_{\rm EC'} &=& 0.2405\pm0.0001, \label{five-MH}
\end{eqnarray}
where the central values are obtained by averaging the two ``${\rm EC'}$" values determined by $\frac{\rho_{5}|_{\rm PAA}}{(1\pm\Delta_{\rho,4})}$.

\begin{figure}[htb]
\epsfysize=9.0cm
\begin{center}
\begin{minipage}[t]{10 cm}
\epsfig{file=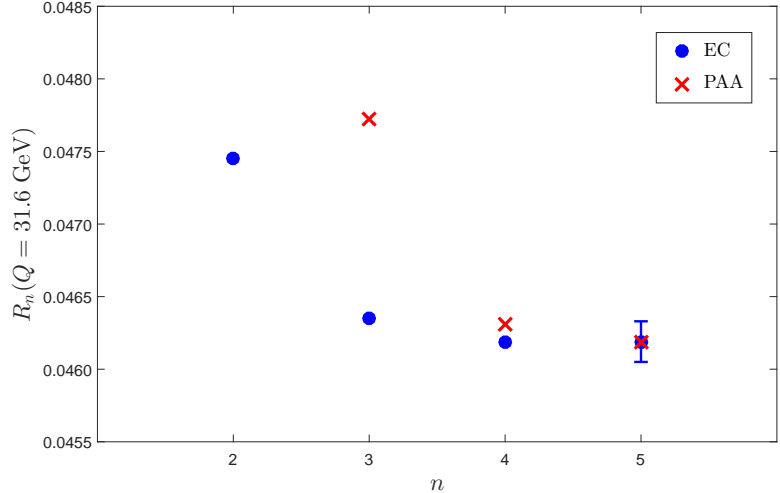,scale=0.6}
\end{minipage}
\begin{minipage}[t]{16.5 cm}
\caption{Comparison of the exact (``EC") and the predicted ([0/n-1]-type ``PAA") pQCD prediction for $R_n(Q=31.6\;{\rm GeV})$ using PMC-s scale-setting~\cite{Du:2018dma}. It shows how the PA predictions change as more loop-terms are included. The five-loop ``${\rm EC'}$" prediction is from Eq.(\ref{five-ee}). \label{Fig:Ree-est}}
\end{minipage}
\end{center}
\end{figure}

\begin{figure}[htb]
\epsfysize=9.0cm
\begin{center}
\begin{minipage}[t]{10 cm}
\epsfig{file=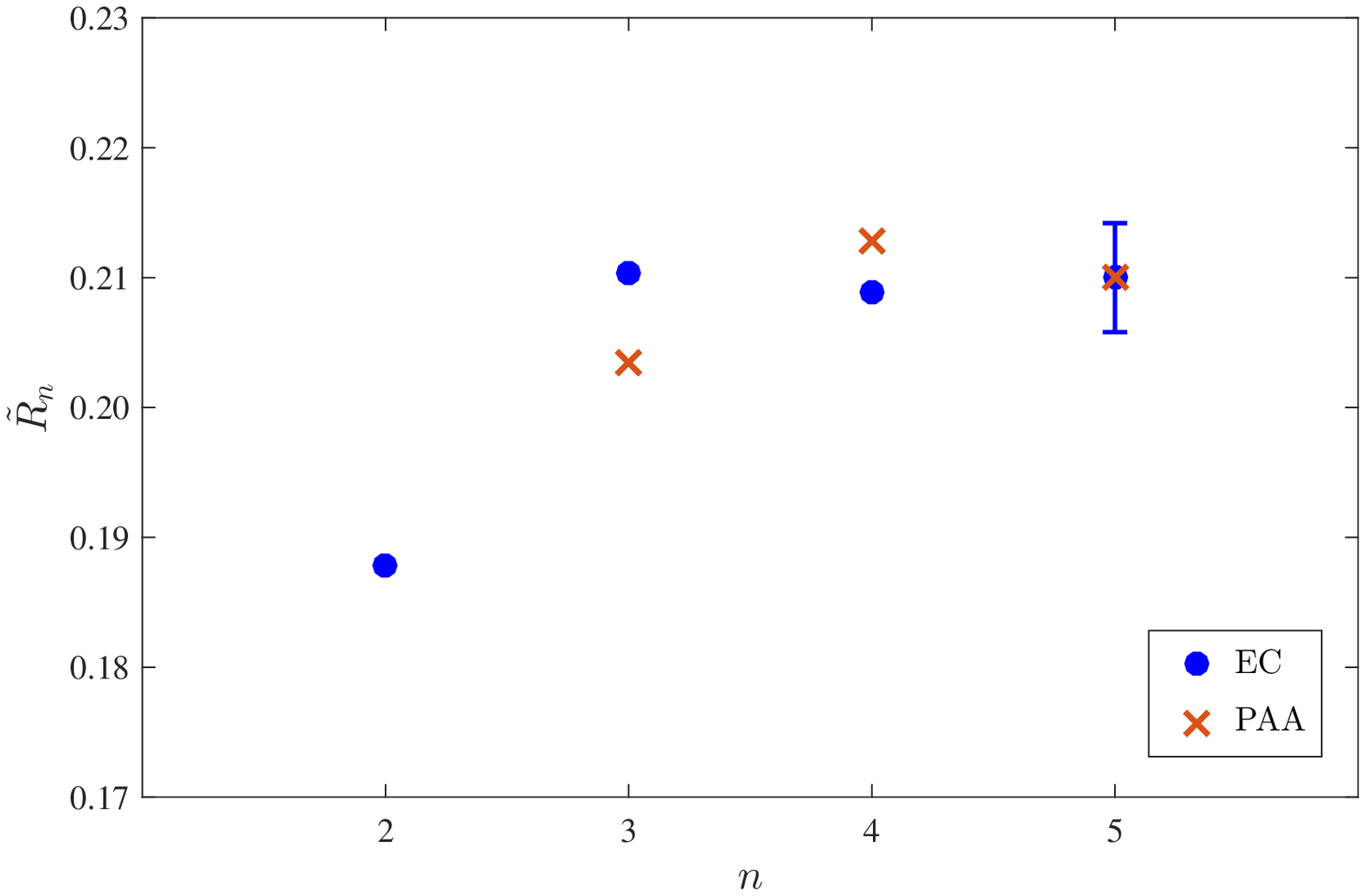,scale=0.6}
\end{minipage}
\begin{minipage}[t]{16.5 cm}
\caption{Comparison of the exact (``EC") and the predicted ([0/n-1]-type ``PAA") pQCD prediction for $\tilde{R}_n$ using PMC-s scale-setting. It shows how the PA predictions change as more loop-terms are included. The five-loop ``${\rm EC'}$" prediction is from Eq.(\ref{five-tau}). \label{Fig:Rtau-est}}
\end{minipage}
\end{center}
\end{figure}

\begin{figure}[htb]
\epsfysize=9.0cm
\begin{center}
\begin{minipage}[t]{10 cm}
\epsfig{file=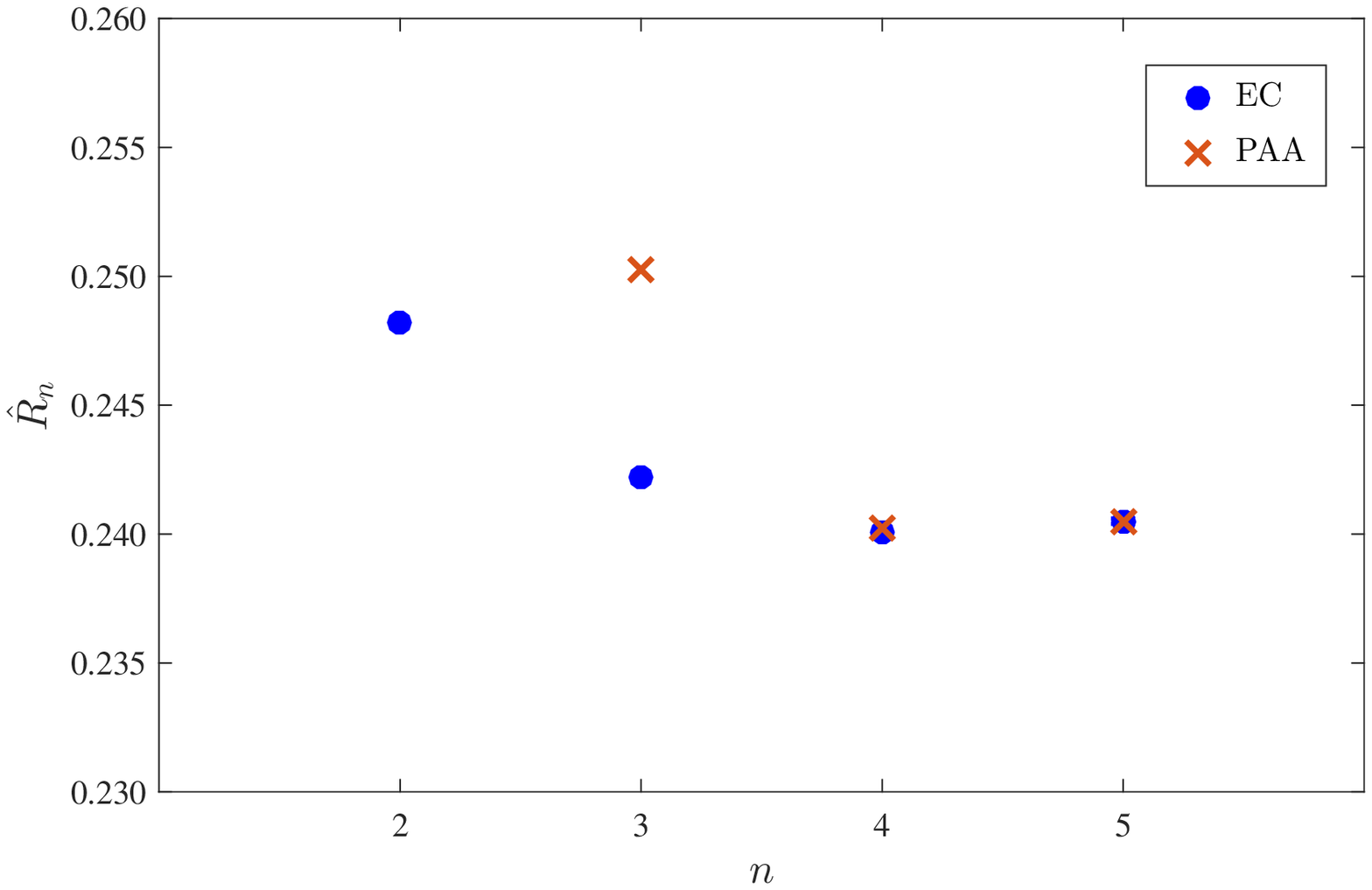,scale=0.6}
\end{minipage}
\begin{minipage}[t]{16.5 cm}
\caption{Comparison of the exact (``EC") and the predicted ([0/n-1]-type ``PAA") pQCD prediction for $\hat{R}_n$ using PMC-s scale-setting. It shows how the PA predictions change as more loop-terms are included. The five-loop ``${\rm EC'}$" prediction is from Eq.(\ref{five-MH}). \label{Fig:MH-est}}
\end{minipage}
\end{center}
\end{figure}

Tables~\ref{estimate-ree}, ~\ref{estimate-rtau} and~\ref{estimate-hbb} show that the difference between the exact and the predicted conformal coefficients at various loops, which decreases rapidly as additional high-order loop terms are included. Table~\ref{Total-observable} shows that the PA approach becomes quantitatively effective even at the NLO level due to the strong $\alpha_s/\pi$-suppression of the conformal series. When using the NLO results $R_2(Q)$, $\tilde{R}_2(M_{\tau})$ and $\hat{R}_2(M_H)$ to predict the N$^2$LO approximants $R_3(Q)$, $\tilde{R}_3(M_{\tau})$ and $\hat{R}_3(M_H)$, the normalized differences between the Pad\'e estimates, and the known results are only about $3\%$.

We show how the PA predictions, $R_n$, $\tilde{R}_n$, $\hat{R}_n$, change when more loop-terms are included in Figure \ref{Fig:Ree-est}, Figure \ref{Fig:Rtau-est}, and Figure \ref{Fig:MH-est}. In these figures, the five-loop ``${\rm EC'}$" predictions are from Eqs.(\ref{five-ee}, \ref{five-tau}, \ref{five-MH}), respectively, which are obtained by setting the normalized difference of the $5_{\rm th}$-loop to match that of the $4_{\rm}$-loop predictions. In some sense the five-loop ``${\rm EC'}$" predictions are infinite-order predictions for those approximants, and they are so far the most precise prediction one can make using the PMC+PA method, given the present knowledge of pQCD.

\section{Summary}
\label{sec:5}

The conventional renormalization scheme-and-scale ambiguities for fixed-order pQCD predictions are caused by the mismatch of the perturbative coefficients and the QCD running coupling at any perturbative order. The elimination of such ambiguities relies heavily on how well we know the precise value and analytic properties of the strong coupling $\alpha_s$. Based on the conventional RGE, an extended RGE has been suggested to determine the $\alpha_s$ scheme-and-scale running behaviors simultaneously. However, those dependences are usually entangled with each other and can only be solved perturbatively or numerically. More recently, a $C$-scheme coupling $\hat \alpha_s$ has been suggested, whose scheme-and-scale running behavior is exactly separated; it satisfies a RGE free of scheme-dependent $\{\beta_{i\ge2}\}$-terms. The $C$-scheme coupling can be matched to a conventional coupling $\alpha_s$ via a proper choice of the parameter $C$. We have demonstrated that the $C$-dependence of the PMC predictions can be eliminated up to any fixed order; since the value of $C$ is arbitrary, it means the PMC prediction is independent of any renormalization scheme. We have illustrated these features for three physical observables $\ree$, $\rtau$ and $\hbb$ which are known up to the four-loop level.

\begin{figure}[htb]
\epsfysize=9.0cm
\begin{center}
\begin{minipage}[t]{10 cm}
\epsfig{file=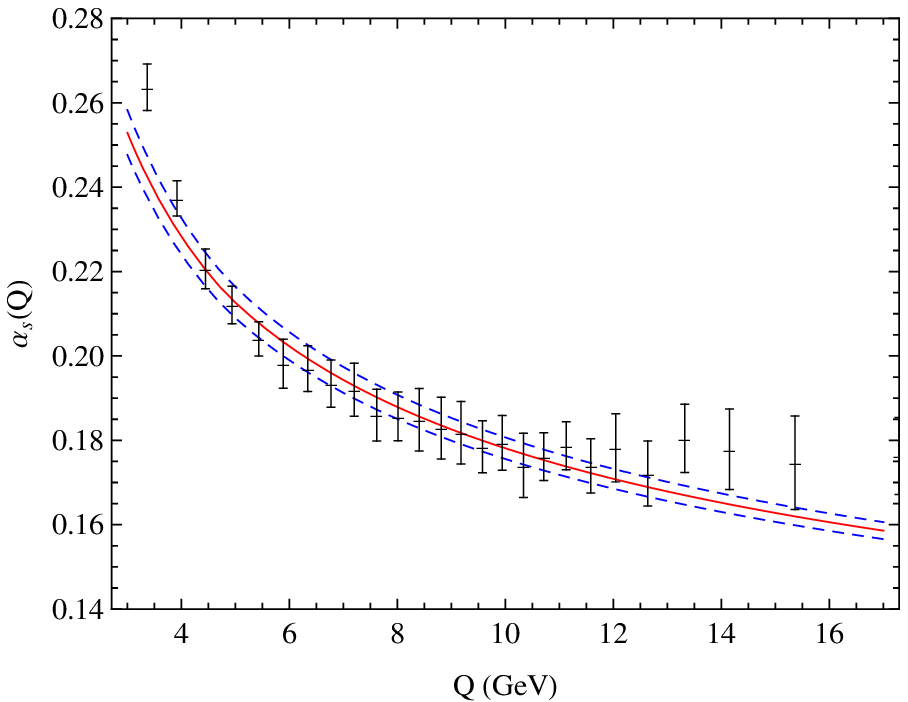,scale=1.0}
\end{minipage}
\begin{minipage}[t]{16.5 cm}
\caption{The extracted $\alpha_s(Q)$ in the $\overline{\rm MS}$-scheme from the comparison of PMC predictions with ALEPH data~\cite{Heister:2003aj}. The error bars are from the experimental data. The three lines are the world average evaluated from $\alpha_s(M_Z)=0.1181\pm0.0011$~\cite{Olive:2016xmw}. }
\label{figasPMCT}
\end{minipage}
\end{center}
\end{figure}

The renormalization scale depends on kinematics such as thrust $(1-T)$ for three jet production via $e^+ e^-$ annihilation. A definitive advantage of using the PMC is that since the PMC scale varies with $(1-T)$, we can extract directly the strong coupling $\alpha_s$ at a wide range of scales using the experimental data at single center-of-mass-energy, $\sqrt{s}=M_Z$. In the case of conventional scale setting, the predictions are scheme-and-scale dependent and do not agree with the precise experimental results; the extracted coupling constants in general deviate from the world average. In contrast, after applying the PMC, we obtain a comprehensive and self-consistent analysis for the thrust variable results including both the differential distributions and the mean values~\cite{Wang:2019ljl}. Using the ALEPH data~\cite{Heister:2003aj}, the extracted $\alpha_s$ are presented in Figure \ref{figasPMCT}. It shows that in the scale range of $3.5$ GeV $<Q<16$ GeV (corresponding ($1-T$) range is $0.05<(1-T)<0.29$), the extracted $\alpha_s$ are in excellent agreement with the world average evaluated from $\alpha_s(M_Z)$.

The PMC provides first-principle predictions for QCD; it satisfies renormalization group invariance and eliminates the conventional renormalization scheme-and-scale ambiguities, greatly improving the precision of tests of the Standard Model and the sensitivity of collider experiments to new physics. Since the perturbative coefficients obtained using the PMC are identical to those of a conformal theory, one can derive all-orders commensurate scale relations between physical observables evaluated at specific relative scales.

Because the divergent renormalon series does not appear in the conformal perturbative series generated by the PMC, there is an opportunity to use resummation procedures such as the PA approach to predict the values of the uncalculated higher-order terms and thus to increase the precision and reliability of pQCD predictions. We have shown that if the PMC prediction for the conformal series for an observable has been determined at order $\alpha^n_s$, then the $[N/M]=[0/n-1]$-type PA series provides an important estimate for the higher-order terms.

\hspace{2cm}

{\bf Acknowledgements}: This work was supported in part by the National Natural Science Foundation of China under Grant No.11625520, No.11847301 and No.11705033, by the Department of Energy Contract No.DEAC02-76SF00515, and by the Fundamental Research Funds for the Central Universities under Grant No.2019CDJDWL0005. SLAC-PUB-17403.

\end{document}